\documentclass[letterpaper,12pt,titlepage,openright,twoside,final]{report}
\pdfoutput=1
\newcommand{\mchapter}[1]{\chapter{#1}\pagestyle{myheadings}  \markboth{#1}{#1}} 
\newcommand{\msection}[1]{\section{#1} \pagestyle{myheadings} \markright{#1}} 

\newcommand{\href}[1]{#1} 

\usepackage{ifthen}
\newboolean{ElectronicVersion}
\setboolean{ElectronicVersion}{true} 

\usepackage{geometry}

\usepackage{makeidx} 
\makeindex

\usepackage{thesis-nomencl}
\usepackage{amsmath,amssymb,amstext,amsthm} 

\usepackage{graphicx}
\usepackage{pgf}

\usepackage[nohug]{diagrams}

\theoremstyle{plain}
\newtheorem{prop}{Proposition}[chapter]
\newtheorem{thm}[prop]{Theorem}
\newtheorem{lem}[prop]{Lemma}
\newtheorem{cor}[prop]{Corollary}
\newtheorem{thmchoi}[prop]{Theorem}

\theoremstyle{definition}
\newtheorem{dfn}{Definition}[chapter]
\newtheorem{qst}{Question}[chapter]

\ifthenelse{\boolean{ElectronicVersion}}{
    \usepackage[letterpaper=true,pdftex,bookmarks,pagebackref,
	plainpages=false, 
        pdfpagelabels=true, 
        pdftitle=Information\ Flow\ at\ the\ Quantum-Classical\ Boundary, 
        pdfauthor=C\'edric\ B\'eny	 
        ]{hyperref}}{}

\let\origdoublepage\cleardoublepage
\newcommand{\clearemptydoublepage}{%
  \clearpage{\pagestyle{empty}\origdoublepage}}
\let\cleardoublepage\clearemptydoublepage

\begin{document}

\pagestyle{empty}
\pagenumbering{roman}

\begin{titlepage}
        \begin{center}
        \vspace*{1.0cm}

        \Huge
        {\bf Information Flow at the Quantum-Classical Boundary}

        \vspace*{1.0cm}

        \normalsize
        by \\

        \vspace*{1.0cm}

        \Large
        C\'edric B\'eny \\

        \vspace*{3.0cm}

        \normalsize
        A thesis \\
        presented to the University of Waterloo \\ 
        in fulfillment of the \\
        thesis requirement for the degree of \\
        Doctor of Philosophy \\
        in \\
        Applied Mathematics \\

        \vspace*{2.0cm}

        Waterloo, Ontario, Canada, 2008 \\

        \vspace*{1.0cm}

        \copyright\ C\'edric B\'eny 2008 \\
        \end{center}
\end{titlepage}

\pagestyle{plain}
\setcounter{page}{2}

\ifthenelse{\boolean{ElectronicVersion}}{
  \noindent

  \bigskip
  
  \noindent
}{
 \noindent

 \smallskip

 \noindent

 \bigskip

 \noindent
}
\newpage


\begin{center}\textbf{Abstract}\end{center}

The theory of decoherence aims to explain how macroscopic quantum objects become effectively classical. Understanding this process could help in the search for the quantum theory underlying gravity, and suggest new schemes for preserving the coherence of technological quantum devices.

The process of decoherence is best understood in terms of information flow within a quantum system, and between the system and its environment. We develop a novel way of characterizing this information, and give a sufficient condition for its classicality. These results generalize previous models of decoherence, clarify the process by which a phase-space based on non-commutative quantum variables can emerge, and provide a possible explanation for the universality of the phenomenon of decoherence. In addition, the tools developed in this approach generalize the theory of quantum error correction to infinite-dimensional Hilbert spaces.

We characterize the nature of the information preserved by a quantum channel by the observables which exist in its image (in the Heisenberg picture). The sharp observables preserved by a channel form an operator algebra which can be characterized in terms of the channel's elements. The effect of the channel on these observables can be reversed by another physical transformation. These results generalize the theory of quantum error correction to codes characterized by arbitrary von Neumann algebras, which can represent hybrid quantum-classical information, continuous variable systems, or certain quantum field theories. 

The preserved unsharp observables (positive operator-valued measures) allow for a finer characterization of the information preserved by a channel. We show that the only type of information which can be duplicated arbitrarily many times consists of coarse-grainings of a single POVM. Based on these results, we propose a model of decoherence which can account for the emergence of a realistic classical phase-space. This model supports the view that the quantum-classical correspondence is given by a quantum-to-classical channel, which is another way of representing a POVM. 

\newpage


\begin{center}\textbf{Acknowledgements}\end{center}

I would like to thank my supervisors Achim Kempf and David Kribs for their help and guidance, as well as my office mates David Campo, William Donnelly, Sasha Gutfraind, Yufang Hao, Rob Martin and Angus Prain for countless discussions and exchange of ideas. 
I also want to thank all the researchers with whom I had discussions related to the material presented in this thesis, including Robin Blume-Kohout and Andreas Winter for providing counter-examples to two hypothesis, Alexandru Nica and Nico Spronk for tips on von Neumann algebras and Raymond Laflamme for introducing me to the concept of noiseless subsystems.
\newpage





\tableofcontents
\newpage

\listoftables
\phantomsection\addcontentsline{toc}{chapter}{\textbf{List of Tables}}
\newpage

\listoffigures
\phantomsection\addcontentsline{toc}{chapter}{\textbf{List of Figures}}
\newpage
 

\pagenumbering{arabic}


\setcounter{page}{0}



\mchapter{Introduction}%



At present, the best of our understanding of the fundamental laws of nature is summarized by the standard model of particle physics, together with the general theory of relativity. 
This understanding is not perfect and fails to account for certain cosmological and astronomical observations.  
In addition, there are purely logical reasons to want to modify general relativity or the standard model. For instance, we do not have a reasonable idea of how the gravitational field, which is modeled classically, should interact with quantum fields. Since all of matter is modeled at the fundamental level by quantum fields, this represents a large hole in our understanding of nature. Experiments are of little help because gravitational effects are extremely weak in the regimes where the quantum nature of fields can be probed.

It is usually assumed on logical grounds that the first necessary step toward a consistent picture of all fundamental interactions would be to replace general relativity by a quantum theory of gravity. 
The problem of quantum gravity can be summarized as the search for a quantum theory which has general relativity as a classical limit.

There do exist several empirical methods to find a ``quantum version'' of a classical theory---a process called {\em \ind{quantization}}---but these methods do not guarantee that the quantum theory obtained will have the classical theory we started from as a classical limit. Instead, quantization methods are most often used as a tool to describe a quantum theory using some form of classical intuition, even if the classical model involved does not relate to any macroscopic phenomenon.

In fact, there is no reason to expect that there exists a well-defined mathematical procedure which would directly yield a quantum theory starting from its classical limit. 
It is the opposite process---namely the derivation of the effective classical limit of a given quantum theory in a given context---which ought to be completely understood. 

Given that, surprisingly, very little is known about the subject,
we may hypothesize that the problem of deriving the classical limit of a quantum theory has never been considered of prime importance in the past. Indeed, there is no real need to be certain that a given quantum theory has the right limit when plenty of experiments are directly testing the candidate quantum theory itself. 
This may explain the usual acceptance of the traditional quantization prescriptions: it does not matter how the quantum theory was arrived at, as long as experiments say it is correct.

However, quantization must be taken more seriously in the context of quantum gravity, since the classical limit is the only prediction that we can use to refute any particular model. 

In this thesis we will study the problem of understanding how a quantum system becomes, or appears, classical. We will do so by generalizing certain results of quantum information processing. Some of these results were published in References \cite{beny07x1, beny07x4, beny08x4}. Our results on decoherence also appeared in Reference \cite{beny08x1}.



\msection{Information, channels and classicality}

Before we explain in more details the problems that we want to solve, let us clarify our use of certain terms which can intuitively have various meanings. 


In Section \ref{section:findimqm} we will define precisely what we mean by {\em quantum} and {\em classical} physical models. This terminology is uncontroversial because we only use the adjectives quantum and classical to qualify a {\em model} and not a physical system itself. For instance the same system could be described with a quantum as well as a classical model. Later in the text we will be less pedantic and simply say ``classical system'' as a shorthand for ``system modeled classically''. If we say that physical systems are fundamentally quantum, we mean that at the microscopic level they are best described by a quantum model. 
Our goal will be to understand how the classical systems that we are familiar with are related to the underlying quantum description of the same physics. Of course, one could find a contrived classical model which is as complete as the quantum description (say Bohm's theory), however, even disregarding the multitude of intrinsic problems associated with such a theory, one would still have to explain how to derive observed classical phenomena as a limiting case.
This is not the path that we are going to follow. We will in fact obtain reasonable conditions under which an effective classical model emerges which is related to the underlying quantum model in a non-contextual way (see Section \ref{section:classicalset}).

 A communication channel, or {\em channel} for short, will be understood as the most general way that the state of a system $B$ can be conditioned on that of a system $A$. If  $A$ and $B$ are both quantum or both classical, this corresponds to the usual notion of a quantum channel (Section \ref{section:channel}) or classical channel (a stochastic map) respectively.

We will use the word ``information'' essentially as a synonym of ``correlation''. Hence two different systems will always be involved in a sentence containing this word. For instance we could say that a system $B$ {\em contains information about} a system $A$. If we possess a joint state for $A$ and $B$, this would mean that some observable of $A$ is correlated with some observable of $B$. More importantly, in the context where the state of $B$ is a function of the state of $A$, i.e. if they are related by a {\em channel}, we will say that $B$ contains information about $A$ if, by performing a measurement on $B$, we learn something also about $A$. More precisely, we will characterize the information that $B$ possesses about $A$ by checking exactly {\em which property} (which observable) we indirectly measure on $A$ when performing a given measurement on $B$. This is the concept that we will analyze in Chapter \ref{chapter:preserved}. 

This qualitative information can be {\em quantified} in certain contexts using generalizations of Shannon's theory (as for instance in Section \ref{section:obsinfo}). However, we will be more interested in certain discrete aspects of the structure of this information, especially whether or not it is {\em classical} (Section \ref{section:classicalset}).


\msection{The emergence of classicality}
\label{section:intro:classicality}


It is widely believed that everything in nature is fundamentally quantum. Hence we need to explain why, and how, most of it actually appears classical in a wide range of situations. 
At first glance, the answer may seem obvious: things appear classical when we do not measure their behaviour accurately enough (compared to $\hbar$) to detect their quantumness. The main problem with this answer is that it does not tell us specifically which observable of the quantum system is being measured when such an ``inaccurate'' observation is performed. Indeed, we know that there are many fundamentally incompatible ways of observing a quantum system. Each way to do it is labeled by an observable. By contrast, there is a single best way to observe a classical system: it suffices to measure its precise state. Everything else can be deduced from it.
We will therefore focus on the following question:
\begin{qst}
\label{question}
Which quantum observable is being measured in the complete observation of a classical system?
\end{qst}
This way of formulating the question of the classical limit is not standard, and, although it is rather natural, it requires some detailed explanations, which are the subject of Section \ref{section:limit}. 



The first thing to note, which might be a source of confusion for the reader, is that there is no need to assume that an observable only takes a single real number as its value. Although this assumption is rather innocent for a classical system, where a non-scalar observable can usually be seen a the joint observation  of several scalar observables, it would be a strong conceptual limitation in the context of quantum mechanics where, as we know, observables cannot in general be measured jointly. This point is to be made in order to make it clear that measuring the state of a classical system amounts to observing one very special observable, namely the observable which gives full information about the classical system. Indeed the state-space of a classical system (i.e. its phase-space) is rarely one-dimensional. Since any observation of a quantum system is specified by an observable, and since every classical system is a quantum system in disguise, this phase-space observation must actually correspond to the observation of a single quantum observable. Hence, our question: which quantum observable defines this classical system?

In order to provide a realistic answer to this question we will need to consider the most general notion of an observable, which is provided by the formalism of Positive-Operator Valued Measures (POVM). For instance, we know that, in general, the various variables which parameterize a classical phase-space correspond to non-commuting quantum observables. This implies that the quantum description of phase-space must require some approximate joint measurement of non-commuting observables, something which is readily characterized by a POVM. Hence we will make extensive use of this notion, although we will specialize it slightly for technical and conceptual reasons (see Chapter \ref{chapter:quantum}). 

Let us remark that there is another important related question, which we will not address here. A system can behave classically, not with respect to any observation, but indirectly through the way that it interacts with an auxiliary quantum system. We are familiar with the description of a quantum system, say an electron, interacting with a system which is assumed classical, for instance the scalar potential of the electromagnetic field. We know that the electromagnetic field is in fact quantum, therefore we must understand why it can be treated classically when we are concerned with the behaviour of that electron. This example involves no act of measurement, and is pertinent to the state of our knowledge of the interaction between quantum fields and gravity, where gravity plays the role of the classical electromagnetic field. 



\msection{Decoherence}
\label{section:intro:decoherence}
\index{decoherence}

Any definition of a quantum measurement refers, explicitly or implicitly, to an interaction between the quantum system and a postulated external classical system (the observer). Note that the standard framework of quantum mechanics will be recovered explicitly in this way in Chapter \ref{chapter:quantum}. However, a quantum system interacting with other quantum systems can also undergo at least part of the phenomenon of collapse associated with a traditional axiomatic measurement. This is the experimentally ubiquitous process of {\em decoherence} characterized, in its simpler form, by an evolution of the form
\begin{equation}
\label{equ:elementarydecoherence}
\rho \mapsto \sum_i \proj{i} \rho \proj{i}.
\end{equation}
for some basis $\ket{i}$, indexed by the integer $i$. 

 
Based on this phenomenon, the {\em \ind{theory of decoherence}} \cite{zurek81, joos85, paz93, giulini96, zurek03, ollivier04, blume-kohout05} 
provides an answer to Question \ref{question}. The premise is that nothing really important happens during the act of observation itself. At the classical level, an observation only involves gathering information which is already present in the environment. The idea is that a system which appears classical is never isolated from the rest of the world---the relevant part of which we call the {\em \ind{environment}}. Uncontrolled interactions constantly spread information about the state of the system into the environment. It is this process itself which selects which observable of the system the environment contains information about. The conscious act of observation then simply amounts to the observer becoming aware of this information. For instance, any macroscopic object constantly bounces into, or emits, photons. Simply gathering a few of these photons certainly has no consequence on the system, but may yield information about the position of the object at the time of the interaction with the photons.

By contrast, what makes an act of observation {\em quantum} is the observer's ability to choose the nature of the interaction between his measurement apparatus and the quantum system,  so as to decide which observable will be measured. Hence, for a system to appear classical, two conditions must be fulfilled: It must be continuously ``monitored'' by its environment, and the observer must learn about it only indirectly through the information which is already contained in the environment.


Although the theory of decoherence has been used to tackle the infamous {\em measurement problem} related to the interpretation of quantum mechanics, we are here interested only in its ability to answer Question \ref{question}, i.e. to tell us which observable of a quantum system defines its classical limit in a given context. 
 
 
Let us briefly summarize this theory here. In elementary presentations of decoherence \cite{giulini96}, 
one considers an open quantum evolution which essentially removes the off-diagonal elements of the density matrix $\rho$ representing a quantum state, as expressed in Equation \ref{equ:elementarydecoherence}.
The basis involved in this expression is selected by the dynamical process itself. Indeed, the initial states of the form $\rho = \proj{i}$ are the only states which stay pure throughout the interaction. They are called {\em \ind{pointer states}}. 
In addition, if this process results from a unitary interaction with an auxiliary quantum system, the joint final state of the two systems must be of the form
\[
U (\ket{\psi} \otimes \ket{0}) = \sum_i  \braket{i}{\psi} \ket{i} \otimes \ket{\phi_i}
\]
where $\ket{\psi}$ is the initial state of the system, $\ket{0}$ the initial state of the auxiliary system (the environment), $\ket{\phi_i}$ some orthonormal basis of the auxiliary system, and $U$ the unitary operator specifying the joint evolution for some interval of time. Such a process has been initially studied by von Neumann \cite{neumann55} as an example of how a measurement apparatus can be modeled by a quantum system. Here the measurement apparatus is the auxiliary system, or environment. This process represents a measurement of the system by the environment in the sense that a further measurement (now in the axiomatic sense) of the environment in the basis $\ket{\phi_i}$ effectively simulates a measurement of an observable of the system with eigenstates $\ket{i}$. 
 Indeed, this measurement yields the outcome labelled by $i$ with probability $|\braket{i}{\psi}|^2$. The state collapses to $\ket{i} \otimes \ket{\psi_i}$, which reduces to $\ket{i}$ when we neglect the environment. 
The observable being measured is fixed only up to its eigenvalues, because we are only interested in the probabilities that it defines. In general we will simply consider a fiducial observable $A = \sum_i \lambda_i \proj{i}$ for some distinct set of eigenvalues $\lambda_i$ which only act as labels for the eigenvectors. We will say that the observable $A$ is the {\em \ind{pointer observable}}. 

This is an example of a purely unitary (and hence purely quantum) process which selects a particular observable of the system, as specified by the orthogonal basis $\ket{i}$, and transfers the information about this observable to the environment. 

Note that the evolution of the system alone is such that any quantum superposition between the states $\ket{i}$ is replaced by a simple statistical mixture. Indeed if the initial state is $\rho = \proj{\psi}$ where $\ket{\psi} = \sum_i \alpha_i \ket{i} $, then the final state of the system is
\begin{equation}
\sum_i \proj{i} \rho \proj{i} = \sum_i |\alpha_i|^2 \proj{i}.
\end{equation}

Although this model introduces some essential ideas, it is too simplistic to be applied to realistic systems. Its main problem is that it can only explain the selection of sharp observables, i.e. traditional observables, and not POVMs, which, as we will argue in more detail in Section \ref{section:limit}, are needed for an understanding of the classical limit of the most common physical systems. 
In addition, this model does not explain why classicality is universal at macroscopic scales, since processes of the above kind are rather special.  
Finally, in this model, the effective classical system represented by the pointer observable does not have any non-trivial dynamics.

These points have all been addressed before (see for instance \cite{giulini96, zurek03}, or the review section in \cite{braun01}). Most of these approaches consider a single non-relativistic particle coupled via its position operator to a thermal bath of harmonic oscillators. One often assumes a Markovian approximation, which allows for the dynamics of the system to be represented by a differential equation in time. This is important as it removes the need for an explicit solution of the dynamics. In this picture, decoherence is always expected to take place in the position basis, due to the local nature of the interaction.
Although such analyzes have shed important light on many aspects of decoherence, they have important limitations. For instance, they provide unsatisfactory, or limited answers to the following general questions. 
If the decoherence takes place in the position basis, what mechanism guarantees the---seemingly contradictory---preservation of momentum information? What singles out momentum as the other phase-space variable?
How would we recognize an interaction which yields the emergence of a non-canonical phase-space structure?
Does this process justify one quantization process over another?
How is this loss of information in the system related to a gain of information in the environement, i.e. a measurement?
Can some quantum information be preserved in the system despite the process of decoherence?






Instead of attempting to improve on these models, we will adopt a new approach which generalizes in a different way the simple model that we first described. We will make no specific physical assumption. Instead we will show that the process of decoherence can be understood at a very fundamental level, in terms of information flow between the system and its environment. 

We can easily generalize von Neumann's analysis of the process of measurement and note that any interaction between a quantum system and another quantum system---the environment---yields correlations which are such that a subsequent measurement of a particular observable of the environment exactly simulates a measurement of some observable of the system. The observable of the system which is indirectly measured is selected by the nature of the interaction between the two systems. Mathematically, if the initial state of the system is $\rho$ and that of the environment $\ket{0}$, and the interaction is specified by the unitary operator $U$, then the final states of the two system is 
\(
U (\rho \otimes \proj{0}) U^\dagger. 
\)
A measurement of a fixed observable with eigenstates $\ket{i}$ on the environment after the interaction yields the probabilities 
\[
\begin{split}
p_i &= \tr((\one \otimes \proj{i})\,U\, (\rho \otimes \proj{0}) \,U^\dagger)\\
&= \tr(U^\dagger \, (\one \otimes \proj{i}) \,U \,(\rho \otimes \proj{0}))\\
&= \tr((\one \otimes \bra{0}) \,U^\dagger\, (\one \otimes \proj{i}) \,U\, (\one \otimes \ket{0}) \rho )\\
&= \tr( A_i \, \rho ).
\end{split}
\] 
where we have only used the cyclicity of the trace, and defined the operators 
\[
A_i := (\one \otimes \bra{0}) \,U^\dagger\, (\one \otimes \proj{i}) \,U\, (\one \otimes \ket{0})
\]
These operators specify a generalized observable on the system, namely a discrete POVM.
Remember that we are trying to answer Question \ref{question}. 
The point of this simple calculation is that the observable represented by the operators $A_i$, which is effectively measured on the system, is fixed by the choice of the unitary operator $U$, and hence by the physical interaction between the two systems. 

The problem is that the POVM which is selected in this process depends also on the observable being measured on the environment. In the special case where the evolution of the system alone is given by Equation \ref{equ:elementarydecoherence}, there is a unique best observable which can be measured via the environment. ``Best'' in the sense that any other observation will simply yield less information about the same observable of the system (a concept which will be made precise in Section \ref{section:classicalset}). However, this is not true of the more general case considered here, as can be seen by changing the basis $\ket{i}$ of the environment in the above example. 
Therefore, the observable being measured in this process still depends on the good will of an observer. The question is only shifted to another system: what compelled the observer to choose this particular observable of the environment?

In fact, we will show in Chapter \ref{chapter:preserved} that there is only one {\em sharp} observable (possibly trivial) which can be measured {\em nondestructively} in any such process. This is the reason why a particular observable is indeed selected in the simpler example defined by Equation \ref{equ:elementarydecoherence}. 
However, these requirements do not uniquely select a general POVM. We will see that for a POVM to be selected uniquely we need to introduce another requirement, namely that the information about the observable be represented redundantly in the environment. 


Redundancy was previously introduced by Ollivier {\it et al.} \cite{ollivier05} 
in a slightly different context. The authors attempted to identify the preferred observable, not via the full dynamical process as above, but only by the form of the final state of the joint system and environment after the interaction. 
The idea is that the preferred observable is that which is correlated with some observable of the environment. They indeed  showed that a sharp observable of the system is uniquely selected if, in addition, these correlations are required to be redundant across many subsystems of the environment. 
Our approach, although it has a different premise, will indeed yield such correlations, hence accounting for their analysis. 

Our approach is based on a new method for characterizing the information preserved in a quantum communication channel (see Chapter \ref{chapter:preserved}). 
The interaction between the system and its environment, for instance, involves two main channels. The first sends the initial state of the system to its final state. Equation \ref{equ:elementarydecoherence} is an example of such a channel. The concept of pointer state and its generalization \cite{zurek92} is one way to characterize the information that this channel preserves. The second important channel is the one which says how the final state of the environment depends on the initial state of the system. This channel has rarely been considered explicitly. However, it is important because it tells us precisely what information the environment learns about the system, and therefore what is being measured in the process.







\msection{Fighting decoherence} 
\index{decoherence}

We motivated the study of decoherence by the need to understand quantization. But understanding decoherence is also important for a completely different purpose: that of fighting it. 

Most introductory textbooks give the impression that quantum mechanics imposes fundamental barriers on what it is possible to do with a physical system. However, the restrictions introduced with quantum mechanics are restrictions on what we can measure, or predict, but they are not restrictions on what a physical system can {\em do}. In fact, these additional constraints on what we can observe allow for a greater freedom in the system's behaviour compared to classical physics. Indeed, quantum systems can behave in ways that were previously thought impossible~\cite{bennett95}. For instance a computer based on the principles of quantum mechanics could solve problems which are classically intractable \cite{deutsch85, shor94}. The exact reason is not clear, although it can be understood intuitively from the fact that a quantum dynamical process seems to explore an infinite number of paths in parallel, in the sense of the path-integral formalism. In fact, a naive attempt at simulating the discrete evolution of a small finite quantum system rapidly yields an exponential blow-up of the memory required to store the classical representation of the quantum state, as well as an exponential slow-down of the simulation.  

If we want to exploit the properties of quantum systems for some practical purpose, we have to make sure that we can maintain their quantumness, and prevent them from undergoing decoherence. This is far from being an easy task given that decoherence is present in every quantum experimental setups, and is typically much faster than thermalization \cite{giulini96}. 


One way to fight decoherence, or any other form of noise, is to simply avoid it. If some form of noise cannot be eliminated by active control of the environment, one may attempt to encode information in a way which will be resistant to this residual noise. For instance, one may be able to find a subspace which is not affected by the noise (i.e. a \ind{decoherence-free subspace}) \cite{zanardi97, palma96, duan97, lidar98}. It is possible to refine this idea and find so-called {\em \ind{noiseless subsystems}} \cite{knill00, zanardi00, kempe01}. A more elaborate strategy is to find some subspaces, or subsystems, which are affected by the noise, but in a way which can be reversed. This yields to the method of {\em quantum error correction} \cite{bennett96,knill97,shor95, steane96,gottesman96}, or to its refinement which allows for subsystem codes (operator quantum error correction) \cite{kribs05, kribs06}.

Clearly these techniques characterize the information which is preserved by the quantum channel representing the noise. For instance, a subspace represents a property of the quantum system (a yes-no observable). If the effect of the noise can be reversed on this subspace, it means in particular that the information represented by this observable has not been destroyed by the noise. 
This is precisely the kind of tool that we are looking for in order to understand the process of decoherence. In Chapter \ref{chapter:preserved}, we will introduce a concept of preserved information which is more general, and better suited for our needs. In Chapter \ref{chapter:qec} we will show how our approach yields, as a special case, a generalization of the theory of quantum error correction for subsystem codes \cite{beny07x1, beny07x4}, which itself generalizes the other theories mentioned above. Importantly, our approach is suitable for a study of error correction on infinite-dimensional Hilbert spaces. In this context, we show that error correcting codes can be represented by any von Neumann algebras. Whereas a general von Neumann algebra represents hybrid quantum-classical information, type II and type III factors represent new exotic types of quantum codes which exist only on infinite-dimensional Hilbert spaces. 


\msection{Summary}

Let us summarize the problem at hand, and our reasoning. We want to understand by what process a quantum system becomes effectively classical, with a focus on the kinematical aspect of the problem. 
We have pointed out in Section \ref{section:intro:classicality}, and will argue in more detail in Section \ref{section:limit}, that a classical limit, at the kinematical level (i.e. everything except the dynamics), must be specified by a generalized observable of the quantum theory. Therefore we have identified the type of object that we are looking for: a POVM. 

We also have a good idea of the type of mechanism which could be at work to select this POVM. The theory of decoherence, which is firmly grounded in theory and experiments, tells us that this must essentially be a measurement of the system by its environment. An observable is indeed what characterizes the information gathered in a measurement. 

However, we need to understand what constitutes a measurement of a quantum system by another quantum system. We expect to be able to do so by studying the nature of the information transmitted from one system to the other.




\mchapter{Quantum theory}%
\label{chapter:quantum}


The role of this chapter is to introduce a number of basic concepts, as well as a certain point of view on quantum mechanics which will be referred to in the rest of this thesis. Our presentation is standard in that it follows, at least at the formal level, the presentation given in Kraus' lecture ``States, Effects and Operations'' \cite{kraus83}, which originates from \cite{ludwig83}. The main difference stems in the fact that we are using a Bayesian interpretation of probabilities rather than a frequentist one. 


The expressions ``quantum mechanics'' and ``quantum theory'' will be used interchangeably, just like the adjectives ``quantum mechanical'' and ``quantum''.  Sometimes, by ``quantum mechanics'' one refers only to Schr\"odinger's model of a non-relativistic quantum particle, so as to distinguish it from relativistic quantum field theory. Here we refer to quantum mechanics in a much more general sense, as in the ``axioms of quantum mechanics''. For example, quantum field theory is an instance of a quantum mechanical theory, or at least tend be so (it is not fully formalized). Quantum theory is concerned with the common aspects of all quantum systems. 






In order to present quantum theory, we will take the view that the primary elements of a physical theory are yes-no questions, i.e. questions with only two possible answers, also called {\em \ind{proposition}s} in the field of logic, or {\em \ind{effects}} in quantum foundations.  We will mostly use the term ``effect'' once we got used to them. However, in the beginning we will call them ``propositions'' in order to remind the reader of their interpretation.
Hence, we will attach a set of propositions \ind{$\aprops$} to any physical system of interest. 

Since we are going to have to deal with quantum mechanics in which uncertainties are unavoidable, even in principle, we will not assume that all questions can always be given an exact answer. Instead we assume that a typical answer will be a degree of confidence between zero and one in the truth of the corresponding proposition, i.e. a \ind{probability}.  Hence a certain \ind{state} of knowledge about the system is an assignment of a probability to each proposition:
\[
\rho: \aprops \rightarrow [0,1].
\]
Since all predictions of the theory are probabilities of the form $\rho(\alpha)$, we will assume that two propositions $\alpha$ and $\beta$ are identical if $\rho(\alpha) = \rho(\beta)$ for all states $\rho$.

  
\msection{Finite-dimensional quantum theory}
\label{section:findimqm}

This section is dedicated to an exposition of finite-dimensional quantum mechanics. This will allow us to introduce most concepts which we will refer to later, while avoiding the technical issues related to the full infinite-dimensional quantum theory, which will be covered in Section \ref{section:formalism}.
 This presentation is not meant to be an introduction to the subject, but rather a way to present a number of concepts which will be used later. Hence we assume that the reader is already familiar with the basic formalism of quantum theory, including the notion of density matrix. 


For a quantum system associated with a finite dimensional Hilbert space $\Hil$, the propositions, which are also called {\em \ind{effects}}, will be assumed to be all the self-adjoint operators with eigenvalues in the interval $[0,1]$:
\[
\qprops \Hil = \{A \in \finops(\Hil) \;|\; 0 \le A \le \one\}
\]
\index{$\qprops \Hil$}
Were we wrote \ind{$\finops(\Hil)$} for the set of all linear operators on $\Hil$. 


The states for this system are the functionals defined by density matrices $\rho$ through 
\[
A \mapsto \tr(\rho A),
\]
for any operator $A \in \qprops \Hil$. 
The set of states therefore is associated with the set of density matrices \index{$\states\Hil$}
\[
\states \Hil = \{\rho \in \finops(\Hil) \;|\; \rho > 0 \text{ and } \tr(\rho) = 1\}
\]

We will say more about the meaning of the mathematical structures which exist in the sets $\qprops \Hil$ and $\qstates \Hil$ in Section \ref{section:effects}.
For now we will assume that they are just given. The purpose of this section is to show that the rest of quantum mechanics mostly follows from interpreting these mathematical objects as propositions and states.

This way of looking at quantum mechanics, although it has a venerable history \cite{ludwig83}, is not widely known. Traditionally one sees an observable as a self-adjoint operator $X$ and postulate that its spectral decomposition $X = \sum_{i=1}^n x_i P_i$ must be used to compute the probabilities $\tr(\rho P_i)$ for each outcome $i=1,\dots,n$. Here, $P_i$ is the projector on the eigenspace of $X$ with eigenvalue $x_i$. Clearly, the projectors $P_i$ are {\em effects}, as defined above, and the probabilities are obtained by the expected formula $\tr(\rho P_i)$. More generally one can introduce the concept of a Positive Operator Valued Measure, or POVM\index{positive operator valued measure}\index{POVM}, which generalizes this traditional concept of observable. Indeed, if we let a quantum system interact with an auxiliary system, and then measure a ``traditional'' observable on that auxiliary system, we obtain probabilities whose dependence on the state $\rho$ of the system is given by the formula $\tr(\rho A_i)$ where the operators $A_i$ need not be projectors. Indeed, suppose that the initial state of the auxiliary system is $\ket{\psi}$, and that the common evolution of the two systems, until we perform the measurement, is given by the unitary time-evolution operator $U$. If the eigen-projectors of the observable measured on the auxiliary system are $P_i$, and if the initial state of the system is $\rho$, then the corresponding probabilities are given by $p_i = \tr(U (\rho \otimes  \proj{\psi}) U^\dagger (\one \otimes  P_i))$, where the first tensor factor corresponds to the system and the second to the auxiliary system. Using the cyclicity of the trace, we easily see that this can be written as $p_i = \tr(\rho A_i)$ where 
\[
A_i := (\one \otimes  \bra{\psi}) (U^\dagger (\one \otimes P_i) U) (\one \otimes  \ket{\psi}).
\]
In this notation, $\bra{\psi}$ is viewed as the linear operator mapping a state $\ket{\phi}$ to the number $\braket{\psi}{\phi}$. This operator, which in finite-dimension can be seen as a matrix of just one row, can be legally tensored with other operators. For instance, $A \otimes \bra{\psi}$ maps a state $\ket{\phi} \otimes  \ket{\phi'}$ of the two systems into a state of the first system, namely $A \ket{\phi} \braket{\psi}{\phi'}$. The ket $\ket{\psi}$ is  the adjoint of $\bra{\psi}$ (a matrix of just one column). It maps a complex number $z$ to the ket $z \ket{\psi}$. Hence, for instance, $\one \otimes \ket{\psi}$ maps a state $\ket{\phi}$ of the system to the composite state $\ket{\phi} \otimes  \ket{\psi}$. To see this, we interpret $\ket{\phi}$ as $\ket{\phi} \otimes  1$, where $1$ is the complex number one. The point is that $\ket{\phi} \otimes  z = z \ket{\phi} \otimes 1$ for any $z$, which means that tensoring with a complex number does not add any more information: it can be done freely, and simply corresponds to a different way of representing the same Hilbert space.

In general these operators $A_i$ representing the discrete  POVM must only be self-adjoint, positive (i.e. with positive eigenvalues), and such that $\sum_i A_i = \one$. This normalization condition in fact implies that the eigenvalues of $A_i$ must also be smaller than $1$, which is what we need to make sure that $\tr(\rho A_i)$ is a probability between $0$ and $1$. These operators $A_i$ are called POVM elements, or effects, and are indeed the operators in our set $\qprops \Hil$. Effects $P \in \qprops \Hil$ which are projectors: $P^2 = P$ will be called {\em sharp}\index{sharp effect}. Effects which are not sharp will be called {\em unsharp}\index{unsharp effect}. 

Instead of referring to the standard notion of observable and POVM, let us derive these concepts from scratch starting with the axioms that propositions are given by the set $\qprops \Hil$ and states by functionals of the form $A \mapsto \tr(\rho A)$. This exercise will allow us to introduce the point of view that an observable is essentially a communication channel from a quantum to a classical system. 


Before we can define precisely what we mean by that, we first need to explain how to model a classical system also in terms of effects and states.

  
\subsection{Classical models}
\label{section:classical}

A classical model is always associated with a phase-space\index{phase-space} \ind{$\Omega$}. For the moment we will ignore any mathematical structure that one may put on a phase-space, like the canonical Poisson structure of mechanical systems. Indeed, we will consider contexts where $\Omega$ would not have any particular structure, for instance in the case where it represents the internal state of a computer, and is therefore discrete. Note that it is not usual to call $\Omega$ a phase-space when it represents a discrete system, however it should be clear that even in that context it plays the same role as that of a mechanical phase-space. Specifically, any element $x \in \Omega$ specifies a complete knowledge about the classical system, a concept which can be understood intuitively, but which will be made precise once we say what the effects and states are.

 Let us assume for simplicity that $\Omega$ is a finite set. The set of propositions \ind{$\cprops \Omega$} will be the set of {\em all} functions $\alpha: \Omega \rightarrow [0,1]$. These functions can be seen simply as vectors with components $0 \le \alpha_i \le 1$ indexed by $i \in \Omega$. The interpretation is the following: if the system is in ``state'' $i \in \Omega$, then the proposition $\alpha$ is true with probability $\alpha_i$. In particular, if the components $\alpha_i$ are either $0$ or $1$ then it selects a subset $\omega \subseteq \Omega$ containing only those ``states'' $i \in \Omega$ which make the proposition $\alpha$ true. This corresponds to the standard notion of a logical proposition. As in the quantum case, these special propositions will be called {\em sharp}. The constant function with value $1$ will be denoted by $\one$ since it plays the same role as the identity operator in the quantum case. Indeed it is the proposition which is always true.

In the quantum case, we referred to the matrix algebra $\finops(\Hil)$ which is the linear span of the set of propositions $\qprops \Hil$. 
Since we will want to consider maps between classical and quantum propositions, it is convenient to consider the equivalent of $\finops(\Hil)$ in the classical case, namely the complex linear span of $\cprops \Omega$. It consists simply of all complex functions on $\Omega$ and will be denoted by \ind{$\finfun(\Omega)$} for the case where $\Omega$ is finite. $\finfun(\Omega)$ has more in common with $\finops(\Hil)$ than the fact of being a complex vector space: it is also an algebra in terms of the standard product of functions. The most important difference with the set of operators $\finops(\Hil)$ is that the product on $\finfun(\Omega)$ is commutative\index{commutative product}. Using this product we can see that what we called classical sharp propositions are also the only functions $\alpha$ which satisfy $\alpha^2 = \alpha$, just like in the quantum case. One can also introduce the concept of an adjoint. The adjoint $f^*$ of a function $f$ is simply its complex conjugate $\overline f$, so that the real functions are also the self-adjoint functions. For instance, the components of an effect $\alpha$ satisfy $(\alpha^*)_i = \overline \alpha_i = \alpha_i$. We see that the range of a function $\alpha \in \finfun(\Omega)$ plays the same role as the set of eigenvalues of an operator $A \in \finops(\Hil)$. Indeed the classical propositions are the self-adjoint, i.e. real, functions with range inside $[0,1]$. In fact this algebra structure together with the operation of taking the adjoint (``star'' operation \index{star operation}), makes $\finfun(\Omega)$ into the mathematical structure called a \ind{C$^*$-algebra}, which captures precisely that which is common between quantum and classical systems. Indeed $\finops(\Hil)$ is a C$^*$-algebra as well. We will see that the systems represented by other types of C$^*$-algebras are in a sense intermediate between being quantum and classical systems, and will come up naturally once we consider the many ways that quantum information can be degraded by noise. 

Classical states are of course represented by probability distributions on $\Omega$, i.e. functions $\mu: \Omega \rightarrow [0,1]$ which are normalized by the condition $\sum_{i \in \Omega} \mu_i = 1$. A probability distribution $\mu$ represents the functional on proposition defined by $\alpha \mapsto \sum_i \alpha_i \mu_i$. Just as in the quantum case, the states $\states\Omega$ can be represented by positive elements of the algebra $\finfun(\Omega)$: \index{$\states\Omega$}
\[
\states \Omega = \{ \mu \in \finfun(\Omega) \;|\; \mu > 0 \text{ and } \sum_{i \in \Omega} \mu_i = 1 \}.
\]

In fact, a classical system can also be represented as a quantum system with some constraints on the states that it can take. Indeed, if $\Omega$ contains $n$ elements, the algebra $\finfun(\Omega)$ can be viewed as the set of all diagonal complex $n$-by-$n$ matrices. The algebra operations, including the star operation (adjoint), are then exactly the quantum ones. In this representation, using the bra-ket notation of quantum mechanics, we can write classical propositions as $\alpha = \sum_{i\in\Omega} \alpha_i \proj{i}$. This is in fact the way that one often deal with classical systems in quantum information theory (for instance this is what is done in \cite{nielsen00}). This notation has the advantage of directly allowing one to deal with the interaction between coexistent quantum and classical systems while working entirely within the formalism of quantum theory.

  


\subsection{Observables as quantum-to-classical maps}
\label{section:observables}
 
\index{observable}
In general we assume that we know how to observe a classical system. This means that if a classical system characterized by the phase-space $\Omega$ is given, it is assumed by default that we are also given an interpretational framework which tells us what the propositions $\cprops \Omega$ actually mean in terms of possible experimental devices. Said differently, we assume that we would know how to conceive experiments which would test the propositions represented by the effect in $\cprops \Omega$. 

In addition, it seems reasonable to say that we only directly observe classical objects, i.e. objects which are understood and modeled in classical terms, rather than quantum objects. This means that if we want to make any kind of observation of a quantum system, we need to somehow couple it to a measurement apparatus, a part of which is a classical system that we directly observe. This classical part of the measurement apparatus which is pertinent in giving us the outcome of the experiment will be called the {\em pointer}. Hence we assume that we can answer any question (i.e. proposition) about the state of the measurement apparatus' classical pointer, and that the answers to these questions somehow convey information about propositions of the quantum systems.  Let $\Omega$ be the phase space for the classical pointer, and $\Hil$ the Hilbert space of the quantum system under observation. We will not need to consider the details of how these two systems might interact. Instead we consider the most general way that propositions about $\Omega$ can be translated to propositions about $\Hil$. 
The translation ``table'' must be given by a map 
\[
\someobs^*: \cprops \Omega \rightarrow \qprops \Hil.
\]
The reason for the star will be explained later. 
This is indeed how we will represent an observable on $\Hil$, with ``spectrum'' $\Omega$ (i.e. the set of values it can take). However we need some additional constraints on this map.

Suppose that we have some knowledge about the quantum system represented by a state $\rho \in \states\Hil$. This knowledge should allow us to make some prediction about the classical system, i.e. it should translate into some knowledge about the measurement apparatus' pointer, represented by a classical state $\mu \in \states\Omega$. Mathematically, this is indeed what we get by using $\someobs^*$ to ``pull back'' the states of the quantum system, which are functionals on $\qprops \Hil$. Remember that the density $\rho$ represents the functional which sends $A$ to $\tr(\rho A)$, for any $A \in \qprops \Hil$. Since a classical proposition $\alpha$ is mapped to the quantum proposition $\someobs^*(\alpha)$, we indirectly get the function $\mu: \alpha \mapsto \tr(\rho(\someobs^*(\alpha))$ on classical observables. However, for this functional to be a valid classical state, it has to satisfy certain conditions. In particular, classical states are linear functionals. This means that for any two effects $\alpha, \beta \in \cprops \Omega$, and complex numbers $x, y$, we must have 
\[
\begin{split}
\tr(\rho(\someobs^*(x \alpha + y \beta)) &= x \tr(\rho(\someobs^*(\alpha)) + y \tr(\rho(\someobs^*(\beta)) \\
&= \tr(\rho(x \someobs^*(\alpha) + y \someobs^*(\beta))).
\end{split}
\]
Since this equation must be true for all quantum states $\rho$, it implies the operator equation
\[
\someobs^*(x \alpha + y \beta) = x \someobs^*(\alpha) + y \someobs^*(\beta)
\]
Hence $\someobs^*$ itself must be linear. Because of this property it extends uniquely to a linear map
\[
\someobs^*: \finfun(\Omega)  \rightarrow \finops(\Hil).
\]
We note this fact because it is convenient to work with linear spaces. However $\someobs^*$ cannot be any linear map, it must also send $\cprops \Omega \subset \finfun(\Omega)$ entirely into $\qprops \Hil \subset \finops(\Hil)$. This is exactly guaranteed if we require that it be {\em positive}, i.e. that it sends positive functions on $\Omega$ to positive operators on $\Hil$, and that in addition $\someobs^*(\lone) \le \one$, i.e. that the eigenvalues of $\someobs^*(\lone)$ be all smaller than $1$. However, we must also require that it sends quantum states to properly normalized functionals on $\Omega$, i.e. $\tr(\rho\someobs^*(\lone)) = 1 = \tr(\rho\one)$ for all $\rho$, which implies
\[
\someobs^*(\lone) = \one.
\]
We summarize this condition by saying that $\someobs^*$ is {\em unital} \index{unital channel (quantum to classical)}. 
Hence, the observables must be represented by linear, positive and unital maps $X^*$.
These conditions are all we need on $\someobs^*$ to make sure that it induces a valid map from quantum state to classical states, which we will write as $\someobs$. Consequently, the map $\someobs$ is also linear: it is indeed the linear \adjoint of $\someobs^*$. Since we are working with finite-dimensional vector spaces, $\someobs^*$ is also the \adjoint of $\someobs$, which justifies the star. The infinite-dimensional case will eventually justify our choice of putting the star on the map which acts on propositions, rather than on that map which acts on states.

Summarizing, the \adjoint map 
$$
\someobs:\states\Hil \rightarrow \states\Omega
$$
which pulls quantum states back to classical states, is the restriction to $\states\Omega \subseteq \finfun(\Omega)$ of the linear map
\[
\someobs: \finops(\Hil) \rightarrow \finfun(\Omega).
\]
The map $\someobs^*$ could be specified simply as the \adjoint of $\someobs$. Just as $\someobs^*$ needs to be positive, $\someobs$ also must be positive. Indeed it must send quantum states, which are positive operators, to classical states which are positive functions. The condition $\someobs^*(\lone) = \one$ however takes a different form in terms of $\someobs$. Recall that we arrived at this equation by requiring that $\someobs$ preserves the normalization of states, i.e. if $\mu = \someobs(\rho)$ then $\sum_i \mu_i = \tr(\rho)$. 

Let us now give a simple representation for the linear maps from $\finfun(\Omega)$ to $\finops(\Hil)$. Since they are linear, they can be characterized by their action on a basis of $\finfun(\Omega)$. We will use the canonical basis formed by the functions $\chi_i(j) := \delta_{ij}$\index{$\chi_i$}. The action of $\someobs^*$ on any classical proposition $\alpha$ is, by linearity, $\someobs^*(\alpha) = \sum_i \alpha_i \someobs^*(\chi_i)$. Hence $\someobs^*$ is entirely specified by the family of effects $\someobs_i := \someobs^*(\chi_i)$. The condition $\someobs^*(\one) = 1$ reads $\sum_i \someobs_i = \someobs^*(\one) = 1$. Reciprocally, any family of propositions $\someobs_i$ such that $\sum_i \someobs_i = \one$ defines an observable through 
\[
\someobs^*(\alpha) :=  \sum_i \alpha_i \someobs_i.
\]
The \adjoint $\someobs$ then sends the quantum state $\rho$ to the classical state $\mu$ which maps a proposition $\alpha$ to the probability $\sum_i \alpha_i \mu_i = \tr(\rho \someobs^*(\alpha)) = \sum_i \alpha_i \tr(\rho \someobs_i)$. Hence $\someobs$ sends $\rho$ to the probability distribution $\mu$ defined by
\[
\mu_i = \tr(\rho \someobs_i).
\]  
The unitality condition $\someobs^*(\lone) = \one$, which is equivalent to $\sum_i \someobs_i = \one$, is also equivalent to $\sum_i \mu_i = \tr(\rho)$. Let us define the classical trace\index{trace (classical)} $\tr(\mu) := \sum_i \mu_i$. Note that it is indeed the trace of the diagonal matrix $\sum_i \mu_i \proj{i}$ which represents the classical state as a quantum state. Hence we can summarize the condition $\sum_i \mu_i = \tr(\rho)$ by saying that $\someobs$ is \ind{trace-preserving}.

This shows, at least in finite dimension and for a finite number of outcomes, that this notion of observable is equivalent to that of Positive Operator-Valued measure (POVM). Indeed, a POVM which represents an observable with only a finite number of outcomes belonging to the set $\Omega$ can be defined as a family of propositions $A_i \in \finops(\Hil)$ such that $\sum_{i \in \Omega} A_i = \one$. The interpretation associated with it is that, if $\rho$ is the state of the system, then the outcome labelled by $i$ is true with probability $\tr(\rho A_i)$, which is precisely what we obtained for our concept of observable.

Note that effects are one-to-one with a particular class of observables, namely observables with two values, i.e. defined for $\Omega = \{0, 1\}$. Indeed, given any effect $A \in \qprops \Hil$, one can define the observable $X: \finfun(\{0,1\}) \rightarrow \qprops \Hil$ by 
\[
X_0 = X^*(\chi_0) = A \quad \text{and} \quad X_1 = X^*(\chi_1) = \one - A.
\]
This gives a particular interpretation for the effect $\one - A$: it is the alternative to $A$ in an observation of $A$. 

Let us also show how we can recover the more conservative notion of observable, namely what we will call {\it sharp} observables\index{sharp observable}. A sharp observable $X$ is an observable which maps sharp effects to sharp effects, i.e. such that for all sharp classical effect \ind{$\chi_\omega$} (the \ind{characteristic function} for a subset $\omega \subseteq \Omega$), the image $X^*(\chi_\omega)$ is a projector, i.e. $(X^*(\chi_\omega))^2 = X^*(\chi_\omega)$. In fact this is equivalent to requiring that $X_i^2 = X_i$ for all $i \in \Omega$. Indeed, if $X_i^2 = X_i$, then $X_i - X_i^2 = (\one-X_i)X_i = 0 $. But $\one -X_i = \sum_{j \neq i} X_j$. Hence also $\sum_{j \neq i} X_i X_j X_i = 0$. Since this is a sum of positive operators, this implies that $X_i X_j X_i = (X_i X_j)^*(X_i X_j) = 0$ for all $j$, which in turn implies $X_j X_i = 0$ for all $i$ and $j$. From the orthogonality of the effects $X_i$ it is easy to see than any sum $X^*(\chi_\omega) = \sum_{i \in \omega} X_i$ is a projector as well. 
 
In fact, this also proves that these projectors are automatically orthogonal. Since also they are complete in the sense that $\sum_i X_i = \one$, then they can be seen as the projectors on the eigenspaces of a \ind{self-adjoint operator} $A = \sum_{i \in \Omega} \lambda_i X_i$, which is indeed the standard notion of an observable. Here $\lambda_i$ can be any family of distinct real numbers. In fact, to be pedantic, we could make sure that $\Omega \subset \mathbb R$ and write $A = \sum_{\lambda \in \Omega} \lambda X_i$. However the actual \ind{eigenvalues} matter little for a discrete observable.

\subsection{Capacity of an observable}
\label{section:obsinfo}
\index{capacity of an observable}

Intuitively, measuring a sharp observable should yield more information about the system than measuring an unsharp one, provided that they have the same number of outcomes. Viewing observables as communication channels yields a natural way to make this concept precise, namely the \ind{information capacity of an observable}.

The idea is the following. Suppose that Alice can prepare a given quantum system in any state she wants, and Bob is going to measure the observable $X$ of this system. If the two can agree of a protocol beforehand, how many bits can Alice communicate in this way to Bob? In order to make this more precise, one must consider that they will actually use an arbitrary large number of systems, and see how many bits Alice can transmit per system, in average, as the number of systems goes to infinity. This is a direct parallel to the definition of capacity for a classical channel. In fact, it can be reduced to it, as shown by the Holevo-Schumacher-Westmorland (HSW) theorem \index{HSW theorem} \cite{nielsen00}.

Indeed, the process by which Alice will encode her classical bits into a quantum state is itself a channel, in this case a classical-to-quantum channel, also called a state-preparation procedure. Let us call such a channel $\phi$. Since $\phi$ is linear and acts on $\finfun(\Omega)$, it can be characterized by its action on the basis states $\chi_i$. Let us write $\rho_i = \phi(\chi_i)$ for the quantum states which are image of these basis states. For a general classical state $\mu = \sum_i \mu_i \chi_i$, we have
\[
\phi(\mu) = \sum_i \mu_i \rho_i.
\]

If we compose this classical-to-quantum channel $\phi$ with the quantum-to-classical channel $X$, we obtain simply a classical channel $X \circ \phi$, for which a Shannon capacity $C(X \circ \phi)$ can be computed. The HSW theorem then states that the classical capacity of the channel $X$ is given by
\[
C(X) = \sup_\phi C(X \circ \phi).
\]

Since the HSW theorem is usually expressed for general quantum channel, this formulation is not usual. Let us therefore show that this is indeed equivalent to the formula given in \cite{nielsen00}, namely
\begin{equation}
\label{equ:HSW}
C(X) = \sup_{\mu_i \rho_i} \Bigl({ S(X(\sum_i \mu_i \rho_i) - \sum_i \mu_i S(X(\rho_i))  }\Bigr)
\end{equation}
Where, by $S(\mu)$, we mean the von-Neumann entropy of the quantum state whose eigenvalues are given by the classical distribution $\mu$, which is simply the Shannon entropy of $\mu$: $S(\mu) = H(\mu)$.

Note that $C(X\circ \phi)$ is the maximum over all initial states $\mu$ of the mutual information for the joint probability $p(i,j)$ of getting outcome $\chi_j$ and input $\chi_i$. This joint probability is given by $p(j|i) \mu_i$, where $p(j|i) = X(\phi(\chi_i))_j$ is the probability to get outcome $\chi_j$ given that the input was $\chi_i$. Therefore $p(i,j) = X(\phi(\chi_i))_j\mu_i$ and has marginals $\sum_j X(\phi(\chi_i))_j = \mu_i $ and $\sum_i X(\phi(\chi_i))_j\mu_i = X(\phi(\mu))_j$. We therefore have
\[
C(X \circ \phi) = \sup_{\mu} [ H(X(\phi(\mu))) + H(\mu) - H(p) ].
\]
But note that 
\[
\begin{split}
H(\mu) - H(p) &= - \sum_i \mu_i \log_2(\mu_i) + \sum_{ij} p(i,j) \log_2(p(i,j)) \\
&= - \sum_i \mu_i \Bigl({  \log_2(\mu_i) - \sum_{j} p(j|i) \log_2(p(i,j)) }\Bigr) \\
&= \sum_i \mu_i \sum_{j} p(j|i) \log_2(p(j|i))\\
& = - \sum_i \mu_i H(X(\phi(\chi_i))) \\
\end{split}
\]
and so,
\[
\begin{split}
C(X) &= \sup_\phi C(X \circ \phi) = \sup_{\phi, \mu} [ H(X(\phi(\mu))) - \sum_i \mu_i H(X(\phi(\chi_i))) ]\\
&= \sup_{\mu_i, \rho_i} \Bigl({ S(X(\sum_i \mu_i \rho_i)) - \sum_i \mu_i H(X(\rho_i)) }\Bigr)\\
\end{split}
\]
which is indeed equal to Equation \ref{equ:HSW}. 



When applied to a discrete sharp observable, this quantity is equal to the logarithm of the number of outcomes. Indeed, if $X$ is sharp, then the optimal strategy for Alice consists in encoding her classical states into eigenstates of $X$. The number of states that she can encode in this way is the number of distinct eigenstates $n$ of $X$. This corresponds to $\log_2(n)$ bits. 


\subsection{Quantum channels}
\label{section:channel}
\index{channel}

We want to consider the most general way that a quantum system $B$ can encode information about another quantum system $A$, as specified by a map
\[
\chan^*: \props {\Hil_B} \rightarrow \props {\Hil_A}
\]
\index{$\chan$}
which translates each proposition on system $B$ into a proposition of system $A$. All the points made in the discussion of observables are still valid here, even though we have replaced the classical system by the quantum system $B$.
Indeed the only algebraic properties of states and effects that we have used in our presentation involved the linearity and the positivity of effects and states, which form the same algebraic structure in the quantum and in the classical case (more will be said about what this structure is in Section \ref{section:formalism}). Hence the most general such map $\chan^*$, which is also required to induce a valid map from states of system $A$ to states of system $B$, must extend to a unital positive linear map
\[
\chan^*: \finops(\Hil_B) \rightarrow \finops(\Hil_A).
\]
This map is the \adjoint of a trace-preserving positive linear map
\[
\chan: \finops(\Hil_A) \rightarrow \finops(\Hil_B)
\]
which maps states of $A$ to states of $B$. 
The relation between the two can be simply expressed from the fact that they yield the same probabilities:
\[
\tr(\chan(\rho) A) = \tr(\rho \chanh(A))
\]
for all states $\rho$ and all effects $A$. 

However we need another important condition which happens to be implicitly satisfied in the case of the quantum-to-classical maps which define observables. We need to make sure that if our systems $A$ and $B$ were only parts of larger systems, then our map could be trivially extended to those larger systems. 
 This means that, if we add a third system $C$ with Hilbert space $\Hil_C$, then the map $\chan^* \otimes \id$ from $\finops(\Hil_A \otimes \Hil_C)$ to $\finops(\Hil_B \otimes \Hil_C)$ defined by $(\chan^* \otimes \id)(X \otimes Y) := \chan^*(X) \otimes Y$, should also be unital, positive and linear. Unitality and linearity come for free. However the positivity must be postulated. If $\chan^* \otimes \id$ is positive no matter what the dimension of the auxiliary system $C$ is, then we say that $\chan^*$ is {\em completely positive}\index{completely positive map}. In this case, the \adjoint $\chan \otimes \id$ is also automatically completely positive. This condition is not only physically reasonable, but also mathematically extremely useful since it allows one to apply a powerful representation theorem by Choi \cite{choi75}, which states that there is a family of operators $E_k$ (the channel's {\em elements}\index{elements of a channel}) such that
\[
\chan^*(X) = \sum_k E_k^* X E_k
\]
from which it is easy to see that also
\[
\chan(\rho) = \sum_k E_k \rho E_k^*.
\]
In fact any map of this form is completely positive (and of course linear). Unitality of $\chan^*$ requires in addition that
\[
\chan^*(\one) = \sum_k E_k^* E_k = \one.
\]
Let us state and prove \ind{Choi}'s theorem for completeness.
\begin{thmchoi}
A linear map $\chan$ from $\finops(\Hil_A)$ to $\finops(\Hil_B)$ is completely positive if and only if it has the form
\(
\chan(\rho) = \sum_k E_k \rho E_k^*
\)
for a family of linear operators $E_k: \Hil_A \rightarrow \Hil_B$. 
\end{thmchoi}
\proof First, it is clear that a map defined by $\chan(\rho) = \sum_k E_k \rho E_k^*$ is positive. Indeed, if $\rho$ is positive then $\sum_k E_k \rho E_k^*$ is a sum of positive operators, which is also positive. In addition, its extension $(\chan \otimes \id)(\rho) = \sum_k (E_k \otimes \one) \rho (E_k \otimes \one)^*$ has the same form and is therefore positive too, which show that $\chan$ is completely positive.
For the converse, suppose that $\chan$ is completely positive. Let us view $\chan$ as going from the complement of its kernel $\ker(\chan)^\perp$ to its range $\ran(\chan)$. Since these subspaces have equal dimensions, they can be both identified with the same space $\Hil$. Consider a basis $\ket{i}$ of $\Hil$. Consider also the Hilbert space $\Hil \otimes \Hil$ and the operator $A := \sum_{ij} \ketbra{i}{j} \otimes \ketbra{j}{i}$ acting on it. Note that $A$ is positive. Indeed for any state $\ket{\Psi} := \ket{\psi}\otimes\ket{\phi}$ with components $\psi_i \phi_j := \braket{i}{\psi}\braket{j}{\phi}$, we have 
\[
\bra{\Psi} A \ket{\Psi} = \sum_{ij} \overline \psi_i \psi_j \overline \phi_j \phi_i =  \left({ \sum_{i} \overline \psi_i \phi_i }\right) \overline{ \left({\sum_j \overline \psi_j  \phi_j }\right) } \ge 0
\]
Since $\chan$ is completely positive, $(\chan \otimes \id)(A)$ is a positive operator on $\Hil_A \otimes \Hil_B$. Therefore it has a complete set of orthogonal eigenvectors $\ket{v_k} \in \Hil_A \otimes \Hil_B$. This means that it can be written as 
$$ 
(\chan \otimes \id)(A) = \sum_{k} \proj{v_k}
$$
where the norm of the vector $\ket{v_i}$ is the square of the corresponding eigenvalue. Explicitly this means that
$$
\sum_{ij} \chan(\ketbra{i}{j}) \otimes \ketbra{j}{i} = \sum_{k} \proj{v_k}
$$
by multiplying by $\one \otimes \bra{j}$ on the left and $\one \otimes \ket{i}$ on the right we get
$$
\chan(\ketbra{i}{j}) = \sum_{k} (\one \otimes \bra{j}) \proj{v_k} (\one \otimes \ket{i})
$$
which is equivalent to
\[
\begin{split}
\chan(\ketbra{i}{j}) &= \sum_{lk} \bra{v_k} (\proj{l} \otimes \ketbra{i}{j}) \ket{v_k}\\
 &= \sum_{lk} \bra{v_k} (\proj{l} \otimes \ketbra{i}{j}) \ket{v_k}\\
 &= \sum_{lk}  \bra{v_k} (\ket{l} \otimes \one) \,\ketbra{i}{j}\, (\bra{l} \otimes \one) \ket{v_k}
\end{split}
\]
But since $\ketbra{i}{j}$ form a basis of all operators on $\Hil$, we have by linearity
$$
\chan(\rho) = \sum_{lk}  \bra{v_k} (\ket{l} \otimes \one) \,\rho\, (\bra{l} \otimes \one) \ket{v_k}
$$
for any operator $\rho$ on $\Hil$. This is the representation that we seeked, with elements
\[
E_{kl} := \bra{v_k} (\ket{l} \otimes \one).
\] 
 \qed
\smallskip

Such a map between states, namely a trace-preserving completely-positive linear map, will be called a {\em quantum channel}, or simply a channel. In order to avoid confusion, we will only call a channel the map $\chan$ on states. $\chan^*$ will simply be called its dual. We may also call an observable $\someobs$ a quantum-to-classical channel\index{channel (quantum to classical)} since it sends quantum states to classical states. 

\subsection{Transformation of observables: the Heisenberg picture}
\label{section:heisenberg}

All observables $X$ are basically made out of effects, namely those in the image of $X^*$: $X^*(\alpha) \in \qprops \Hil$. For convenience, we will refer to them as the {\em effects of $X$}\index{effects of an observable}.
Since the \adjoint of a channel maps effects to effects, it also maps observables to observables. Indeed, consider the observable $Y^*$ defined by 
\(
Y^*(\alpha) := \chan^*(X^*(\alpha))
\)
for all classical effects $\alpha$. To see that $Y^*$ is indeed an observable, simply note that it is unital, positive and linear, because both  $\chan^*$ and $X^*$ are. This equation can also be written as 
\[
Y^* = \chan^* \circ X^* \quad \text{ or } \quad Y = X \circ \chan
\]
or simply in terms of the elements $Y_i = Y^*(\chi_i)$ and $X_i = X^*(\chi_i)$ as
\[
Y_i = \chan^*(X_i)
\]
An alternative way of writing this equation, which is intuitively helpful, is as a commutative diagram showing the flow of information between states of the three systems involved, namely the source quantum system $A$, the target quantum system $B$, and the classical pointer $\Omega$.
\begin{diagram}
  \finops(\Hil_A) &\rTo^{\quad \chan \quad}  & \finops(\Hil_B)\\
  & \rdTo_{X} & \dTo_{Y} \\
  & & \finfun(\Omega)
\end{diagram}
Each object can be seen as a physical system, be it quantum or classical, and each arrow as a transfer of information between systems. An arrow from a quantum to a classical system corresponds to an observation.
We will see later that this can be formally seen as a diagram in the category where the objects are the pre-dual of von Neumann algebras and the morphisms the trace-preserving completely-positive linear maps between them. 
The correct way to read such a diagram is as a set of equations stating that each possible path between two objects are equal. 

This is how the channel $\chan$ acts on observables. In the case where $\chan$ represents some time evolution for a certain interval of time, this is simply the \ind{Heisenberg picture}. Indeed, in the case of a unitary evolution $U_t$, remember that we get the Heisenberg picture by noting that what matters are expectation values of the form $\bra{\psi(t)} A \ket{\psi(t)}$ for a self-adjoint operator $A$ representing a sharp observable, and for the time-dependant state $\ket{\psi(t)} = U_t \ket{\psi}$. These expectation values can alternatively be written as $\bra{\psi} U_t^* A U_t \ket{\psi} = \bra{\psi} A(t) \ket{\psi}$, where we defined $A(t) := U_t^* A U_t$. Hence we can either assume that our knowledge represented by the states evolves as a function of time (the Schr\"odinger picture), or alternatively assume that our knowledge is fixed, but that it is our interpretation of what the observables are which changes with time: a point of view called the Heisenberg picture. A channel $\chan$ specifies a more general type of evolution, however the scheme is the same. We took the more general expectations values (probabilities) of the form $\tr(\rho A)$ as the fundamental predictions of the theory, and noted that a given map $\chan^*$ on effects also induces a map $\chan$ on states through the relation $\tr(\chan(\rho) A) = \tr(\rho \chan^*(A))$. Hence the map $\chan^*$, which says how effects evolve in the case of a fixed state, specifies the evolution in the Heisenberg picture. However, writing an observable $X$ as a channel, rather than in terms of its elementary effect $X_i$, we can also write the Heisenberg evolution using the ``Schr\"odinger'' channel $\chan$, as $X \mapsto Y = X \circ \chan$. 

However there is something that this more general picture renders apparent which is normally overlooked in the unitary case. It should be clear by now that if the channel maps {\em states} of a system $A$ to states of another system $B$, then the effect of the channel in the dual Heisenberg picture is to map observables of $B$ to observable of $A$, i.e. it goes ``backward''. This can also be seen by looking at what happens if we compose two channels $\chan$ and $\chan'$. We get the new channel $\chan' \circ \chan$. But in the Heisenberg picture the action is given by $(\chan' \circ \chan)^* = \chan^* \circ (\chan')^*$. Hence in the Heisenberg picture we have to apply the latest channel first. If the channel represents a time evolution, one way to interpret this is to think of the observable $Y = X \circ \chan$ as that which, if measured at the initial time, would yield the same result as $X$ measured at the later time. 


\subsection{Measurements}
\label{section:measurement}


We have argued that the most general way that a classical system with phase-space $\Omega$ can hold information about a quantum system defined by the Hilbert space $\Hil$, is specified by a trace-preserving linear map $X: \finops(\Hil) \rightarrow \finfun(\Omega)$. But if we actually perform this measurement, we may expect that the quantum system that we have measured still exists afterward. This means that the process of measurement should map the state of our system not only to the state of our measurement apparatus' pointer, but also to the later state of the system itself as well. 
Hence it should be a channel
\[
\chan: \finops(\Hil) \rightarrow \finops(\Hil) \otimes \finfun(\Omega)
\]
with \adjoint
\[
\chanh: \finops(\Hil) \otimes \finfun(\Omega) \rightarrow \finops(\Hil)
\]
where $\Hil$ is the Hilbert space of our quantum system and $\Omega$ the phase-space of the measurement apparatus' pointer. 

In order to see what the tensor product between a quantum and a classical system is, consider the representation of classical states and classical effects as diagonal quantum states and diagonal quantum effects. For instance, the classical effect $\alpha$ can be represented as $\alpha = \sum_{i \in \Omega} \alpha_i \proj{i}$. Its tensor product with a quantum effect $A$ then is simply $A \otimes \alpha = \sum_{i \in \Omega} \alpha_i A \otimes \proj{i}$. This shows that in this representation, the space $\finops(\Hil) \otimes \finfun(\Omega)$ consists of block-diagonal matrices. It is in fact a subalgebra of the algebra of full matrices. It defines in its own right a physical system which is not quantum nor classical, but \ind{hybrid quantum-classical}. We will see more general structures of this type in Section \ref{section:formalism} where they will be formalized as C$^*$-algebras.



For the process represented by $\chan$ to represents the measurement of a specific observable $\someobs$ we require that
\begin{equation}
\label{equ:measobs}
\chan^*(\one \otimes \alpha) = \someobs^*(\alpha)
\end{equation}
for any classical effect $\alpha \in \cprops \Omega$. Indeed, $\one \otimes \alpha$ is an effect on the joint system which simply ignores the quantum system, since $\one$ is true no matter what the state of the quantum system is. Note that the dual of the completely positive map $\alpha \mapsto \one \otimes \alpha$ is the \ind{partial trace} over the quantum system. Indeed, given a basis $\ket{i}$ of the quantum system, we can write this channel explicitly in terms of its elements:
\[
\alpha \mapsto \sum_i \proj{i} \otimes \alpha = \sum_i (\ket{i} \otimes \one) \,\alpha\, (\bra{i} \otimes \one).
\]
We therefore immediately obtain the dual map on states by swapping the elements:
\[
\rho \otimes \mu \mapsto \sum_i  (\bra{i} \otimes \one)( \rho \otimes \mu )(\ket{i} \otimes \one) = \sum_i \bra{i} \rho \ket{i}\, \mu = \tr(\rho) \mu = \mu.
\]

Equation \ref{equ:measobs} implies that $\chan^*(\one \otimes \chi_i) = \someobs_i$ for all $i \in \Omega$. Note that the maps $\mathcal F_i^*$ defined by
\[
\mathcal F_i^*(A) := \chan^*(A \otimes \chi_i)
\]
from $\finops(\Hil)$ to itself are completely positive, although not unital.
These maps can be anything apart from the requirement given by Equation \ref{equ:measobs}, which is
\[
\mathcal F^*_i(\one) = \someobs_i.
\]
Hence the most general form of a measurement of the observable $\someobs$ is as a map
\[
\chan^*(A \otimes \alpha) = \sum_i \alpha_i \mathcal F^*_i(A)
\]
for some completely positive maps $\mathcal F^*_i$ which satisfy $\mathcal F^*_i(\one) = \someobs_i$. The dual $\chan$ is given by
\[
\chan(\rho) = \sum_i \mathcal F_i(\rho) \otimes \chi_i.
\]
If we represent the classical system as a diagonal quantum system, we could write this as $\chan(\rho) = \sum_i \mathcal F_i(\rho) \otimes \proj{i}$.

Hence, if our knowledge of the system was represented by $\rho$, then after it has interacted with the measurement device, our joint knowledge of the system and the pointer is represented by the state $\sum_i \mathcal F_i(\rho) \otimes \chi_i$. In order to see what this implies, let us consider a second measurement after this one.


\subsection{Repeated measurements and collapse}

Suppose that we perform a measurement of $\someobs$ as specified above, followed by a measurement of a second observable $\someobs'$ associated with the phase-space $\Omega'$. By composing the measurement $\chan$ with the channel $\someobs'$ applied on the final state of the quantum system, we obtain the purely classical state $\mu$ of the two pointers:
\[
\mu = \sum_i \someobs'(\mathcal  M_i(\rho)) \otimes \chi_i = \sum_{ij} \tr(\mathcal F_i(\rho)\someobs_j') (\chi_j \otimes \chi_i)
\]
This is simply the functional way of writing the joint probability distribution
\[
\mu_{ij} = \tr(\mathcal F_i(\rho)\someobs_j')
\]
where $i \in \Omega$ is the outcome of the first measurement and $j \in \Omega'$ the outcome of the second. From this joint probability distribution we can compute the conditional probability $p(j|i)$ of obtaining outcome $j \in \Omega'$ in the second measurement, knowing that the outcome of the first measurement was $i \in \Omega$:
\[
\begin{split}
p(j|i) &= \frac{\mu_{ij}}{\sum_j \mu_{ij}} = \frac{\tr(\mathcal F_i(\rho)\someobs_j')}{\tr(\mathcal F_i(\rho))}\\
&= \tr(\rho_i \someobs_j') 
\end{split}
\]
where we have introduced the quantum states
\[
\rho_i := \frac{\mathcal F_i(\rho)}{\tr(\mathcal F_i(\rho))}.
\]
Since $\someobs'$ is not involved in the definition of $\rho_i$, we see that after the measurement of $\someobs$, and the finding of the measurement apparatus' pointer in state $i \in \Omega$, our knowledge about the quantum system can be updated to the new state $\rho_i$, which has the usual form given by the collapse postulate\index{collapse of states}. 

For instance, if the observable $\someobs$ is sharp, in the sense that all the effects $\someobs_i = P_i$ are sharp: $P_i^2 = P_i$, then one possible choice of measurement apparatus is given by 
\[
\mathcal F_i(\rho) := P_i \rho P_i
\]
This indeed defines a measurement of $\someobs$, since $\mathcal F_i^*(\one) = P_i^2 = P_i$. We then obtain the usual state update formula
\[
\rho_i = \frac{P_i \rho P_i}{\tr(P_i \rho)}.
\]


\msection{Effect algebras}
\label{section:effects}

This section is meant to give some logical motivations for the mathematical structures that we postulated on both the quantum and classical sets of propositions. It is interesting to note that a physical transformation (channel) is not required to completely preserve all of these structures. For instance, the product of effects need not be preserved. The main algebraic structure which is preserved is the addition of effects. We introduce here one program of axiomatization of quantum mechanics which is based on this structure \cite{foulis94}. 


Let us suppose that the set $\aprops$ represents the propositions associated with some physical system. We assume that we also have a set of states \ind{$\astates$} on it, which assign a probability to each proposition. 
We want to define on $\aprops$ a generalization of the logical ``or''\index{logical or} between disjoint propositions. We write $\gamma = \alpha + \beta$ for three propositions $\alpha$, $\beta$ and $\gamma$ whenever $\rho(\gamma) = \rho(\alpha) + \rho(\beta)$ for all states $\rho$. Addition of probabilities is indeed what we expect for the logical \ind{disjunction} of two propositions. In general, the sum $\alpha + \beta$ cannot be defined for all pairs of propositions $\beta$ and $\alpha$ since $\rho(\alpha) + \rho(\beta)$ might be greater than $1$. Pairs of propositions for which the sum exists will be said to be valid alternatives. The sum $\alpha+\beta$, when it exists, can always be assumed to be unique, given that the corresponding probabilities are defined uniquely by the relation $\rho(\alpha+\beta) = \rho(\alpha) + \rho(\beta)$ for any state $\rho$. 

One property that we get for free from this definition is that $\alpha+\beta = \beta + \alpha$.
Many interesting results can be obtained with this structure if we postulate a few useful axioms\index{axioms of quantum theory} which make $\aprops$ into an {\em \ind{effect algebra}} \cite{foulis94}. In order to get a sort of associativity, we assume that if $\alpha+\beta$ and $\gamma + (\alpha + \beta)$ are defined, then $(\gamma+\alpha) + \beta$ is defined too. From the definition of the addition as it relates to states, we get $(\gamma+\alpha) + \beta = \gamma + (\alpha + \beta)$. 
Next we postulate that there are two special propositions: the proposition $0$ which is always false, i.e. $\rho(0) = 0$ for any state $\rho$, and the proposition $\lone$ which is always true, i.e. $\rho(\lone) = 1$ for any state $\rho$. 
We then assume that for any proposition $\alpha$, there is a proposition $\beta$ such that $\alpha+\beta = \lone$. 
$\beta$ can
be written as $\beta = 1-\alpha$. It can be interpreted as the \ind{negation} of $\alpha$. Indeed for any state $\rho$ such that $\rho(\alpha) = 1$, we have $\rho(\lone-\alpha) + \rho(\alpha) = \rho(\lone) = 1$. Hence $\rho(\lone-\alpha) + 1 = 1$, which implies $\rho(\lone-\alpha) = 0$. This shows that the complement $1-\alpha$ is false whenever $\alpha$ is true. Conversely we also have that $1-\alpha$ is true whenever $\alpha$ is false, which is what we expect of the negation of $\alpha$. 

The last axiom needed to define an effect algebra is that $\lone + \alpha$ exists only for $\alpha = 0$. 
In addition, we assume for the moment that there are no other restrictions on states other than those implied in the definition of $+$, $0$ and $1$. This means that the set of states consists of {\em all} the functions $\rho$ from $\aprops$ to the interval $[0,1]$ which are such that $\rho(\alpha) + \rho(\beta) = \rho(\alpha+\beta)$, $\rho(0) = 0$ and $\rho(\lone) = 1$. 

 
An interesting property of an effect algebra is that it automatically comes with a \ind{partial order} defined by $\alpha < \beta$ whenever $\gamma + \alpha = \beta$ for some proposition $\gamma$. This relation can be understood as a form of logical implication from $\alpha$ to $\beta$. 
Indeed for any state $\rho$ such that $\rho(\alpha) = 1$, i.e. any state for which $\alpha$ is true, we have $0 \le \rho(\beta-\alpha) = \rho(\beta) - \rho(\alpha) = \rho(\beta) - 1$. Hence $1 \ge \rho(\beta) \ge 1$, which implies $\rho(\beta) = 1$, i.e. $\beta$ is true.  Hence $\alpha$ implies $\beta$. 

\subsection{Linear effect algebras}

\index{linear effect algebras}

The effects algebras of quantum and classical systems have an additional simple structure in common which makes the above axioms more concrete and easier to understand. 

Suppose that we are given an ordered vector space $V$ with a special element $e > 0$. An \ind{ordered vector space} is a vector space with a partial order $\le$ satisfying $u \le v$ if and only if $0 \le v-u$, and $0 \le \lambda v$ for all $0 \le v$ and $\lambda \in \mathbb R_+$. Then an effect algebra is automatically formed by those elements of $V$ which are larger than the zero vector $0$ and smaller than $e$:
\[
\props V := \{ \alpha \in V \;|\; 0 \le \alpha \le e \}.
\]\index{$\props V$}
This set $\props V$ is an effect algebra for the partially-defined addition defined by the addition of vectors, with the false proposition given by the zero vector and the true proposition given by the specially chosen vector $e$. The natural partial order of this effect algebra is of course the one which we started from. Hence what we have done was to use the partial order to tell which vectors can be added. Indeed we have that $\alpha+\beta$ exists exactly when $\alpha+\beta \le \lone$. Such a structure is called a {\em linear effect algebra} \cite{gudder99}.
Note that since we are only interested in the vector space in the extent that it defines the effect algebra, we will always assume that $V$ is the span of the effect algebra $\props V$. 

Note that the states are automatically given by linear maps on $V$. Indeed, a state $\rho$ is required to satisfy $\rho(\alpha + \beta) = \rho(\alpha) + \rho(\beta)$ for any effects $\alpha$ and $\beta$. This means that for any positive integer $n$ such that $n \alpha \le \lone$, we must have $\rho(n \alpha) = n\rho(\alpha)$. In addition, for any positive integer $m$ we have $\alpha = m \frac{1}{m} \alpha$, hence $\rho(\alpha) = m \rho(\frac{1}{m}\alpha)$, which implies $\rho(\frac{1}{m} \alpha) = \frac{1}{m}\rho(\alpha)$. Combining the two results, for any positive rational $q = \frac{n}{m}$ and effect $\alpha$ such that $q \alpha \le \lone$, we have $\rho(q \alpha) = q \rho(\alpha)$. In addition, note that $\alpha < \beta$ implies $\rho(\alpha) \le \rho(\beta)$. Indeed, $\alpha < \beta$ means that there exists $0 \le \gamma \le \lone$ such that $\alpha + \gamma = \beta$, from which we have $\rho(\alpha) + \rho(\gamma) = \rho(\beta)$, and hence, $\rho(\alpha) \le \rho(\beta)$. Now for any real $0 \le r \le 1$, consider a sequence $0 \le q_n \le r \in \mathbb Q$ and a sequence $1 \ge q_n' \ge r \in \mathbb Q$ both converging to $r$. We have $q_n \rho(\alpha) = \rho(q_n \alpha) \le \rho(r \alpha) \le \rho(q_n' \alpha) = q_n'\rho(\alpha)$ for all $n$. In the limit, this implies $r \rho(\alpha) = \rho( r \alpha)$.   


\subsection{Morphisms between effect algebras} 

We want to consider maps between two effect algebras:
\[
\morph: \aprops_B \rightarrow \aprops_A
\]
We have seen that once we specialize to quantum and classical systems, the interesting maps are those which preserve the addition of effects, i.e. $\morph(\alpha) + \morph(\beta)$ is defined whenever $\alpha+\beta$ is, and
\[
\morph(\alpha+\beta) = \morph(\alpha) + \morph(\beta)
\] 
This automatically implies that $\morph(0) = 0$. But in addition we require that 
\[
\morph(\lone) = \lone.
\]
These assumptions are what we need for the map $\morph$ to make sense as a transformation of states. Indeed, consider a state $\rho_A$ of $\aprops_A$. We automatically obtain a function $\rho_B$ on $\aprops_B$ defined by $\rho_B(\alpha_B) = \rho_A(\morph(\alpha_B))$, i.e.
\[ 
\rho_A \mapsto \rho_B = \rho_A \circ \morph
\]
Although the function $\rho_B$ is always defined, it may not be a state. It will be a state if it satisfies $\rho_B(\alpha+\beta) = \rho_B(\alpha) + \rho_B(\beta)$ and $\rho_B(\lone) = 1$. The first condition implies that $\rho_A(\morph(\alpha+\beta)) = \rho_A(\morph(\alpha)) + \rho_A(\morph(\beta))$ for all $\rho_A$, which implies that 
\(
\morph(\alpha+\beta) = \morph(\alpha) + \morph(\beta).
\)
From the second condition we have $1 = \rho_B(1) = \rho_A(\morph(1)) = \rho_A(1)$ for all $\rho_A$, which implies 
\( 
\morph(\lone) = \lone.
\) 
Maps $\morph$ having these properties will be called {\em morphisms}\index{morphism (of effect algebras)} as they preserve all the structures of the effect algebras. They correspond to the most general transformations which maps states of $A$ to states of $B$ {\em and} effects of $B$ to effects of $A$. 

For both quantum and classical systems, these morphisms almost define what the possible physical transformations are. In the quantum case we will just need one additional assumption which will be that $\phi$ must also be a morphism if the system of interest is only part of a larger system. However, we have not introduced the structure needed in order to define a subsystem in this abstract framework. In fact, the question of what structure one needs to add to an abstract effect algebra, if any, in order to define this concept, is an open research problem. 

If $\props {V_A}$ and $\props {V_B}$ are two linear effect algebras built respectively from the vectors spaces $V_A$ and $V_B$, then the  morphisms are the linear maps $\phi$ from $V_B$ to $V_A$ which are {\em positive}\index{positive linear map}, and such that $\phi(\lone) = \lone$. The map $\phi$ is called {\em positive} if $\phi(\alpha) \ge 0$ whenever $\alpha \ge 0$. This is all that is required for $\phi$ to preserve the order structure and therefore the induced effect algebra structure.
 

\subsection{States}
\label{section:effalgstates}

Let us say a little bit more about states. 
Note that the real interval $[0,1]$ itself is a linear effect algebra in terms of the addition of real numbers, and the natural complete order of real numbers. Hence the set of states on an effect algebra $\aprops$ is simply the set of morphisms from $\aprops$ to $[0,1]$. 

If $\aprops$ is a linear effect algebra spanning the vector space $V$ then the states are elements of the dual space $V^*$. For instance, when $V$ is finite-dimensional, $V^*$ is a vector space of same dimension as $V$. Note that $V^*$ is an ordered vector space like $V$. Indeed, for an elements $\rho \in V^*$, we have $\rho \ge 0$ if $\rho(\alpha) \ge 0$ for all $\alpha \ge 0$. This is of course the kind of positivity which is required of a state $\rho$. Therefore states are positive elements of the dual space $V^*$. In addition we have to require $\rho(\lone) = 1$. In order to make connection with what follows, let us define the trace of an element $\rho \in V^*$ as 
\[
\tr(\rho) := \rho(\lone).
\]
Then we can write the set of states as
\[
\states V = \{\rho \in V^* \;|\; \rho \ge 0 \text{ and } \tr(\rho) = \lone \}
\]

Remember that a morphism $\phi$ between two effect algebras acts on states through $\rho \mapsto \rho \circ \phi$. This map defines the linear dual $\phi^*(\rho) = \rho \circ \phi$. 
The positivity of $\phi$, which is defined by $\phi(\alpha) \ge 0$ for all $\alpha \ge 0$ , is equivalent to $\rho(\phi(\alpha)) = \phi^*(\rho)(\alpha) \ge 0$ for all states $\rho$ and all $\alpha \ge 0$, which in turns means $\phi^*(\rho) \ge 0$ for all states $\rho$, or simply for all positive functionals $\rho \ge 0$. This means that positivity of $\phi$ is equivalent to the positivity of $\phi^*$ in terms of the dual order structure. 
The last defining property of $\phi$ is the property $\phi(\lone) = \lone$. Let us see what this means in terms of the dual. This condition is equivalent to $\rho(\phi(\lone)) = \phi^*(\rho)(\lone) = \rho(\lone)$ for all functionals $\rho$. This simply means that $\phi^*$ preserves our notion of trace: $\tr(\phi^*(\rho)) = \tr(\rho)$ for all $\rho$. 
Therefore the action of a morphism on states takes the form of a trace-preserving positive linear map.

\subsection{Effect algebra of a C$^*$-algebra}

Both quantum and classical sets of propositions have a number of properties in common beyond the fact that they are effect algebras.
First, as suggested above, their effect algebras are both defined from a vector space, i.e. are linear effect algebras. Again this means that we are given a vector space $V$, a partial order $<$ and a special element $e > 0$, and that the effect algebra is defined by
\(
\aprops = \{ \alpha \in V \;|\; 0 \le \alpha \le e \}.
\)
The additional structure that we will fix can be seen as a way of specifying the partial order relation on our vector space. 

We will assume that $V$ is a complex vector space, and also has a product between elements which make it into an algebra, an involution $\alpha \mapsto \alpha^*$ which makes it into a $*$-algebra, and a norm which satisfies the axioms of a \ind{C$^*$-algebra}, namely
\[
\|\alpha\, \beta\| \le \|\alpha\|\|\beta\| \quad \text{and} \quad \|\alpha^* \alpha\| = \|\alpha\|^2.
\]
In addition we assume that $e$ is the algebraic unit of $V$.
Since it is an algebra, let us call our vector space \ind{$\mathcal A$} rather than $V$. Also we will write $\one$ instead of $e$.

Let us now explain how the partial order is derived from the C$^*$-algebraic structure. We can define the \ind{spectrum} $\sigma(\alpha)$ of an element $\alpha \in \mathcal A$ by
\[
\sigma(\alpha) = \{ \lambda \;|\;\one - \lambda \alpha \text{ has an inverse} \}
\] 
We then say that $\alpha > 0$ if $\alpha^* = \alpha$ and $\sigma(\alpha) \subseteq [0,\infty)$. This defines a partial order between vectors through
\[
\alpha < \beta \quad \text{when}\quad \beta-\alpha > 0.
\]
This is the partial order which defines the effect algebra associated with our C$^*$-algebra $\mathcal A$. 

The main theorem of the theory of C$^*$-algebra states that all C$^*$-algebras can be represented as $*$-subalgebras of the set \ind{$\mathcal B(\Hil)$} of bounded operators acting on a Hilbert space $\Hil$. This is what makes it an appropriate axiomatization for quantum mechanics. 


There is an important physical concept naturally encoded in the structure of concrete C$^*$-algebras which is not present in the pure effect algebra formalism, namely the notion of independent subsystems. Given two C$^*$-algebras represented as operators on a Hilbert space: $\mathcal A \subseteq \mathcal B(\Hil)$ and $\mathcal A' \subseteq \mathcal B(\Hil')$, one has the tensor product $\mathcal A \otimes \mathcal A' \subseteq \mathcal B(\Hil \otimes \Hil')$ which describes the physical system composed of $\mathcal A$ and $\mathcal A'$. It is the algebra spanned by all pairs of the form $\alpha \otimes \alpha'$ where $\alpha \in \mathcal A$ and $\alpha' \in \mathcal A'$, with the product defined pairwise: $(\alpha \otimes \alpha')(\beta \otimes \beta') = (\alpha \beta) \otimes (\alpha' \beta')$. The algebra $\mathcal A$ exists within $\mathcal A \otimes \mathcal A'$ in the form of the operators $\alpha \otimes \one$ where $\alpha \in \mathcal A$. Reciprocally, the elements $\alpha' \in \mathcal A'$ are represented in $\mathcal A \otimes \mathcal A'$ through the operators $\one \otimes \alpha'$. 



\msection{Infinite-dimensional quantum theory}
\label{section:formalism}

Let us present a possible way of extending the quantum formalism presented in Section \ref{section:findimqm} to infinite-dimensional Hilbert spaces. The formalism that we will present is standard, in its most general form, in algebraic quantum field theory, but it also reduces easily to a convenient formalism for non-relativistic quantum mechanics.


A lot of subtleties associated with infinite-dimensional quantum mechanics can be understood as coming from the fact that an infinite-dimensional vector space is not isomorphic to its double-dual. 
This means that we have to be careful in choosing whether we want the states to be functionals on effects, or the effects to be functionals on states.

In the following chapters, we will make essential use of sharp effects, i.e. projectors. This implies that we should work not with general C$^*$-algebras, but instead with their closure in the weak topology, which are von Neumann algebras. Indeed, \ind{von Neumann algebras} are spanned by their projectors, which implies that many of their properties can be deduced from the properties of the projectors themselves.  

The Von Neumann algebras which represent typical classical systems are however rather large. This implies that if we were to assume that states are linear functionals on the effects, which span the algebra, there would be an awful lot of them. We will see that it is in fact more convenient to assume that the effects are functionals on states, i.e. that the von Neumann algebra is the dual of the span of the set of states. This is where another advantage of von Neumann algebras appears: they are always the dual of something, namely a \ind{Banach algebra} (i.e. a C$^*$-algebra without the axiom which states that $\|\alpha^* \alpha\| = \|\alpha\|^2$). If $\mathcal A$ is a von Neumann algebra, we will write \ind{$\mathcal A_*$} for its {\em \ind{pre-dual}}, i.e. the Banach algebra which is such that $(\mathcal A_*)^* = \mathcal A$.

For instance, \ind{non-relativistic quantum mechanics} can be formalized in a satisfactory way by using for $\mathcal A$ the set of bounded operators on a separable Hilbert space $\Hil$, which we write as $\mathcal A = \mathcal B(\Hil)$. Its pre-dual can be identified with the Banach algebra of \ind{trace-class operators} $\mathcal A_* = \mathcal B_t(\Hil)$\index{$\mathcal B_t(\Hil)$}. The trace-class operators are the elements $A \in \mathcal B(\Hil)$ for which the expression 
\[
\sum_i \bra{i} \sqrt{A^* A} \ket{i}
\]
converges for any basis $\ket{i}$. 
For self-adjoint elements $\rho \in \mathcal B_t(\Hil)$, this defines the {\em \ind{trace}} 
\[
\tr(\rho) := \sum_i \bra{i} \rho \ket{i}.
\]
In fact, the product of an element of $\rho \in \mathcal B_t(\Hil)$ with any operator $A \in \mathcal B(\Hil)$ is also trace-class, which implies that we can define $\tr(\rho A)$. This is how $\mathcal B_t(\Hil)$ is identified with $(\mathcal B(\Hil))_*$. Indeed, $\rho \in \mathcal B_t(\Hil)$ defines the functional
\[
A \mapsto \tr(\rho A).
\]

A general von Neumann algebra is equipped with a \ind{weak-$*$ topology}, which amounts to defining the convergence of a sequence $A_n \in \mathcal B(\Hil)$ in terms of expectation values. I.e. we say that $A_n$ converges to $A$ as $n \rightarrow \infty$ if $\rho(A_n)$ converges to $\rho(A)$ for all states $\rho \in \mathcal A_*$, which makes some physical sense. In the case of $\mathcal B(\Hil)$, this means that the sequence converges to $A$ if the numbers $\tr(\rho A_n)$ converge to $\tr(\rho A)$ as $n \rightarrow \infty$ for all states $\rho \in \mathcal B_t(\Hil)$. 

This implies that the states represented by elements of $\mathcal B_t(\Hil)$, seen as linear functionals of effects, are continuous with respect to the weak-$*$ topology. Indeed, if the sequence $\{A_n\}_{n=1}^\infty$ converges to $A$ in this topology, then, by definition $\tr(\rho A_n) \rightarrow \tr(\rho A)$. Hence the map $A \mapsto \tr(\rho A)$ is continuous. Conversely, those are {\em all} the weak-$*$ continuous positive linear functionals. Therefore, our choice of states corresponds to restricting the natural set of all linear functionals on effects to only those which are weak-$*$ continuous, or {\em normal}\index{normal states} for short. 



Note that a von Neumann algebra, when seen as a subalgebra of $\mathcal B(\Hil)$, also has another weak topology, which is simply called the {\em weak topology}. It corresponds to defining the limit of operators only in terms of pure states, i.e, we say that $A_n$ converges weakly to $A$ if $\bra{\psi} A_n \ket{\psi}$ converges to $\bra{\psi} A \ket{\psi}$ for all $\ket{\psi} \in \Hil$. Hence convergence in the weak-$*$ topology is a stronger condition.



\subsection{Classical, quantum and hybrid systems}
\label{section:hybrid}




A classical system will be assumed to be specified by a \ind{commutative von Neumann algebra}, which can always be represented as the algebra \ind{$L^\infty(\Omega)$} of (almost everywhere) bounded functions defined on a measure space $\Omega$. 
This set $\Omega$ can be interpreted as the phase-space. The fact that $\Omega$ is a measure space means that functions on it can be integrated.
The pre-dual, which contains the states, is then identified with the algebra \ind{$L^1(\Omega)$} of absolutely integrable functions on $\Omega$. These are the functions $\mu$ which are such that their absolute value $x \mapsto |\mu(x)|$ has a finite integral over $\Omega$.  
The state represented by the positive function $\mu \in L^1(\Omega)$ is then simply given, for $\alpha \in L^\infty(\Omega)$, by 
\[
\alpha \mapsto \int_\Omega \mu(x) \, \alpha(x) \,dx.
\]
The normalization of the state $\mu$ is therefore given by
\[
\int_\Omega \mu(x) \, dx = 1.
\] 
Hence the states are standard continuous probability distributions. We see that this formalism only handles probability measures on $\Omega$ which are absolutely continuous with respect to the default measure on $\Omega$. 
If we need a more exotic classical state, we may need to take a different $\Omega$. For instance if we choose $\Omega = \mathbb R$ with the standard Lebesgue measure, we will not have any discrete probability measure (i.e. no delta function).
However, we can include them by using instead the disjoint union $\Omega = \mathbb R \cup \mathbb N$.



A general von Neumann algebra $\mathcal A$ has a {\em \ind{center}}, denoted by $\mathcal Z(\mathcal A)$, which is the commutative subalgebra containing the elements of $\mathcal A$ which commute with all other elements of $\mathcal A$. In general, the center of an algebra can be understood as the ``classical'' part of the algebra, since it characterizes observables which commute with everything else, and are therefore not subject to any uncertainty relation. In the case of a commutative algebra, this center is obviously the whole algebra. 

From this point of view, it makes sense to define a ``pure'' quantum system as being defined by an algebra whose center is trivial. The center is trivial if it contains only elements of the form $z \one$,  $z \in \mathbb C$. Such an algebra is called a {\em \ind{factor}}. 
There exists three main types of factor. 
Most of the time we will assume that we are working with $\mathcal B(\Hil)$ which is the most general form of a factor of {\em type I}. We note that factors of type II are rarely used in physics, and factors of type III appear in algebraic quantum field theory. Remember however that all von Neumann algebras can be seen as subalgebras of some $\mathcal B(\Hil)$. 


Note that this formalism, with the physical interpretation based on states and effects, works fine even if the von Neumann algebra $\mathcal A$ is not a factor, nor commutative. In fact we will see that these systems can be decomposed into factors labelled by a classical variable. In this sense, they represent {\em hybrid quantum-classical} information\index{hybrid systems} \cite{kuperberg02}.

Let us summarize this in a definition:
\begin{dfn}
A physical model characterized by a von Neumann algebra $\mathcal A$ will be called {\em classical} if $\mathcal A$ is commutative, {\em quantum} if $\mathcal A$ is a factor and {\em hybrid} otherwise.
\end{dfn}

Since all commutative von Neumann algebras are of the form $L^\infty(\Omega)$, we will always explicitly use this representation for classical systems. Also note that we will in general only be working with factors of type $I$ which are of the form $\mathcal B(\Hil)$. In addition, we will often assume that $\Hil$ is separable. 

A special case to point out is the von Neumann algebra $\mathbb C$ which is both commutative and a factor. It has only one state, and therefore represents a system with no degree of freedom. It is such that $\mathbb C \otimes \mathcal A = \mathcal A \otimes \mathbb C = \mathcal A$ for any von Neumann algebra $\mathcal A$. Indeed, $z \otimes A = 1 \otimes z A$ for any $z \in \mathbb C$, which means that $\mathbb C \otimes \mathcal A$ can be identified with $\mathcal A$ through the map $z \otimes A \mapsto z A$ and its inverse $A \mapsto 1 \otimes A$. The same argument works for $\mathcal A \otimes \mathbb C$.


Let us list a few facts about von Neumann algebras that we will use
\begin{itemize}
\item The commutant of any ``self-adjoint'' set of operators (i.e. closed under the \adjoint operation) is a von Neumann algebra \cite{jones}. 
\item A von Neumann algebra is the norm-closure of the span of its projectors \cite{jones}. This also implies that it is the weak-$*$ closure of the span of its projectors. 
\item A commutative von Neumann algebra is always of the form $L^\infty(\Omega)$ for some measure space $\Omega$ \cite{dixmier81}.
\item A factors of type I is isomorphic to $\mathcal B(\Hil)$ for some Hilbert space $\Hil$ \cite{dixmier81}.
\item If $\Hil$ is separable, then $\mathcal B_t(\Hil)$ is separable \cite{goldberg59}. 
\end{itemize}



\subsection{Channels}
\label{section:infchan}

A communication \ind{channel} from a system represented by the algebra $\mathcal A_A$ to a system represented by $\mathcal A_B$ can be defined, as in the finite-dimensional case, by a unital completely positive linear map
\[
\chanh: \mathcal A_B \rightarrow \mathcal A_A
\]
on effects.
However, since we asked for the states to be weak-$*$ continuous functions, we also need the channels to be weak-$*$-weak-$*$ continuous, which we will summarize by saying that they are {\em normal}\index{normal map}. This implies that $\chan^*$ is always the \adjoint of a linear map
\[
\chan: \mathcal (\mathcal A_A)_* \rightarrow  \mathcal (\mathcal A_B)_*
\]
which is meant to act on states. As in the finite-dimensional case, it is the map on state $\chan$ that we will call a channel, while $\chanh$ will only be called its dual.

If $\mathcal A_A = \mathcal B(\Hil_A)$ and $\mathcal A_B = \mathcal B(\Hil_B)$ for some separable Hilbert spaces $\Hil_A$ and $\Hil_B$, then these maps can be represented as 
\[
\chanh(A) = \sum_{k=1}^\infty E_k^* A E_k
\]
and
\[
\chan(\rho) = \sum_{k=1}^\infty E_k \rho E_k^*
\]
where the sum can now be infinite \cite{kraus83}. We will call this the \ind{Choi-Kraus form} of the channel $\chan$. 
The elements $E_k$ are bounded linear operators from $\Hil_A$ to $\Hil_B$. This can be understood starting from the \ind{Stinespring dilation} theorem for completely positive maps between C$^*$-algebras, which states that there is a representation $\pi$ of $\mathcal A_B$ on some Hilbert space $\Hil$, and an isometry $V: \Hil_B \rightarrow \Hil$, such that
\[
\chanh(A) = V^* \pi(A) V
\] 
for all $A \in \mathcal A_B$. Since $\mathcal A_B = \mathcal B(\Hil_B)$, and $\chanh$ is normal, the representation $\pi$ on $\Hil$ is of the form $\pi(A) = A \otimes \one$. 
Also, if $\Hil_B$ is separable, then so is $\Hil$ \cite{paulsen02}. Therefore the subsystem on which $\pi(A)$ acts trivially is also separable, and possesses a discrete basis $\ket{i}$. This implies that
\[
\begin{split}
\chanh(A) &= V^* (A \otimes \one) V = \sum_i V^* (A \otimes \proj{i}) V  \\
&= \sum_i (V^* \otimes \ket{i})A  (V \otimes \bra{i}).
\end{split}
\]
Therefore the elements\index{elements of a channel} of the channel $\chan$ can be chosen to be $E_i = V \otimes \bra{i}$. 
Note that this makes it clear that there is a large ambiguity in the choice of the elements $E_i$. Indeed, any orthonormal basis $\ket{i}$ would potentially yield a different set of elements.

This definition applies for arbitrary von Neumann algebras $\mathcal A_A$ and $\mathcal A_B$. In particular, if both are commutative, this simply defines a \ind{stochastic map} between classical probability distributions. 


\subsection{Observables}

Given that a classical system is also modeled by a von Neumann algebra, we will simply define an observable to be a {\em channel} (in the above sense) from a classical system to a quantum system, i.e.
\begin{dfn}
An {\em \ind{observable}} $X$ on the system specified by the algebra $\mathcal A$ and which takes values in the measure space $\Omega$, is specified by a unital positive normal map
\(
\someobs^*: L^\infty(\Omega) \rightarrow \mathcal A, 
\)
or equivalently by a trace-preserving positive linear map
\[
\someobs: \mathcal A_* \rightarrow L^1(\Omega).
\]
of which $X^*$ is the \adjoint.
\end{dfn}
In this case the complete positivity comes for free as in the finite-dimensional case \cite{paulsen02}. Note that in general, $\mathcal A_*$ is not equipped with a trace like the one defined for $\mathcal B_t(\Hil)$. Hence the trace which is preserved in the general case is $\tr(\rho) := \rho(\one)$. 

\begin{figure}
\centering
\includegraphics[width=0.35\textwidth]{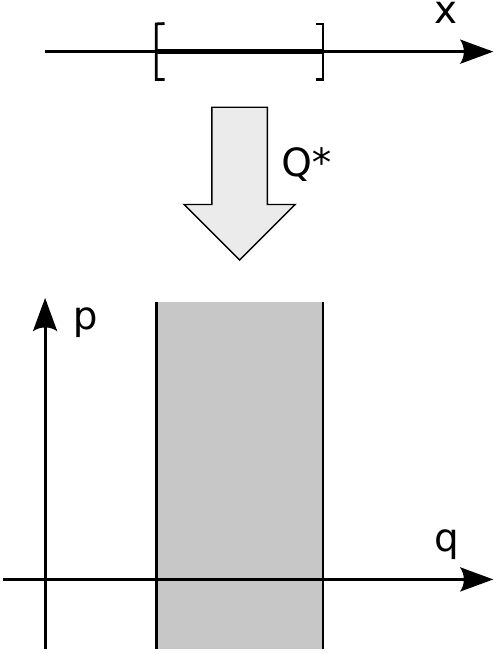}
\caption[Classical position observable]{\small The classical position observable is defined by the positive map $Q^*$ which sends a function on $\mathbb R$, which represents a proposition about the position, to a function on $\mathbb R^2$ representing a proposition about the full phase-space. The interval on $\mathbb R$ here represents the characteristic function for this interval, which is a sharp effect. It is mapped to the characteristic function for an infinite stripe in phase-space. }
\label{figure:position}
\end{figure} 

This definition also holds when $\mathcal A$ is commutative, where it defines a classical observable\index{observable (classical)}. Note that a classical observable is identical to a classical channel, i.e. a stochastic map. 
For instance, consider the case where $\mathcal A = L^\infty(\mathbb R^2)$, which could describe a single classical particle with phase-space $\mathbb R^2$ parameterized by position-momentum pairs $(q,p) \in \mathbb R^2$. The position observable is then represented by the map 
\[
Q^*: L^\infty(\mathbb R) \rightarrow L^\infty(\mathbb R^2)
\]
which sends a function $\alpha$ on $\mathbb R$ to the function $\beta = Q(\alpha)$ defined by $\beta(q,p) := \alpha(q)$. Hence it says how a question about position can be translated into a question about phase-space (see Figure \ref{figure:position}).



This notion of an observable corresponds only to a subset of the positive operator valued measures\index{positive operator valued measure}\index{POVM}. In order to obtain a POVM $E$ from the observable $X$, it suffices to apply $X^*$ to the characteristic functions $\chi(\omega)$ of each subsets $\omega \subseteq \Omega$. Hence the POVM $E$ corresponding to $X$ is given by
\[
E(\omega) = X^*(\chi_\omega)
\]
for any $\omega \subseteq \Omega$. This is indeed a POVM since
\[
E(\bigcup_k \omega_k) = X^*(\sum_k \chi_{\omega_k}) = \sum_k  X^*(\chi_{\omega_k}) = \sum_k(\omega_k)
\]
for any disjoint sequence of subsets $\omega_k \subseteq \Omega$, and also
\[
E(\Omega) = X^*(\lone) = \one.
\]
In order to see that this does not yield every POVMs, consider the case where $\mathcal A$ is one-dimensional, which is to say $\mathcal A = \mathbb C$. Then a POVM $E$ on $\Omega$ with values in $\mathcal A = \mathbb C$ is any positive measure on $\Omega$. Now suppose that $X$ is an observable on $\mathcal A = \mathbb C$ as defined above. Then it is a linear map from $\mathbb C$ to $L^1(\Omega)$, which is just an element $\alpha \in L^1(\Omega)$ such that $X^*(\chi_\omega) = \int_\omega \alpha(x) \,dx$. Hence $X$ can only represent measures which are absolutely continuous with respect to the measure associated with $\Omega$, contrary to a POVM which could represent any positive measure. This means in particular that, if we were to take $\Omega = \mathbb R$ then we could not represent any discrete POVM. However this can be overcome simply by choosing instead the disjoint union $\Omega = \mathbb R \cup \mathbb N$. 

Note that we could have modeled all POVMs if we did not require that the channel representing the observable be weak-$*$ continuous. Indeed, a POVM can be seen as a unital positive linear map
\[
E: C_0(\Omega) \rightarrow \mathcal A
\]
where $C_0(\Omega)$ is the C$^*$-algebra of continuous functions on $\Omega$ with compact support \cite{paulsen02}. 
 
The fact that our notion of observables is more restrictive than the usual definition in terms of POVM is desired, because it makes the {\em set of all observables} easier to handle. In particular, if $\mathcal A$ is finite-dimensional, then the set of all the observables with value in a fixed set $\Omega$ is separable.


\subsection{Sharp observables}

As in the finite-dimensional case, an observable $X$ will be called {\em sharp}\index{sharp observable (continuous)} if it maps all the sharp classical propositions $\chi_\omega$, $\omega \subseteq \Omega$ to sharp quantum propositions $(X^*(\chi_\omega))^2 = X^*(\chi_\omega)$. This implies that  the corresponding POVM is in fact a \ind{projection valued measure} (PVM)\index{PVM}. A PVM is also known as an {\em orthogonal \ind{partition of the unity}}. It can always be seen as representing the {\em spectral measure}\index{spectral measure} of a self-adjoint operator $A$. If $\Omega = \mathbb R$, and we define the PVM $\Pi(\omega) := X^*(\chi_\omega)$, we have
\[
A = \int_{x \in \mathbb R} x \, \Pi(dx)
\] 
This notation can be understood intuitively as meaning that we define the integral as the limit of discrete sums with intervals $\Delta x$. $\Pi(dx)$ can then be conceived intuitively as the limit of the operators $\Pi(\Delta x)$. 

If the sharp observable takes value in a set $\Omega \neq \mathbb R$, representing it as the spectral measure of a self-adjoint operator $A$ would require a measure-preserving map $f: \Omega \rightarrow \mathbb R$. We could then write
\[
A := \int_{x \in \Omega} f(x) \Pi(dx).
\]
This is due to the fact that an observable represented by a self-adjoint operator takes values in its spectrum, which must belong to $\mathbb R$. It suffices to think about the case $\Omega = \mathbb R^2$ to realize that the map $f$ can be pretty messy. This shows that the formalism of PVMs can be more convenient than that of self-adjoint operators in that it easily allows to talk about observables with non-scalar values.


Conversely, if we are given the self-adjoint operator $A$, then the corresponding PVM is given by the $L^\infty(\Omega)$ functional calculus of $A$, where $\Omega$ is the spectrum of $A$. This means that for any classical proposition $\alpha \in L^\infty(\Omega)$, we obtain the effect $X^*(\alpha)$ through the relation
\[
X^*(\alpha) = \alpha(A).
\]

The map $X^*$ is a von Neumann algebra isomorphism. In particular it satisfies
\[
X^*(\alpha \beta) = (\alpha \beta)(A) = \alpha(A) \beta(A) = X^*(\alpha) X^*(\beta).
\]
The image of $X^*$ is a commutative subalgebra of the von Neumann algebra on which it is defined, namely the von-Neumann algebra generated by $A$, which is a faithful representation of the algebra $L^\infty(\Omega)$. 





\subsection{A category}
\label{section:category}

One advantage of the formalism chosen is that classical systems and quantum systems are both modeled in the same way, as von Neumann algebras, and all the maps of interest are normal completely-positive unital maps between these algebras. In the dual picture, we are working with the pre-duals of von Neumann algebras and the trace-preserving completely positive maps between them. 
We will generally make use of the maps between pre-duals since they directly say how {\em states} evolve. Intuitively, the right direction of the flow of information is from states to states. Indeed, states represent the information that we, the observers, possess about the systems, and it is our information which flows from one to the other. 

Hence most of our discourse will take place in the \ind{category} whose objects\index{objects (category)} are the pre-duals of von Neumann algebras, and whose morphisms\index{arrows (category)}\index{morphism (category)} are trace-preserving completely positive maps between them \cite{kuperberg05}. For instance, an observable $X$ of a quantum system represented by the Hilbert space $\Hil$ is a morphism
\begin{diagram}
\mathcal B_t(\Hil) & \rTo^{X}  & L^1(\Omega)\\
\end{diagram}
where $L^1(\Omega)$ contains the states of the classical pointer which represents the value of the observable. 
Of course, quantum channels are also morphisms:
\begin{diagram}
\mathcal B_t(\Hil_A) & \rTo^{\chan}  & \mathcal B_t(\Hil_B)\\
\end{diagram}
States themselves can be represented by the morphisms
\begin{diagram}
\mathbb C & \rTo^{\rho} & \mathcal B_t(\Hil)
\end{diagram}
Indeed, all that this map $\rho$ can do is send each complex number $z$ to $z$ times a fixed element, say $\hat \rho$, of $\mathcal B_t(\Hil)$: $z \mapsto z \hat \rho$. Since $\rho$ is positive, $\hat \rho$ must be a positive operator. In addition, the fact that $\rho$ is trace-preserving implies $\tr(z \hat \rho) = z\tr(\hat \rho) = \tr(z) = z$, i.e. $\tr(\hat \rho) = 1$.


As an exercise, note that the trace is the only morphism from $\mathcal B_t(\Hil)$ to $\mathbb C$. It amounts to erasing all information.
\begin{diagram}
\mathcal B_t(\Hil) & \rTo^{\tr} & \mathbb C\\
\end{diagram}
Effects can be represented as observables with pointer $\Omega = \{0,1\}$, i.e. with target $L^1(\{0,1\}) = L(\{0,1\}) \approx \mathbb C^2$:
\begin{diagram}
 \mathcal B_t(\Hil) & \rTo^{A} & L^1(\{0,1\})\\
\end{diagram}

At this point we may wonder about the possible meaning of a morphism $\phi:  L^1(\Omega) \rightarrow \mathcal B_t(\Hil)$, i.e. a flow of information from a classical system to a quantum system.
Since it is linear it must be defined by its action on a basis of $L^1(\Omega)$. Consider for instance the case where $\Omega$ is finite. Then a basis is given by the pure states $\chi_i(j) = \delta_{ij}$ for all $i,j \in \Omega$. Since these functions are states, the images are quantum states $\rho_i := \phi(\chi_i)$ which are completely arbitrary. 
Hence $\phi$ is a state-preparation device. It encodes classical information into a quantum system by directly mapping the pure classical states into arbitrary quantum states:
\begin{diagram}
L^1(\Omega) & \rTo^{\phi} & \mathcal B_t(\Hil)\\
\end{diagram}

The composition of a state preparation and an observation yields a channel between two classical systems:
\begin{diagram}
   L^1(\Omega_A) & \rTo^{\phi} & \mathcal B_t(\Hil) & \rTo^{X} &  L^1(\Omega_B)\\
\end{diagram}
which is simply a stochastic process mapping a probability distribution on $\Omega_A$ into a probability distribution on $\Omega_B$. For instance we get the \ind{classical capacity} $\chan$ by maximizing the Shannon capacity of the classical channel
\begin{diagram}
   L^1(\Omega_A) & \rTo^{\phi} & \mathcal B_t(\Hil_A)  & \rTo^{\chan} & \mathcal B_t(\Hil_B) & \rTo^{X} &  L^1(\Omega_B)\\
\end{diagram} 
over all states preparations $\phi$ and all observables $X$.




\msection{Summary}

\begin{table}
\begin{center}
\begin{tabular}{|l|c|c|c|c|c|}
\hline
 & Hybrid  & \multicolumn{2}{c|}{Classical} & \multicolumn{2}{c|}{Quantum} \\
\hline 
Algebra is ... &  arbitrary & \multicolumn{2}{c|}{commutative}  & \multicolumn{2}{c|}{factor of type I} \\
\hline
dimension &  arbitrary & finite & arbitrary & finite &  arbitrary \\
\hline
Algebra & $\mathcal A, \mathcal B, \dots$ & $\finfun(\Omega)$ & $L^\infty(\Omega)$ & $\finops(\Hil)$ & $\mathcal B(\Hil)$ \\
Pre-dual & $\mathcal A_*, \mathcal B_*, \dots$ & $\finfun(\Omega)$ & $L^1(\Omega)$ & $\finops(\Hil)$ & $\mathcal B_t(\Hil)$\\
\hline 
Effects & $\props {\mathcal A}, \props {\mathcal B}, \dots$ & \multicolumn{2}{c|}{$\cprops \Omega$} & \multicolumn{2}{c|}{$\qprops \Hil$}\\
States & $\states {\mathcal A}, \states {\mathcal B}, \dots$ & \multicolumn{2}{c|}{$\cstates \Omega$} & \multicolumn{2}{c|}{$\qstates \Hil$}\\
\hline
\end{tabular}
\end{center}
\caption{\small Summary of the symbols used for sets of propositions and states.}
\label{table:notation}
\end{table}


In this chapter, we have seen that we can model quantum and classical systems in terms of their set of effects, or equivalently in terms of the dual set of states. Effects represent elementary propositions and states assign a probability to each effects. In both the quantum and classical cases, the effects can be seen as elements of a C$^*$-algebra. The essential aspect of these algebras, for our purpose, is the fact that they are ordered vector spaces, i.e. vector spaces with a suitably compatible partial order. The effects are those vectors which are larger than zero and smaller than $\one$. For our purposes, it is convenient to assume that the C$^*$-algebras have a pre-dual, which means that they are von-Neumann algebras. We then restrict states to those functionals which live in the pre-dual. 
Table \ref{table:notation} summarizes some of our notations.

A transfer of information, or channel, from system $A$ to system $B$ is represented by a map from the states of $A$ to the states of $B$, whose dual is a unital completely-positive map from effects of $B$ to effects of $A$. 
For clarity, it is always the map on {\em states} which will be referred to as the {\em channel}, since it goes in the direction which we intuitively associate with the direction of the flow of information. 

A quantum observable, or observable for short, is a channel from a quantum to a classical system. The classical system can be interpreted as that part of a measurement apparatus which displays the result of the measurement. This notion of an observable is essentially equivalent to that of a POVM, although slightly more restrictive. 


\mchapter{Preserved information}%
\label{chapter:preserved}





Recall that our aim is to characterize decoherence through the type of information flowing between a system and its environment, or within the system. Therefore, we need a way to formally talk in detail about the nature of the information preserved by a quantum channel. In this chapter, we introduce the main new concept that we will be using, and derive a number of basic results about it. 
Although our results are based on many known results, mostly in the theory of quantum error correction, we will present them in a logical, rather than a historical order. Hence we will not mention the known results here, but instead show how to derive them from our more general point of view in the next chapter.

The main definition that we will be playing with is the following:
\begin{dfn}
An observable $X$ is said to be {\em preserved}\index{preserved observable} by a channel $\chan$ if there exists an observable $Y$ such that $X = Y \circ \chan$.
We write \ind{$\preserved \chan$} for the set of observables preserved by $\chan$. 
\end{dfn}
Recall that the equation $X = Y \circ \chan$ means that $X$ is the image of $Y$ in the Heisenberg picture (see Section \ref{section:heisenberg}).
For a discrete observable with elements $X_i$ and $Y_i$, this means that $X_i = \chanh(Y_i)$ for all $i$. If $\Omega$ is the set of values of $X$, we have:
\begin{diagram}
  {\mathcal A_1} &\rTo^{\quad \chan \quad}  & {\mathcal A_2}\\
  & \rdTo_{X} & \dTo_{Y} \\
  & & L^1(\Omega)
\end{diagram}
Hence, for any state of system $A$, measuring $Y$ on system $B$ yields the same result (i.e. the same probability distribution on $\Omega$) as measuring $X$ on the initial system $A$. This justifies the idea that the information about $A$ represented by $X$ has been preserved by the channel $\chan$: we can perfectly simulate an observation of $X$ on the initial state by observing $Y$ after the action of the channel. 

If the observable $X$ is sharp, we know that it can also be represented by a self-adjoint operator $\widehat X$. As an example, suppose that this observable has a discrete spectrum. This implies that it is of the form $\widehat X = \sum_i x_i P_i$ where $x_i$ is the eigenvalue corresponding to the spectral projection $P_i$. In this case, $X$ is preserved if $P_i = \chanh(Y_i)$ for all $i$, where the effects $Y_i$ form a discrete observable, i.e. $\sum_i Y_i = \one$.

 It is not sufficient to require that $\widehat X = \chanh(\widehat Y)$ for some self-adjoint operator $\widehat Y$. Indeed, this equation only means that the first moment\index{preserved first moment} of $\widehat X$ is preserved for any state $\rho$:
\(
\tr(\rho \widehat X) = \tr(\chan(\rho) \widehat Y).
\)
On the other hand, if we require that $P_i = \chanh(Y_i)$ for all $i$, then all the moments of $X$ are preserved. Indeed, for any $n$, $\widehat X^n = \sum_i x_i^n P_i$, from which it is easy to see that also 
\[
\tr(\rho \widehat X^n) = \tr(\chan(\rho) \widehat Y^n)
\] 
for any $n$.

Note that we can also define the moments of any scalar (and possible unsharp) observable $X$ defined by
\[
X^*(\alpha) = \int_{\mathbb R} \alpha(x) \, X_x \, dx
\]
where $X_x$ are operators, and $\alpha$ a real function. 
The $n$th moment of $X$ can be represented by the operator
\[
\widehat {X^n} := \int_{\mathbb R} x^n X_x \,dx.
\]
Indeed, we have
\[
\tr(\rho \widehat {X^n}) =  \int_{\mathbb R} x^n \tr(\rho X_x) \,dx = \int_{\mathbb R} x^n X(\rho)(x) dx 
\]
This equation means that the expectation value of $\widehat {X^n}$ on $\rho$ yields the $n$th moment of the probability distribution $X(\rho)$.
If $X$ is preserved, then so are its moments, in the sense that $\widehat {X^n} = \chanh(\widehat {Y^n})$ for all $n$. 



Consider the case of yes-no observables, i.e. observables with two outcomes, which are one-to-one with effects. If $X_0$ and $X_1$ are the two elements characterizing a yes-no observable, we must have $X_0 + X_1 = \one$, which implies $X_1 = \one - X_0$. This observable is preserved by the channel $\chan$ when there exists another yes-no observable $Y$ such that $X_0 = \chanh(Y_0)$ and $X_1 = \chanh(Y_1)$. However the second equation is redundant. Indeed, this first equation implies $\chanh(Y_1) = \chanh(\one - Y_0) = \chanh(\one) - \chanh(Y_0) = \one - X_0 = X_1$. Therefore, the preserved yes-no observables are characterized by what we define as the {\em \ind{preserved effects}}, namely the effects which are image of another effect under $\chanh$.
\begin{dfn}
An effect $A \in \props {\mathcal A_1}$ is {\em preserved} by $\chan: {\mathcal A_1} \rightarrow {\mathcal A_2}$ if there exists $B \in \props {\mathcal A_2}$ such that $A = \chanh(B)$. Hence, the \ind{set of preserved effects} is $\chanh(\props {\mathcal A_2})$. 
\end{dfn}

A general preserved observable $X$ satisfies $X^* = \chanh \circ Y^*$, which means that for any classical proposition $\alpha$, $X^*(\alpha) = \chanh(Y^*(\alpha))$. Hence all the effects $X^*(\alpha)$ associated with the observable $X$ are individually preserved by the map $\chan$. For a discrete observable $X$, this means that any sum of its elements belongs to $\chanh(\props {\mathcal A_2})$.


\msection{A set of observables?}
\label{section:preservedeff}

We introduced a notion of preserved observable, and therefore also a notion of preserved effect. Since preserved observables are made of preserved effects, we may wonder if we could limit the discussion to the set of preserved effects, which are simpler objects. In addition, the set of preserved effects is an effect algebra, which can be understood as defining an effective physical system of its own, as discussed in Section \ref{section:effects}.

However, the set of preserved effects, although it certainly gives interesting information about the channel, does not always contain as much information as the set of preserved observables. Indeed, it is not always true that a preserved observable is characterized by preserved effects. A counter example is given in Section \ref{section:counterexample} below. 


However it is true in the ``generic case'', when the channel has no kernel:
\begin{prop}
\label{prop:invertible}
If the channel $\chan: {\mathcal A_1} \rightarrow {\mathcal A_2}$ is invertible, then the two following statements are equivalent
\begin{enumerate}
\item The observable $X$ is preserved by $\chan$,
\item $X^*(\alpha) \in \chanh(\props {\mathcal A_2})$ for all $\alpha$.
\end{enumerate}
\end{prop}
\proof We have already shown above that Statement $1$ implies Statement $2$. Conversely, suppose that $X^*(\alpha) \in \chanh(\props {\mathcal A_2})$ for all $\alpha$. Let $\phi := (\chanh)^{-1} \circ X^*$. We have to show that the linear map $\phi$ is positive and unital. Since $\chanh$ is invertible, its range is the whole positive cone of $\mathcal A_2$ and its inverse maps the positive cone of $\mathcal A_2$ to that of $\mathcal A_1$, which means that it is also positive. Given that $X^*$ itself is positive, $\phi$ is positive. In addition we have that $(\chanh)^{-1}(\one) = \one$, which implies $\phi(\one) = \one$.
\qed


\subsection{A counter-example}
\label{section:counterexample}

Let us now give an example of a channel, together with an observable which is not preserved but which is made of preserved effects.
We consider a channel of the form
\[
\chan: \finops(\mathbb C^3) \rightarrow \finops(\mathbb C^4).
\]
Let $\ket{i}$, $i = 1,\dots,4$ denotes a basis of $\mathbb C^4$. For convenience, we will use its first three elements $\ket{i}$ for $i=1,\dots,3$ to define a basis of $\mathbb C^3$ too. Also, we write $P_i := \proj{i}$. 
We define $\chan$ by
\[
\chan(\rho) := \frac{1}{3} \sum_{i=1}^3 \proj{i} \rho \proj{i} + \frac{2}{3} \tr(\rho) P_4
\]
It is easy to see that this map is completely positive. Indeed the first term already has the Choi-Kraus form, and the second term can be written as $\frac{2}{3} \sum_{i=1}^3 \ket{4}\bra{i} \rho \ket{i}\bra{4}$. Also it is trace-preserving. Indeed, 
\[
\begin{split}
\tr(\chan(\rho)) &=  \frac{1}{3} \sum_{i=1}^3 \bra{i} \rho \ket{i} + \frac{2}{3}\tr(\rho) \\
&= \frac{1}{3} \tr(\rho) + \frac{2}{3} \tr(\rho) = \tr(\rho).
\end{split}
\]
The \adjoint is
\[
\chanh(A) =  \frac{1}{3} \sum_{i=1}^3 \proj{i} A \proj{i} + \frac{2}{3} \bra{4}A\ket{4} \one_3
\]
Now consider the discrete observable $X$ on $\mathbb C^3$ with elements
\begin{align*}
X_1 &= \frac{1}{3}(P_2+P_3) + \frac{1}{3} \one \\
X_2 &= \frac{1}{3}(P_1+P_3)\\
X_3 &= \frac{1}{3}(P_1+P_2)
\end{align*}
These elements are clearly positive and sum to the identity:
\[
\sum_i X_i = \frac{2}{3} \sum_i P_i + \frac{1}{3} \one_3 = \one_3.
\] 
In addition, any sums of the elements $X_i$ are inside $\chanh(\qprops {\mathbb C^4})$. To see this it is useful to note that $\chanh$, viewed as a linear map, has a kernel which is generated by 
\[
K = P_1 + P_2 + P_3 - \frac{1}{2}P_4.
\]
Since $\chanh(K) = 0$, we can subtract any multiple of $K$ in the argument of $\chanh$ in order to make sure that the argument is an effect. If we define the effects
\begin{align*}
Y^0_1 &:=  P_2 + P_3 + \frac{1}{2}P_4\\
Y^0_2 &:= P_1 + P_3\\
Y^0_3 &:= P_1 + P_2
\end{align*}
then we have $X_i = \chanh(Y^0_i)$ and also
\begin{align*}
X_1 + X_2 &= \chanh( Y^0_1 + Y^0_2 - K ) = \chanh( P_2 + P_4  )  \\ 
X_2 + X_3 &= \chanh( Y^0_2 + Y^0_3 - K ) = \chanh( P_1 + \frac{1}{2} P_4)  \\ 
X_1 + X_3 &= \chanh( Y^0_1 + Y^0_3 - K ) = \chanh(  P_2 + P_4  )  \\ 
\end{align*}
This suffices to show that $X^*(\alpha) \in \chanh(\qprops {\mathbb C^4})$ for any $\alpha$. 
However, the observable $X$ as a whole is {\em not} preserved by the channel $\chan$.
In order for it to be preserved, we would need to find another observable $Y$ with elements $Y_i$, $i = 1,\dots,3$ such that $X_i = \chanh(Y_i)$ for all $i$. We note that the effects $Y^0_i$ cannot serve this purpose because they do not form an observable. Indeed they fail to sum to the identity: $\sum_i Y^0_i = 2 P_1 + 2 P_2 + 2 P_3 + \frac{1}{2} P_4 = 2 \one_4 - \frac{1}{2} P_4$. If we wish to modify these three operators while maintaining their images under $\chanh$, the only thing we can do is add or subtract different multiples of $K$ to each of them. However it is easy to see that this cannot be done without getting them out of the set of effects $\qprops {\mathbb C^4}$. Indeed, note that they are all diagonal operators, hence the diagonal coefficient are the eigenvalues. It is easy to see that adding or subtracting any amount of $K$ to any of them would set some diagonal elements either smaller that $0$ or larger than $1$, which would mean that they are not effects anymore. 

Note that this example is entirely classical since all the operators involved are codiagonal. Therefore the counter example also applies to classical channels.

\subsection{Preserved effective theory}

As mentioned above, the set of effects preserved by a channel is an effect algebra. It can be understood as defining an effective physical system, namely a ``subsystem''---in a very general sense---of the initial quantum system. When the channel is invertible, this subsystem has a true operational existence as it effectively models the experimentally accessible degrees of freedom of the system filtered through the noise represented by the channel.


 However, 
the above counter-example shows that for more general channels, some of the observables represented by this set of effects cannot in fact be measured (are not themselves preserved). This implies that, in principle, we should use a more general framework than that of an effect algebra, and define a physical theory directly by a set of general observables. 

This explains why, in the rest of this thesis, we will always take the set of preserved observable as the fundamental object of study, and reduce it to effects only when possible.


\msection{Sharp preserved observables}
\label{section:sharpobs}

In this section we will consider a quantum channel $\chan: \mathcal B_t({\Hil_1}) \rightarrow \mathcal B_t({\Hil_2})$. We will also assume that $\Hil_1$ is separable. 
We saw in Section \ref{section:infchan} that such a channel can always be represented in terms of a countable family of channel elements $E_k$, as
\[
\chanh(B) = \sum_k E_k^* B E_k
\]
where the sum converges in the weak-$*$ topology. Note that from now on, all sums of operators will be assumed to be at most countably infinite, and convergent in the weak-$*$ topology. 

Let us consider the sharp preserved effects, i.e. those which belong to $\chanh(\qprops{\Hil_2})$. A sharp effect $P \in \qprops{\Hil_1}$ is a an effect which is also a projection, i.e. $P^2 = P$. If it is preserved, then there is an effect $B \in \qprops{\Hil_2}$ such that
\begin{equation}
\label{equ:preseff}
P = \chanh(B).
\end{equation}
In order to characterize the sharp preserved effects, we want to obtain an equivalent condition which involves only the channel elements $E_k$. 

First, let us state a simple technical lemma that we will use several times:
\begin{lem}
\label{lemma:trick}
If $\sum_k F_k^* B F_k = 0$ for some positive operator $B$ and a family of operators $F_k$, then $B F_k = 0$ for all $k$. 
\end{lem}
\proof
For any vector $\ket{\psi}$, we have 
\[
\sum_k \bra{\psi} F_k^* B F_k \ket{\psi} = \sum_k \bra{\psi} F_k^* \sqrt{B}^* \sqrt{B} F_k \ket{\psi} = \sum_k \| \sqrt B F_k \ket{\psi}\|^2 = 0.
\]
This implies that each positive term $\| \sqrt B F_k \ket{\psi} \|^2 $ in the sum is zero. Therefore $\sqrt B F_k \ket{\psi} = 0$ for any vector $\ket{\psi}$, which means that $\sqrt B F_k = 0$.
\qed

We could apply this lemma to Equation \ref{equ:preseff} if we had a zero on the left-hand side of the equation. This is the case if we multiply this term by $P^\perp = \one - P$. Indeed $P^\perp P = (\one - P) P = P - P^2 = P - P = 0$. But we want the right-hand side term to be positive for the trick to work, therefore we multiply by $P^\perp$ from both sides:
\[
P^\perp \chanh(B) P^\perp = P^\perp P P^\perp = 0
\]
Applying Lemma \ref{lemma:trick} we get $B E_k (\one - P) = 0$ for all $k$, i.e. 
\begin{equation}
\label{equ:preseff1}
B E_k = B E_k P.
\end{equation}
There is another similar equation that we can use. For every preserved effect $P$ we get another preserved effect for free, namely its orthocomplement: $\one - P$. Indeed, $\chanh(\one - B) = \one - \chanh(B) = \one - P$, and $\one - B$ is a valid effect. Using the same argument as above, we obtain $(\one - B) E_k P = 0$, i.e.
\begin{equation}
\label{equ:preseff2}
E_k P = B E_k P. 
\end{equation}
Combining Equations \ref{equ:preseff1} and \ref{equ:preseff2}, we get
\begin{equation}
\label{equ:preseff3}
B E_k = E_k P.
\end{equation}
This means than the existence of an effect $B$ which is such that $B E_k = E_k P$ for all channel elements $E_k$ is a necessary condition for $P$ to be preserved by the channel. In fact, it is easy to see that this condition is sufficient. Indeed, if $B E_k = E_k P$, then
\[
\chanh(B) = \sum_k E_k^* B E_k = \sum_k E_k^* E_k P = \chanh(\one) P = P.
\]

We achieved obtaining a condition which depends on each channel elements $E_k$ independently, rather than on the whole channel. However, if we really want a useful characterization of the sharp preserved effects, we need to eliminate any reference to the unknown operator $B$. First note that by taking the adjoint of Equation \ref{equ:preseff3} we obtain $E_k^* B = P E_k^*$. Together with Equation \ref{equ:preseff3}, this implies
\[
E_k^* E_j P = E_k^* B E_j = P E_k^* E_j.
\]
This is a necessary condition for $P$ to be preserved, which does not involve the unknown effect $B$. In fact, this condition is sufficient. 

\begin{thm}
\label{thm:preseff}
A sharp effect $P$ is preserved by the channel $\chan(\rho) = \sum_k E_k \rho E_k^*$ if and only if
\begin{equation}
\label{equ:preseff4}
[P, E_i^* E_j] = 0 \text{ for all $i$,$j$.} 
\end{equation}
\end{thm}
\proof We have already proven the necessity. In order to prove the sufficiency, we will use this condition to build the effect $B$ of which $P$ is the image. We will need the completely positive map $\chan_\lambda$ defined by
\begin{equation}
\label{equ:chanlambda}
\chan_\lambda(A) := \sum_{i=0}^\infty \lambda_i E_i A E_i^*
\end{equation}
where $\lambda_i := 2^{-i}$. This choice of $\lambda$ guarantees that the sum converges in norm for any effect $A$ since $\|E_i A E_i^*\| \le 1$. In fact, any choice of components $\lambda_i > 0$ which makes this sum weak-$*$ convergent for any effect $A$ would be sufficient for our purpose. Note that if $\lambda_i = 1$ for all $i$, then $\chan_\lambda = \chan$, which is defined only on trace-class operators.

We will try the ansatz
\begin{equation}
\label{equ:presansatz}
B := (\chan_\lambda(\one))^{-1} \chan_\lambda(P).
\end{equation}
First, we have to show that $\chan_\lambda(\one)$ can indeed be inverted.  
Note that $\chan_\lambda(\one) \ket{\psi} = 0$ if and only if $\bra{\psi} \chan_\lambda(\one) \ket{\psi} = 0$ since it is positive. If we write the channel in terms of the elements $E_k$, we obtain a sum of positive terms which must all be equal to zero: $\bra{\psi} E_i E_i^* \ket{\psi} = 0$. Hence $E_i^* \ket{\psi} = 0$ for all $i$. This means that if $\ket{\psi}$ is in the kernel of $\chan_\lambda(\one)$, it must be in the kernel of each $E_i^*$, and therefore be orthogonal to the range of each $E_i$. This shows that we can invert $\chan_\lambda(\one)$ on the range of any of the operators $E_i$. In the following we will always be able to assume that $(\chan_\lambda(\one))^{-1}$ operates on the span of the ranges of the operators $E_i$. In particular, Equation \ref{equ:presansatz} is well-defined. 

Now that we have defined $B$, let us check that $\chanh(B) = P$.
We have
\[
\begin{split}
B E_i &= (\chan_\lambda(\one))^{-1} \chan_\lambda(P) E_i \\
&= (\chan_\lambda(\one))^{-1} \sum_k \lambda_k E_k P E_k^* E_i \\
&= (\chan_\lambda(\one))^{-1} \sum_k \lambda_k E_k E_k^* E_i P  \\
&= (\chan_\lambda(\one))^{-1} \chan_\lambda(\one) E_i P  = E_i P  \\
\end{split}
\]
which is Equation \ref{equ:preseff3} and proves that $\chanh(B) = P$. However we have to check that $B$ is an effect. 
Note that $B E_i = E_i P$ implies $\sum_i \lambda_i B E_iE_i^* = \sum_i \lambda_i E_i P E_i^*$, i.e.,  $B \chan_\lambda(\one) = \chan_\lambda(P)$, which implies $B = \chan_\lambda(P) (\chan_\lambda(\one))^{-1}$.
 From the definition of $B$, we also have $B = (\chan_\lambda(\one))^{-1} \chan_\lambda(P)$. Hence $[(\chan_\lambda(\one))^{-1}, \chan_\lambda(P)] = 0$, which implies that
\begin{equation}
\label{equ:almostR}
B = \chan_\lambda(\one)^{-\frac{1}{2}} \chan_\lambda(P) \chan_\lambda(\one)^{-\frac{1}{2}} \ge 0
\end{equation}
In addition, $B \le \chan_\lambda(\one)^{-\frac{1}{2}} \one \chan_\lambda(\one)^{-\frac{1}{2}} = \one$, which shows that it is an effect. \qed


The dependence of $B$ on $P$ in the above proof is in fact given by the dual of a completely positive trace-preserving map, i.e. a channel. Explicitly, this channel is
\begin{equation}
\label{equ:R}
\mathcal R(\rho) = \chanh_\lambda(\chan_\lambda(\one)^{-\frac{1}{2}} \, \rho \, \chan_\lambda(\one)^{-\frac{1}{2}})
\end{equation}
and its dual
\[
\mathcal R^*(A) = \chan_\lambda(\one)^{-\frac{1}{2}} \chan_\lambda(A) \chan_\lambda(\one)^{-\frac{1}{2}}.
\]
\index{$\mathcal R$}
We will see in Chapter \ref{chapter:qec} that this channel $\mathcal R$ is  the ``correction'' channel of the theory of quantum error correction. This particular way of writing it appeared before in \cite{blume-kohout08} for the case $\lambda_i = 1$ which works when $\Hil_1$ is finite-dimensional. 

Theorem \ref{thm:preseff} yields also some unsharp preserved effects. Indeed, given the linearity of $\chanh$, if two effects are preserved, then so is any of their convex combinations (which are also effects). Therefore, the convex hull of the preserved projectors is entirely preserved. In fact, the continuity of this channel implies that the weak-$*$ closure of this convex hull is also preserved. 

Consider the commutant $\mathcal A_\chan$ of the operators $E_i^*E_j$, i.e. the set of operators which commute with them. This set is clearly an algebra. In fact, it is a von Neumann algebra, which always has the property that the set of effects it contains is the closed convex full of its projectors \cite{davidson96}.  
Since all the projectors in $\mathcal A_\chan$ are preserved, then so are all the effects it contains. This proves the following:
\begin{cor}
The set of effects spanning the von Neumann algebra
\[
\mathcal A_\chan = \{A \in \mathcal B(\Hil_1) \;|\; [A,E_i^* E_j] = 0 \text{ for all $i$,$j$}\}
\]
are all preserved by $\chan$. In addition, this algebra contains all the preserved sharp effects. 
\end{cor}

We can now easily generalize Theorem \ref{thm:preseff} to all sharp observables. 
\begin{thm}
\label{thm:presobs}
A sharp observable $X$ is preserved by the channel $\chan$ if and only if 
\begin{equation}
\label{equ:presobs}
X^*(\alpha) \in \mathcal A_\chan \quad \text{for all $\alpha$}
\end{equation}
\end{thm}
\proof The necessity of this condition follows from the fact that if $X$ is preserved, then there is an observable $Y$ such that $X^*(\alpha) = \chanh(Y^*(\alpha))$ for all $\alpha$, which implies in particular that all the sharp effects $X^*(\chi_\omega)$ for $\omega \subseteq \Omega$ are preserved by $\chan$, and therefore belong to $\mathcal A_\chan$. 
Since the unsharp effects $X^*(\alpha)$ all belong to the von Neumann algebra generated by the sharp effects $X^*(\chi_\omega)$, we also have $X^*(\alpha) \in \mathcal A_\chan$ for all $\alpha$.
Conversely, consider the channel $Y := X \circ \mathcal R$, which is clearly an observable. If $X^*(\chi_\omega) \in \mathcal A_\chan$ for all $\omega$, then we know that $X^*(\chi_\omega) = \chanh(Y^*(\chi_\omega))$ for all $\omega$. By linearity and continuity of $\chanh$, $X^*(\alpha) = \chanh(Y^*(\alpha))$ for all $\alpha$. 
\qed


This theorem yields a direct interpretation of the whole algebra $\mathcal A_\chan$ if we represent the sharp observables by self-adjoint operators in the traditional manner. Indeed, remember that we can represent a sharp observable $X$ by the (possibly unbounded) self-adjoint operator $\widehat X$ whose spectral projection-valued measure is $\omega \mapsto X^*(\chi_\omega)$. They are related through the equation
\[
X^*(\alpha) = \alpha(\widehat X)
\]
which involves the functional calculus on $\widehat X$. 
\begin{cor}
\label{thm:presobs2}
A sharp observable represented by the bounded self-adjoint operator $\widehat X$ is preserved by the channel $\chan$ if and only if 
\[
\widehat X \in \mathcal A_\chan.
\]
\end{cor}
\proof
If $X$ is preserved, then $\mathcal A_\chan \ni X^*(\id) = \id(\widehat X) = \widehat X$. Conversely, if $\widehat X \in \mathcal A_\chan$, then $X^*(\alpha) = \alpha(\widehat X) \in \mathcal A_\chan$ for all $\alpha$.
\qed


We have not yet discussed the properties of the operators $B$ obtained by Equation \ref{equ:almostR}. In fact they are projectors. Indeed, remember that $B = \chan_\lambda(\one)^{-1} \chan_\lambda(P) = \chan_\lambda(P) \chan_\lambda(\one)^{-1}$. Therefore 
\[
\begin{split}
B^2 &=  \chan_\lambda(\one)^{-1} \chan_\lambda(P) \chan_\lambda(P) \chan_\lambda(\one)^{-1} \\
&=  \chan_\lambda(\one)^{-1} \chan_\lambda(\one) \chan_\lambda(P^2) \chan_\lambda(\one)^{-1} \\
&=  \chan_\lambda(P^2) \chan_\lambda(\one)^{-1} \\
&=  \chan_\lambda(P) \chan_\lambda(\one)^{-1} =  B. \\
\end{split} 
\]
This can be used to show that
\begin{prop}
\label{prop:sharpsharp}
A preserved sharp observable is always the image of a sharp observable.
\end{prop}
\proof
Remember that if the sharp observable $X$ is preserved, then it is the image of the observable $Y$ given by $Y^*(\alpha) = \mathcal R^*(X^*(\alpha)) = \chan_\lambda(\one)^{-\frac{1}{2}} \chan_\lambda(X^*(\alpha)) \chan_\lambda(\one)^{-\frac{1}{2}}$. From the above observation, we know that since $X^*(\chi_\omega)$ is a projector for any subset $\omega \subseteq \Omega$, then so is $Y^*(\chi_\omega)$, which proves that $Y$ is sharp. 
\qed



\msection{Outgoing information}
\label{section:outinfo}


If the evolution in time of a quantum system is given by a channel 
\[
\chan: \mathcal B_t({\Hil_1}) \rightarrow \mathcal B_t({\Hil_2})
\]
which is not unitary, then it means that the system is interacting with an ``environment''. The \ind{environment} can be any other quantum system, the state of which we have ignored in our description. 
This fact is reflected in the Stinespring dilation theorem \cite{paulsen02}, which states that a completely positive map as above can always be represented (via its dual) as
\begin{equation}
\label{equ:stinespring}
\chanh(A) = V^* (A \otimes \one) V
\end{equation}
where $V$ is a bounded operator
\[
V: \Hil_1 \rightarrow \Hil_2 \otimes \Hil_E
\]
such that $V^* V = \one$, i.e. an isometry. Here $\Hil_E$ can be interpreted as the Hilbert space of the environment. The Choi-Kraus representation of the channel follows from this theorem by noting that, for any orthonormal basis $\ket{i}$ of the environment,
\[
\chanh(A) = \sum_i V^* (A \otimes \proj{i}) V = \sum_i V^* (\one \otimes \ket{i}) A (\one \otimes \bra{i}) V
\]
Hence the channel elements of $\chan$ are 
\begin{equation}
\label{equ:stinekraus}
E_i = (\one \otimes \bra{i}) V
\end{equation}
which means that
\[
\begin{split}
\chan(\rho) &= \sum_i E_i \rho E_i^* = \sum_i (\one \otimes \bra{i}) V A V^* (\one \otimes \ket{i})\\
&= \tr_E(V A V^*)
\end{split}
\]
where we have introduced the channel
\[
\tr_E:  \mathcal B_t(\Hil_2 \otimes \Hil_E) \longrightarrow \mathcal B_t(\Hil_2)
\]
defined by
\[
\tr_E(\rho) = \sum_i (\one \otimes \bra{i}) \rho (\one \otimes \ket{i}).
\]
This channel is called a {\em \ind{partial trace}}, here taken over the environment $\Hil_E$. It corresponds to the action of ``erasing'' any information about the system $\Hil_E$. Its dual, which we implicitly used above, is the {\em \ind{ampliation map}}
\begin{align*}
\tr_E^*: \mathcal B(\Hil_2)  &\longrightarrow \mathcal B(\Hil_2 \otimes \Hil_E)\\
A &\longmapsto  A \otimes \one
\end{align*}
It says how an effect of $\Hil_2$ can be also seen as an effect of $\Hil_2 \otimes \Hil_E$.

\begin{figure}
\begin{center}
\begin{pgfpicture}{0}{0}{5in}{2.5in}
 \pgfputat{\pgfxy(0,-2)}{
   \pgfputat{\pgfxy(0,0)}{
     \pgfbox[left,bottom]{\includegraphics[width=7cm]{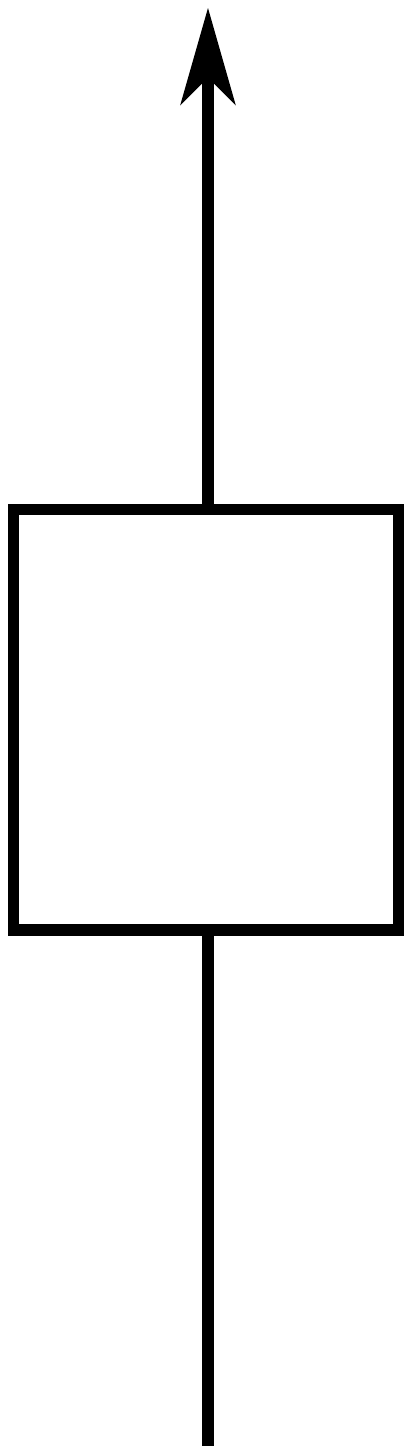}}
     \pgfputat{\pgfxy(1.55,5.3)}{\pgfbox[left,bottom]{$\chan$}}
     \pgfputat{\pgfxy(1.7,2.5)}{\pgfbox[center,bottom]{Sys}}
     \pgfputat{\pgfxy(1.7,7.9)}{\pgfbox[center,bottom]{Sys}}
   }
   \pgfputat{\pgfxy(2.8,5.3)}{\pgfbox[left,bottom]{$=$}}
   \pgfputat{\pgfxy(2.5,0)}{
     \pgfbox[left,bottom]{\includegraphics[width=7cm]{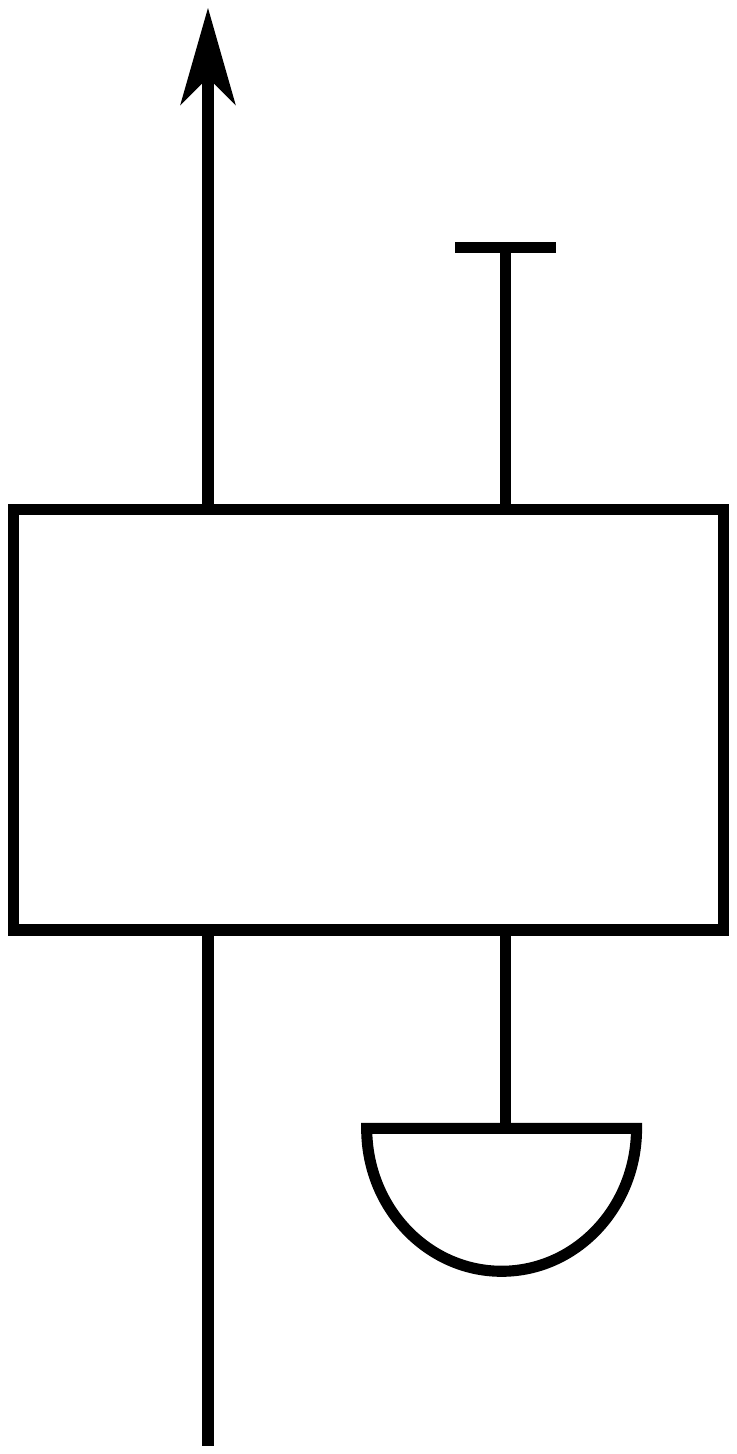}}
     \pgfputat{\pgfxy(2,5.3)}{\pgfbox[left,bottom]{$U$}}
     \pgfputat{\pgfxy(2.55,3.65)}{\pgfbox[left,bottom]{$\psi$}}
     \pgfputat{\pgfxy(1.7,2.5)}{\pgfbox[center,bottom]{Sys}}
     \pgfputat{\pgfxy(1.7,7.9)}{\pgfbox[center,bottom]{Sys}}
     \pgfputat{\pgfxy(2.7,7.2)}{\pgfbox[center,bottom]{(Env)}}
   }
   \pgfputat{\pgfxy(6,0)}{
     \pgfbox[left,bottom]{\includegraphics[width=7cm]{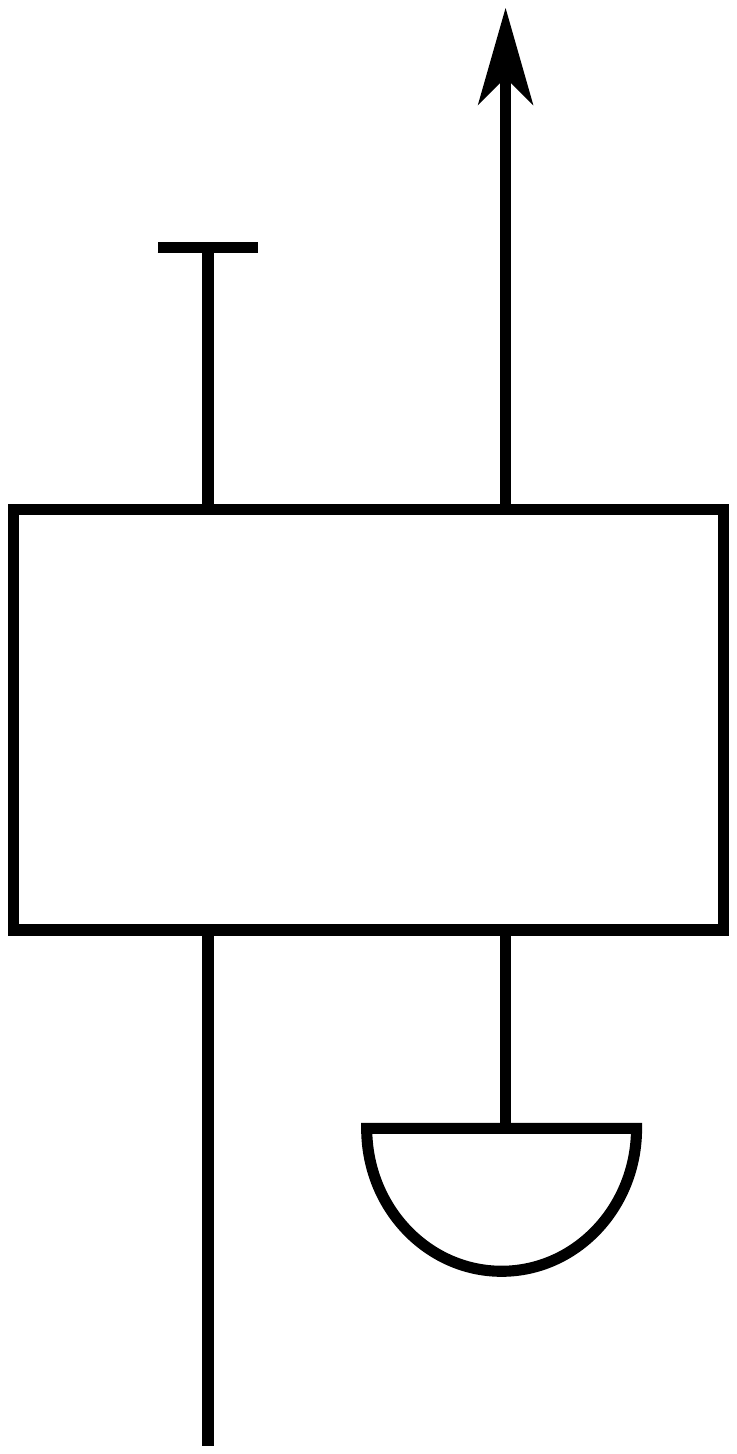}}
     \pgfputat{\pgfxy(2,5.3)}{\pgfbox[left,bottom]{$U$}}
     \pgfputat{\pgfxy(2.55,3.65)}{\pgfbox[left,bottom]{$\psi$}}
     \pgfputat{\pgfxy(1.7,2.5)}{\pgfbox[center,bottom]{Sys}}
     \pgfputat{\pgfxy(1.7,7.2)}{\pgfbox[center,bottom]{(Sys)}}
     \pgfputat{\pgfxy(2.7,7.9)}{\pgfbox[center,bottom]{Env}}
   }
   \pgfputat{\pgfxy(9.85,5.3)}{\pgfbox[left,bottom]{$=$}}
   \pgfputat{\pgfxy(9,0)}{
     \pgfbox[left,bottom]{\includegraphics[width=7cm]{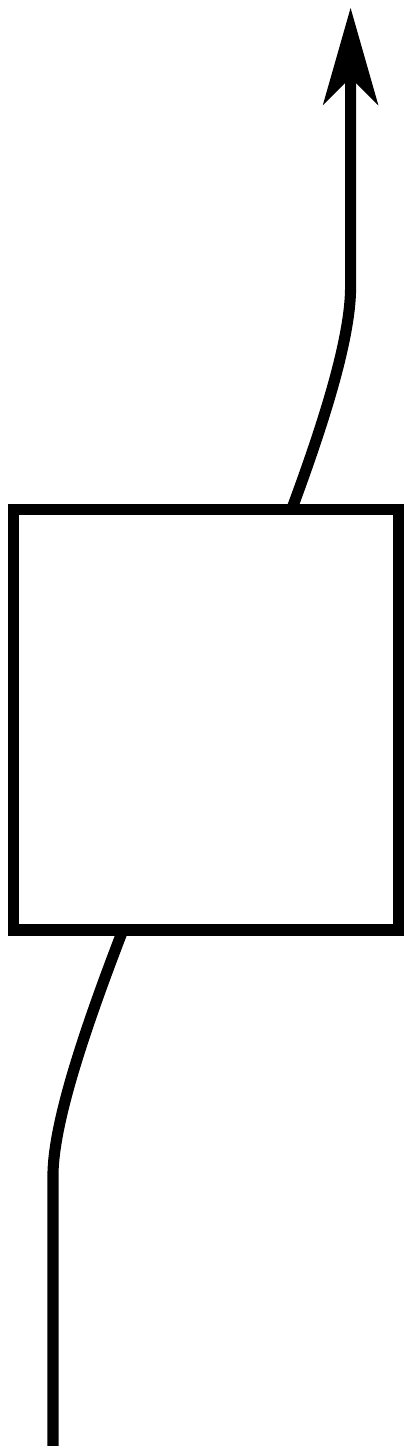}}
     \pgfputat{\pgfxy(2.1,5.3)}{\pgfbox[left,bottom]{$\chan_c$}}
     \pgfputat{\pgfxy(1.7,2.5)}{\pgfbox[center,bottom]{Sys}}
     \pgfputat{\pgfxy(2.7,7.9)}{\pgfbox[center,bottom]{Env}}
   }
}
\end{pgfpicture}
\end{center}
\caption[Dilation and complementary channel]{ \small Dilation of the channel $\chan$, and of the complementary channel $\chan_c$, represented as circuit diagrams. $\ket{\psi}$ is the initial state of the environement, and $U$ the unitary operator representing the joint evolution of the system and its environement. The horizontal bar ending a vertical line represents a partial trace. }
\label{fig:dilation}
\end{figure}

Physically, suppose that the environment starts in a known state $\ket{\psi_E}$, and that the two evolve according to the joint unitary operator $U$, i.e. $\rho \otimes \proj{\psi_E}$ evolves to $U (\rho \otimes \proj{\psi_E}) U^*$. Since the initial state of the environment is fixed, and uncorrelated with $\rho$, we can view this simply as a map on $\rho \in B_t(\Hil_1)$, with target in $\mathcal B_t(\Hil_2 \otimes \Hil_E)$:
\[
\rho \longmapsto U (\rho \otimes \proj{\psi_E}) U^* = U (\one \otimes \ket{\psi_E}) \rho (\one \otimes \bra{\psi_E}) U^*
\]
(See figure \ref{fig:dilation}.)
This gives a physical interpretation to the isometry 
\[
V := U (\one \otimes \ket{\psi_E}).
\]
To check that this is indeed an isometry, note that $V^* V =  (\one \otimes \bra{\psi_E}) U^*  U (\one \otimes \ket{\psi_E}) = \one \braket{\psi_E}{\psi_E} = \one$. 

The operation which consists in taking the partial trace over the environment means that we ignore its state for the calculation of any further expectation value, or probability. Indeed, if, after the interaction, we only plan to measure observables of the system, which are of the form $A \otimes \one$, then all expectation values take the form
\[
\tr( V \rho V^* (A \otimes \one)) = \tr( \tr_E(V \rho V^*) A ).
\]

This interpretation is also straightforward if we stick to the Heisenberg picture, where the form $\chanh(A) = V^* (A \otimes \one) V$ directly implies that we are only worrying about observables of the form $A \otimes \one$ on the final joint state of the system and environment.

The characterization of the preserved sharp observables, derived in the previous section, can also naturally be expressed in terms of the isometry $V$ rather than in terms of the channel elements $E_k$. Remember that the preserved observables are those which commute with the operators $E_i^* E_j$ for all $i$ and $j$. Using Equation \ref{equ:stinekraus}, we have 
\[
E_i^* E_j = V^* (\one \otimes \ketbra{i}{j}) V.
\]
Hence the span of the operators $E_i^* E_j$ is simply $V^* (\one \otimes \mathcal B(\Hil_E)) V$. 
This means that a sharp effect $P$ is preserved by $\chan$ if and only if 
\begin{equation}
\label{equ:presalg}
[P , V^* (\one \otimes B) V] = 0 \quad \text{ for all $B \in B(\Hil_E)$}.
\end{equation}



In fact, the effects of the form $V^* (\one \otimes B) V$ have an important physical meaning.
Indeed, they are preserved by the channel $\chan_c$ defined by its dual
\[
\chan^*_c(B) := V^* (\one \otimes B) V.
\]
This channel $\chan_c$ is called the {\em \ind{complementary channel}} to $\chan$. As $\chan$ maps $\Hil_1$ to $\Hil_2$, $\chan_c$ maps $\Hil_1$ to $\Hil_E$. It tells what information flows from the initial states of the system $\Hil_1$ to the environment $\Hil_E$. Indeed, the effects of the form $\chan^*_c(B) \in \mathcal B(\Hil_1)$
can be simulated by measuring $B$ on the final state of the environment. 

This gives us directly an interpretation for Equation \ref{equ:presalg}:
\begin{cor}
\label{cor:outgoing}
A sharp observable is preserved by $\chan$ if and only if it commutes with all the effects preserved by the complementary channel $\chan_c$. 
Hence, the preserved algebra $\mathcal A_\chan$ is given by 
\[
\mathcal A_\chan = \{A \;|\; [A, B] = 0 \text{ for all $B \in \chan_c^*(\qprops {\Hil_E})$  } \}.
\]
\end{cor}

In particular, this implies that the sharp observables preserved by $\chan$ all commute with the sharp observables preserved by $\chan_c$. Hence the algebra 
\[
\mathcal C_\chan := \mathcal A_\chan \cap \mathcal A_{\chan_c}
\]
 is commutative and therefore characterizes an effective classical system with phase-space $\Omega$ such that
\[
\mathcal C_\chan = L^\infty(\Omega).
\]

This is a detailed form of the \ind{no-broadcasting} theorem, which is a stronger version of the \ind{no-cloning} theorem. 
Indeed, the algebra $\mathcal C_\chan$ characterizes the sharp observables representing information which has been preserved in the system and at the same time transfered to the environment. Therefore this information is accessible from two different systems. This means that it has been at best broadcast (if there are correlations) and at worst cloned (if there are no correlations). Hence, the fact that $\mathcal C_\chan$ is commutative means that the sharp information which is broadcast must always be classical. 


We will argue in Section \ref{section:sharpdecoh} that this gives a way in which a classical system can emerge within a completely quantum system. Here the classical system is characterized by the only sharp information which is broadcast by an arbitrary quantum interaction. 
A generalization of this result to unsharp observables will be the basis of our picture of decoherence, presented in Section \ref{section:fulldecoh}.

\msection{Example: measurements}
\label{section:exmeas}



\newcommand{\pfff}{\psi}

Examples involving sharp observables will be given in the context of quantum error correction in Chapter \ref{section:qec}. Since most results presented in this chapter concern only sharp observables, let us give here some examples involving unsharp preserved observables.


Consider a channel on $\mathcal B(\Hil)$ of the form\footnote{This is an example of an {\em \ind{entanglement-breaking channel}}}
\[
\chan(\rho) = \sum_k \lambda_k \ketbra{\psi_k}{\pfff_k} \rho \ketbra{\pfff_k}{\psi_k}
\]
where the states $\ket{\psi_k}$ are normalized, but need not be orthogonal to each other. 
This channel maps an observable $Y$ with values in $\Omega$ to the observable $X = Y \circ \chan$. Explicitly,
\[
\begin{split}
X^*(\alpha) &= \chanh(Y^*(\alpha)) \\
&= \sum_i \lambda_i \ket{\pfff_i}\bra{\psi_i} Y^*(\alpha) \ket{\psi_i} \bra{\pfff_i}\\
&= \sum_i \lambda_i \bra{\psi_i} Y^*(\alpha) \ket{\psi_i}  \proj{\pfff_i}
\end{split}
\]
for all classical effects $\alpha \in L^\infty(\Omega)$. 
By construction, $X^*$ is preserved by $\chan$. In fact, it is also preserved by the complementary channel $\chan_c$.
Using a basis $\{\ket{i}\}_i$ of $\Hil$, we can write the complementary channel as
\[ 
\begin{split}
\chan_c^*(B) &= \sum_{ijk} \sqrt{\lambda_i \lambda_j} \ket{\pfff_i}\braket{\psi_i}{k}\bra{i} B \ket{j} \braket{k}{\psi_j} \bra{\pfff_j}\\
&= \sum_{ij} \sqrt{\lambda_i \lambda_j}  \bra{i} B \ket{j} \ket{\pfff_i} \braket{\psi_i}{\psi_j}  \bra{\pfff_j}.\\
\end{split}
\]
If we define an observable $Z$ of the environment by
$$
Z^*(\alpha) := \sum_{i} \bra{\psi_i} Y^*(\alpha) \ket{\psi_i}  \proj{i}
$$
then we have
\[
\begin{split}
\chan_c^*(Z^*(\alpha)) &=  \sum_{i} \lambda_i \bra{\psi_i} Y^*(\alpha) \ket{\psi_i}  \proj{\pfff_i} = X^*(\alpha)
\end{split}
\]
This proves that all the observables preserved by $\chan$ are also preserved by $\chan_c$.  

In this case, this can also be seen by the fact that we can write\footnote{In the literature, this property is summarized by saying that $\chan_c$ {\em \ind{degradable}}, while  $\chan$ is said to be {\em \ind{anti-degradable}}}
\[
\chan = \mathcal F \circ \chan_c
\]
for some channel $\mathcal F$, which is given by the expression that we used to define $Z$ from $Y$:
\[
\mathcal F^*(A) = \sum_{i} \bra{\psi_i} A \ket{\psi_i} \proj{i}.
\]
Therefore, for a channel of this form, all the information which is preserved in the system also flows to the environment. This means that the whole of the information preserved by $\chan$ is {\em broadcast}, in the sense that this information is accessible from both the system and the environment after the interaction. We saw in Section \ref{section:outinfo}, that the sharp observables broadcast in this way must form a commutative algebra. 

Clearly, this is not true of these broadcast unsharp effects. Indeed, it is easy to see that the operators $\lambda_k \proj{\pfff_k}$, which need not commute with each other, are preserved by $\chan$ and therefore also by $\chan_c$. We will see in Section \ref{section:classicalset} that this information is ``classical'', but in a more general way. This is not surprising as the channel $\chan$ can be relayed by a classical system. Indeed, it could be simulated by first measuring the observable $\Gamma$, with elements
\[
\Gamma_k = \lambda_k \proj{\pfff_k},
\]
and then preparing the state $\ket{\psi_k}$ corresponding to the outcome  $k$ of the measurement.
Note however that, as we will also see in Section \ref{section:nonclasscloning}, the information broadcast between only two systems need not even be classical in that more general sense.

Another way of looking at this channel $\chan$ is as a {\em \ind{pre-measurement}} of the discrete observables $\Gamma$ defined above. This means that it describes the effect on the state $\rho$ of a certain measurement of $\Gamma$, if the actual outcome of the measurement is traced-over or ``forgotten''. In the notation of Section \ref{section:measurement}, the measurement is defined by the partial channels $\mathcal F_k(\rho) = \lambda_k \ketbra{\psi_k}{\pfff_k} \rho \ketbra{\pfff_k}{\psi_k}$. 
Note that $\Gamma$ itself is at best approximately preserved by $\chan$, in the sense that coarse-grainings (approximate versions of) $\Gamma$ are preserved, as for instance the observable with elements
$$
\widetilde \Gamma_k = \sum_i \lambda_k |\braket{\psi_i}{\pfff_k}|^2 \Gamma_i = \chanh(\Gamma_k).
$$
$\widetilde \Gamma$ can be considered an approximation of $\Gamma$ because the distribution $|\braket{\psi_i}{\psi_k}|$ is ``peaked''at $i=k$. In the limit where the vectors $\ket{\psi_i}$ are orthogonal, $|\braket{\psi_i}{\psi_k}| \rightarrow  \delta_{ik}$. However, note that there exists non-trivial observables for which $|\braket{\psi_i}{\psi_k}|$ is independent of $k$ and $i$ (which are the SIC-POVMs described in Section \ref{section:exsymclone}).


The fact that the observable being measured is not preserved in the system illustrates the well-known fact that two successive measurements of a POVM do not yield the same outcome for sure. 
However, $\Gamma$ is exactly preserved by the complementary channel $\chan_c$. Indeed, we have
\[
\begin{split}
\chan_c^*(\proj{k}) &= \sum_{ij} \sqrt{\lambda_i \lambda_j}  \braket{i}{k}  \braket{k}{j} \ket{\pfff_i} \braket{\psi_i}{\psi_j}  \bra{\pfff_j}\\
 &= \lambda_k\proj{\pfff_k} = \Gamma_k\\
\end{split}
\]
This is what is important for $\chan$ to be a measurement of $\Gamma$: the apparatus (environment) must gain full information about $\Gamma$. 

It is always true that, if a channel represents the effect of a measurement, then its complement exactly preserves the observable being measured. Indeed, recall that in Section \ref{section:measurement} we introduced a measurement by postulating that the channel describing the evolution of the system be complementary to the quantum-to-classical channel defining the observable being measured, namely $\Gamma$. The quantum-to-classical channel $\Gamma$ can always be seen as a fully quantum channel by embedding its target algebra into the full quantum algebra $\mathcal B(\Hil_E)$. In addition $\Gamma$ is always preserved by $\Gamma$ itself, since it is the image of the ``identity'' classical observable. Let us illustrate this when $\Gamma$ is discrete, with elements $\Gamma_i$. If we represent $\Gamma$ as a full quantum channel, its action is
\[
\Gamma(\rho) = \sum_i \tr(\rho \Gamma_i) \proj{i}
\]
for some orthonormal basis $\ket{i}$ of the Hilbert space $\Hil_E$. 
The dual is
\[
\Gamma^*(A) = \sum_i \bra{i} A \ket{i} \Gamma_i 
\]
which implies that
\[
\Gamma^*(\proj{i}) = \Gamma_i  
\]
Hence $\Gamma$ is the image under $\Gamma$ of the sharp observable with eigenstates $\ket{i}$, which is the complete ``universal'' observable on the classical system $\Omega$. In fact, $\Gamma$ represented as a quantum channel is the minimal channel preserving $\Gamma$ itself, which makes sense. We will say more about this type of preserved information in Section \ref{section:classicalset} where we will show that the information preserved by an observable represents in a sense the most general way that classical information can be embedded in a quantum system. 



An important example of a channel of this type is given by replacing the states $\ket{\psi_i}$ by the continuous set of coherent states. This ``coherent-state channel'' will be studied in Sections \ref{section:quantization} and \ref{section:sysdecoh}. 




\begin{figure}
\begin{center}
\begin{tabular}{ccc} 
\includegraphics[width=0.32\columnwidth]{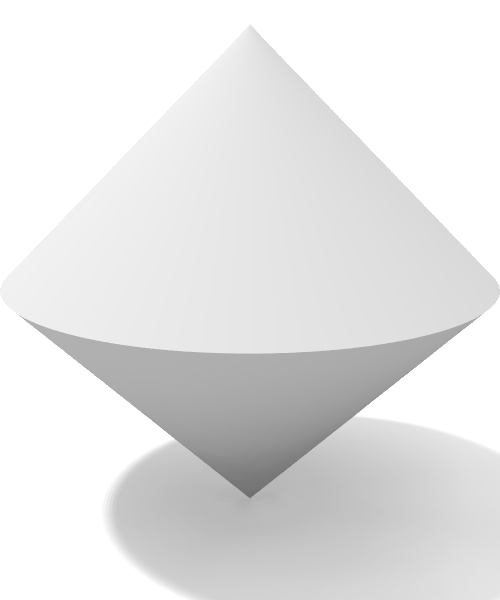} &\includegraphics[width=0.32\columnwidth]{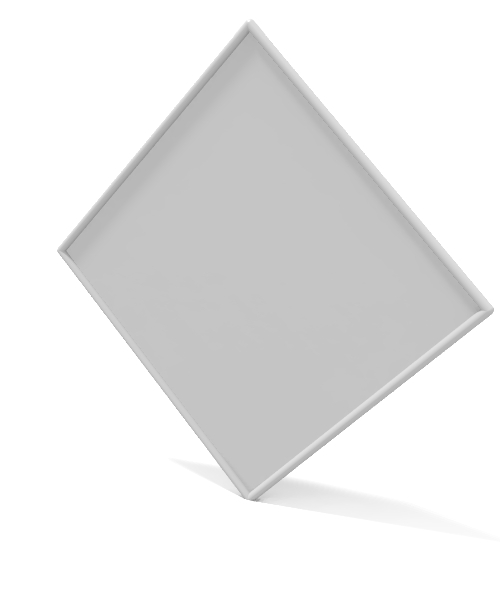} &\includegraphics[width=0.32\columnwidth]{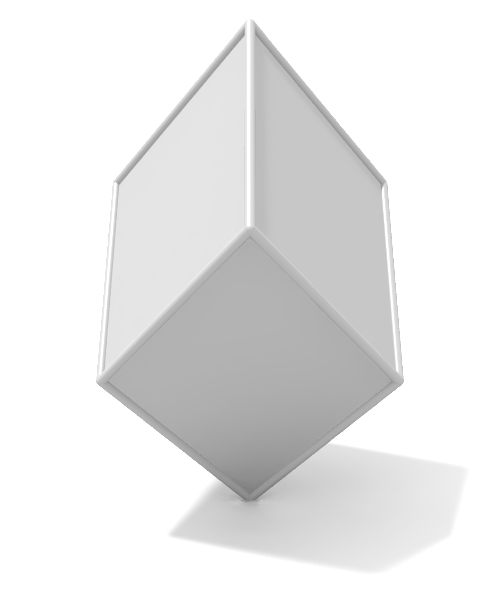}\\
All effects & $n=2$ (PVM) & $n = 3$  \\
\includegraphics[width=0.32\columnwidth]{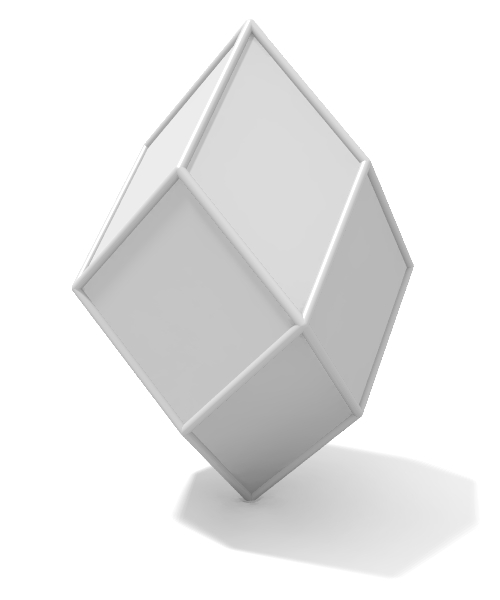} &\includegraphics[width=0.32\columnwidth]{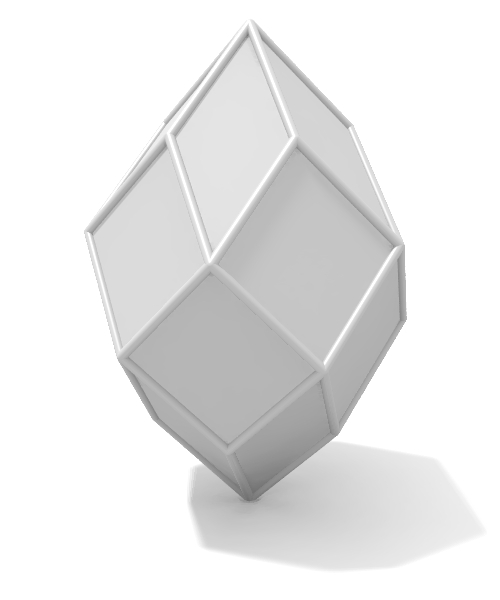}&\includegraphics[width=0.32\columnwidth]{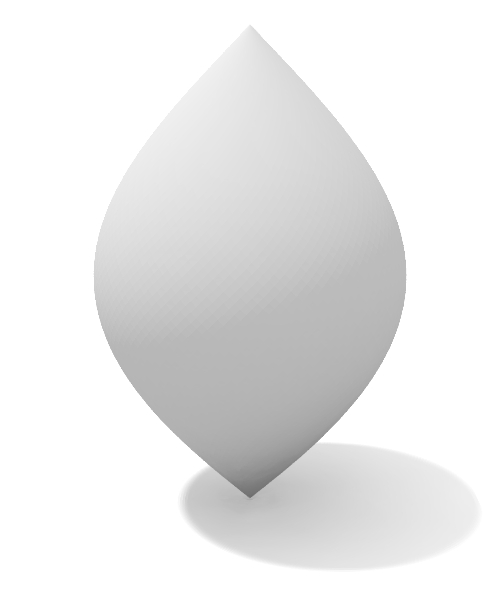} \\
$n = 4$ & $n = 5$ & $n = \infty$\\
\end{tabular}
\caption[Preserved effect algebras]{\small Representations of subsets of effects on a qubits in terms of the components $\one, \sigma_x, \sigma_z$. The first image is the full set of effects. The lower tip is the zero operator, the upper tip is the identity operator. All other shapes represent the set of preserved effects for the channels defined in Equation \ref{equ:symqubitchan} for $n=2,3,4,5$, and Equation \ref{equ:allstateschan} for $n=\infty$. These channels also have the property of being fully decoherent (see Section \ref{section:sysdecoh}). 
}
\label{fig:diamonds}
\end{center}
\end{figure}

Figure \ref{fig:diamonds} represents graphically the set of preserved effects for a few channels of the form
\begin{equation}
\label{equ:symqubitchan}
\chan_n(\rho) = \sum_{k=1}^n \frac{2}{n} \ketbra{k}{\psi_k^{(n)}} \rho \ketbra{\psi_k^{(n)}}{k}
\end{equation}
where 
\[
\proj{\psi_k^{(n)}} = \frac{1}{2} \one + \frac{1}{2} \cos(2\pi k/n) \, \sigma_x + \frac{1}{2} \sin(2\pi k/n) \, \sigma_z,
\]
$\sigma_x$ and $\sigma_z$ are two Pauli matrices, and the vectors $\ket{k}$ form an orthoginal basis in the channel's target space.
The case labelled $n=\infty$ represents the channel
\begin{equation}
\label{equ:allstateschan}
\chan_n(\rho) = \frac{1}{2\pi}\int_0^{2\pi} \ketbra{\theta}{\psi_\theta} \rho \ketbra{\psi_\theta}{\theta}\,d\theta
\end{equation}
where $\ket{\theta}$ are pseudo-eigenstates of a multiplication operator on the circle, and
\[
\proj{\psi_\theta} := \frac{1}{2} \one + \frac{1}{2} \cos(\theta) \sigma_x + \frac{1}{2} \sin(\theta) \sigma_z.
\]

\msection{Example: uncertainty relations}
\label{section:exuncertainty}

We will use the concept of preserved observable to show that an approximate measurement of the position of a particle preserves an approximate momentum observable, where the relationship between the precision of both is given by Heisenberg's uncertainty relations. 

Consider the position $\hat x$ of a one-dimensional non-relativistic particle, which is the multiplication operator on $\Hil = L^2(\mathbb R)$. It will be convenient to follow the tradition and use the improper eigenstates $\ket{x}$ of $\hat x$.
An approximate measurement of this sharp observable can be represented by the unsharp observable $X$ defined by
\[
X^*(\alpha) = \int_{\mathbb R} \alpha(y) X_y \, dy
\]
where the operators $X_y$ are 
\[
X_y = \int dx \, |G(x-y)|^2 \proj{x} 
\]
for a smearing function $G(x)$, which could for instance be chosen to be a Gaussian centered around zero. Note that since the observable $\hat x$ does not in fact have eigenstates, the proper way of defining $X_y$ would be using functional calculus, as
\[
X_y = |G(\hat x - y \one)|^2.
\]
Normalization of the observable $X$ implies 
\[
\int dy |G(x,y)|^2 = 1.
\]
The effect on the particle of a measurement of this observable could be given by the channel
\[
\chan(\rho) = \int E_y \rho E_y^* dy
\]
where the operators $E_y$ are defined by
\[
E_y = \int dx \, G(x-y) \proj{x} = G(y \one - \hat x).
\]

Note that since this channel represents the effect of a measurement, we already know that the information which flows to the environment is precisely the classical information represented by the observable $X$ being measured.  
The channel itself however preserves a lots of information. In particular, it preserved exactly the sharp position observable. Indeed, we have
\[
\begin{split}
\chanh(\proj{z}) &= \int dy \, E_y^* \proj{z} E_y\\
&= \int dy \, dx \, dx'\, \overline{G(x,y)} G(x',y) \delta(x-z) \delta(x'-z) \ketbra{x}{x'}\\
&= \left({ \int dy |G(z,y)|^2 }\right) \proj{z} \\
&= \proj{z} \\
\end{split}
\]



Now suppose that we measure the momentum observable $\hat p = i \hbar \frac{\partial}{\partial x}$ after the channel has acted. If $\ket{p}$ represents the improper eigenstates of $\hat p$, we have
\[
\begin{split}
\chanh(\proj{p}) &= \int dy \, E_y^* \proj{p} E_y\\
&= \int dy \, \int dx \, \overline{G(x-y)} e^{i x p} \int dx'\, G(x'-y) e^{-ix'p} \ketbra{x}{x'}\\
&= \int dq dq' dy dz dz' \, \overline{G(z)} e^{i (z+y) (p - q)} G(z') e^{-i(z'+y)(p - q')} \ketbra{q}{q'}\\
&= \int dq dq' dy \, \overline{\widetilde G(p-q)}  \widetilde G(p-q') e^{i y (q'-q)} \ketbra{q}{q'}\\
&= \int dq \, \overline{\widetilde G(p-q)}  \widetilde G(p-q) \ketbra{q}{q}\\
&= \int dq \, |\widetilde G(p-q)|^2 \ketbra{q}{q}\\
\end{split}
\]
where $\widetilde G(p-q)$ is the Fourier transform of $G(x-y)$.
This shows that the channel preserves an approximate momentum observable $P$ defined by
\[
P^*(\alpha) = \int_{\mathbb R} \alpha(p) P_p \, dp
\]
where
\[
P_p := \int dq \, |\widetilde G(p-q)|^2 \ketbra{q}{q}.
\]

Since $\widetilde G(p)$ is the Fourier transform of $G(x)$, the second moment of the distributions $|G(x)|^2$ and $|\widetilde G(p)^2|$ satisfy the Heisenberg uncertainty relation. This gives an operational formulation of the uncertainty relation which works for a single measurement, i.e. which does not rely on expectation values as in the original formulation. It shows that if an observable is measured with a finite accuracy, the information about the canonical conjugate is preserved up to an accuracy given by the uncertainty relation.

Note that the measurement which yields the channel $\chan$ is naturally generated by an interaction of the type 
\[
H = \nu \, \hat x \otimes \hat p.
\]
between two Schr\"odinger particles. 
Indeed, defining 
\[
K_x := \nu x \hat p
\]
we obtain
\[
\begin{split}
e^{-\frac{i}{\hbar} t H} &= \sum_{n} \frac{(-it/\hbar)^n}{n!} \hat x^n \otimes (\nu \hat p)^n\\
&= \sum_{n} \int dx \, \frac{(-it)^n}{n!} x^n \proj{x} \otimes (\nu \hat p)^n\\
&= \int dx \, \proj{x} \otimes \sum_{n} \frac{(-it)^n}{n!} x^n (\nu \hat p)^n\\
&= \int dx \, \proj{x} \otimes e^{-\frac{i}{\hbar}t K_x}.\\
\end{split}
\]
Therefore, if the initial state of the second system is $\ket{\psi}$, the channel describing the evolution of the first system is
\[
\begin{split}
\chanh(A) &= (\one \otimes \bra{\psi}) e^{\frac{i}{\hbar} t H}(A \otimes \one) e^{-\frac{i}{\hbar} t H}(\one \otimes \ket{\psi}) \\
 &= \int dy\, (\one \otimes \bra{\psi}) e^{\frac{i}{\hbar} t H}(A \otimes \proj{y}) e^{-\frac{i}{\hbar} t H}(\one \otimes \ket{\psi}) \\
&= \int dy\, E_y^* A E_y \\
\end{split}
\]
where
\[
E_{y} = (\one \otimes \bra{y}) e^{-\frac{i}{\hbar} t H} (\one \otimes \ket{\psi}) = \int dx \, \braket{(y - x) \nu t}{\psi} \proj{x}.
\]
By $\ket{(y-x) \nu t}$, we mean the eigenstate of position with eigenvalue $(y-x) \nu t$.
This has the form of the approximate position measurement where the smearing function 
\[
G(y-x) = \braket{(y - x) \nu t}{\psi}
\] 
depends on the initial state $\ket{\psi}$ of the apparatus. 
Note that the second moment of $|G(x)|^2$ decreases linearly with time. Hence the measurement's accuracy increases continuously with time. However this fact would largely depend on the self-Hamiltonian of the second system, which we neglected here.


\mchapter{Correctable information}%
\label{chapter:qec}




In the previous chapter, we studied observables which are {\em preserved} by a channel $\chan$, in the sense that they can be precisely simulated by another observable after the channel has acted, no matter what the state was. Formally, $X$ is preserved if there exists $Y$ such that
\[
X = Y \circ \chan.
\]
We have seen that when $X$ is sharp, $Y$ can be obtained from $X$ via a channel $\mathcal R$:
\[
Y = X \circ \mathcal R.
\]
For these observables, we have 
\[
X = X \circ \mathcal R \circ \chan.
\]
In this chapter, we will study in greater depth the class of observables which have this property, i.e. which are {\em correctable}. Correctable observables are rare compared to preserved observables, because they require the channel to have special symmetry properties.




We will characterize classes of observables correctable by a common channel $\mathcal R$, and show that they generalize the known {\em quantum error correcting codes} \cite{bennett96,knill97,shor95, steane96, gottesman96}.
Our main result is a generalization of the Knill-Laflamme conditions \cite{knill97} to infinite-dimensional codes, and hybrid quantum-classical codes. Our approach is close in spirit to the first concept of noiseless subsystem introduced in \cite{knill00}, and technically relies on later results on the correctability of noiseless subsystems \cite{kribs05, kribs06}. Some of the new results presented were published in \cite{beny07x1, beny07x4, beny08x4}.

\msection{Quantum error correction}
\label{section:qec}





Let us suppose that some information is sent through a channel $\chan$. The aim of error correction is to find certain degrees of freedom, the {\em error correcting code}\index{error correcting code} (whose exact nature we deliberately keep imprecise for the moment), on which the effect of the channel can be inverted. Since the inversion must be implemented physically, it must be a valid physical transformation, i.e. a channel. The inverse channel $\mathcal R$ is called the {\em  \ind{correction channel}}. 
The fact that a channel $\chan$ can always be written as
\[
\chan(\rho) = \sum_i E_i \rho E_i^*
\]
means that we can always assume that the noise is given by a discrete family of individual errors $E_i$. 

In fact we will see that if $\mathcal R$ corrects this channel on some code, then it would correct also any channel whose elements span the same operator space as the elements of $\chan$.  
This fact is important because often one does not know the precise channel elements $E_i$. Indeed, suppose that the system interacts continuously with its environment via a Hamiltonian $H$. The most general form of $H$ would be
\[
H = \sum_i J_i \otimes K_i
\]
where the {\em \ind{interaction operators}} $J_i$ act on the system, and the operators $K_i$ act on the environment.

Let us assume that the initial state of the environment is given by some vector $\ket{\psi}$. It is unlikely that we know much about $K_i$ or $\ket{\psi}$ given that the environment may be very large and complex. However, it is generally conceivable that we have a good knowledge of what the operators $J_i$ can be. If $\chan_t$ is the channel describing the evolution of the system alone up to time $t$, we have
\[
\begin{split}
\chanh_t(A) &= (\one \otimes \bra{\psi}) e^{it H} (A \otimes \one)e^{-it H}(\one \otimes \ket{\psi})\\
 &= \sum_k (\one \otimes \bra{\psi}) e^{it H}(\one \otimes \ket{k}) A (\one \otimes \bra{k})e^{-it H}(\one \otimes \ket{\psi})\\
&= \sum_k E_k^*(t) A E_k(t)
\end{split}
\]
where the channel elements are
\[
\begin{split}
E_k(t) &= (\one \otimes \bra{k})e^{-it H}(\one \otimes \ket{\psi})\\
&= \sum_n \frac{(-it)^n}{n!} (\one \otimes \bra{k}) H^n (\one \otimes \ket{\psi})\\
&= \sum_n \sum_{j_1 \dots j_n} \frac{(-it)^n}{n!} \bra{k} K_{j_1} \cdots K_{j_n}\ket{\psi} J_{j_1}\cdots J_{j_n}.  \\
\end{split} 
\]
Hence the channel elements $E_k(t)$ all belong to the algebra generated by the interaction operators $J_i$.
At this point, we could try to find correctable codes for the class of channels whose elements span this algebra. This is possible, but would in fact result in codes which need no correction at all (i.e. noiseless subsystems \cite{knill00}). 

However, if the time $t$ at which we aim to perform the correction is small enough, we can do better. Indeed, suppose that $\lambda$ is some interaction parameter with unit of energy, then the above series is expressed in powers of $t\lambda$. We see that to the $n$th order in $t \lambda$, the elements of the channel are in the span of the $n$th order products $J_{j_1}\cdots J_{j_n}$. Hence, if we correct often enough (in order to limit the value of $t \lambda$), then we only need to find a correction channel and a code for channel elements in the span of the operators $J_{j_1}\cdots J_{j_n}$, $n < N$, for a fixed $N$. In this context, the operators $J_i$ can be seen as representing independent errors. The goal is to find a code as large as possible which corrects as many of these errors as possible.


For clarity of the presentation we will stick to the simple picture where the channel $\chan$ is given. However we will keep the more realistic situation in mind and check that the correction procedure we devised works not just for the given channel, but also for any channel whose elements span the same space.

\msection{Correctable observables}

Let us introduce a notion of correctability which fits our framework, and then later show how we recover, as a special case, the known results of quantum error correction.
In this section we consider again a channel
\[
\chan: \mathcal B_t({\Hil_1}) \rightarrow \mathcal B_t({\Hil_2})
\]

\begin{dfn}
We say that an observable $X$ is {\em correctable}\index{correctable observable} for $\chan$ if there is a channel $\mathcal R$, called the {\em \ind{correction channel}}, which is such that
\[
X = X \circ \mathcal R \circ \chan.
\]
We then say that $X$ is {\em fixed}\index{fixed observable} by the channel $\mathcal R \circ \chan$.
\end{dfn}
The operational meaning of this definition is clear. If $X$ is fixed by $\mathcal R \circ \chan$, then a measurement of $X$ after the application of the channel $\chan$ followed by $\mathcal R$ will yield the same outcome as if nothing had happened to the system.

It directly follows from this definition that if $X$ is correctable, then it is also preserved. Indeed it suffices to define the observable $Y := X \circ \mathcal R$ to observe that $X = Y \circ \chan$. Preserved observables however are rarely correctable. Indeed, if $X$ is correctable then, in particular, the effect $X/\|X\|$ is also correctable. This suffices to show, for instance, that none of the preserved effects are correctable in the examples shown in Figure \ref{fig:diamonds}, except for the case $n=2$ where they are all correctable.


We have already identified a class of preserved observables which are correctable, namely the preserved sharp observables. 
Indeed, in order to prove Theorem \ref{thm:presobs}, in Section \ref{section:sharpobs}, we introduced the channel defined by Equation \ref{equ:R}, namely
\[
\mathcal R^*(A) = \chan_\lambda(\one)^{-\frac{1}{2}} \chan_\lambda(A) \chan_\lambda(\one)^{-\frac{1}{2}}
\]
which was shown to correct all the preserved sharp observables.
\begin{thm}
\label{thm:corrcond}
Given a channel $\chan$ with elements $E_i$, and a sharp observable $X$ represented by the self-adjoint operator $\widehat X$, the following conditions are all equivalent
\begin{enumerate}
\item $X$ is preserved by $\chan$,
\item $X$ is correctable for $\chan$,
\item $[\widehat X,E_iE_j] = 0$ for all $i$, $j$.
\end{enumerate}
\end{thm}
Note the similarity between this proposition and the one obtained in \cite{blume-kohout08}, for an initially different concept of ``preserved information''. 

\subsection{Simultaneously correctable observables}
\label{section:simultcorr}

We know that, in fact, the sharp preserved observables can all be corrected with the same correction channel. Let us explicitly define this concept.
\begin{dfn}
A set of observables is said to be {\em simultaneously correctable} for $\chan$ if the observables it contains are all correctable in terms of the same correction channel $\mathcal R$, which means that they are all fixed by $\mathcal R \circ \chan$.
\end{dfn}
Before we go further, let us remark that, contrary to what happened for preserved observables, the correctable observables are entirely characterized by the simultaneously correctable {\em effects}. Indeed, it is clear that an observable $X$ is fixed by the channel $\mathcal F$ if and only if its effects $X^*(\alpha)$ are all fixed by $\mathcal F$. Indeed, by definition, $X = X \circ \mathcal F$ if and only if $X^*(\alpha) = \mathcal F^*(X^*(\alpha))$. More generally, any set of observables having simultaneously correctable effects are simultaneously correctable. 

Let us state what we already know:
\begin{prop}
\label{prop:corralg}
The effects in the algebra \[
\mathcal A = \{ A \in \mathcal B(\Hil_1) \;|\; [A,E_i^* E_j] = 0 \text{ for all $i$, $j$}\}
\]
are all simultaneously correctable by the channel defined in Equation \ref{equ:R}. 
\end{prop}
As noted above, this implies that the observables $X$ with $X^*(\alpha) \in \mathcal A$ are correctable by the same channel $\mathcal R$. 
We will call $\mathcal A$ {\em the \ind{correctable algebra}} for $\chan$. 
This is justified by the fact that any other algebra spanned by correctable effects is inside $\mathcal A$. Indeed, such an algebra would be spanned by its projectors, but $\mathcal A$ already contains all the correctable projectors. 


In general, a set of simultaneously preserved effects need not be characterized by an algebra. However, if a set of effects is simultaneously correctable, then so are any linear combinations which also form effects. Since, in addition, effects are self-adjoint, this means that a set of simultaneously correctable effects will always be characterized by an {\em \ind{operator system}} \ind{$\mathcal S$}, i.e. a linear subspace $\mathcal S \subseteq \mathcal B(\Hil_1)$ which is closed under the $*$-operation. Hence, if we say that an operator system $\mathcal S$ is correctable\index{correctable operator system}, we mean that all the effects in it are correctable. 
In fact, given that we only consider normal channels, then also any weak-$*$ limit of such linear combinations will be correctable by the same correction channel. Hence we only need to consider weak-$*$ closed operator systems. 

We will now show that many correctable operator systems can be obtained thank to Proposition \ref{prop:corralg}.
Consider any subspace $\Hil_0 \subseteq \Hil_1$, and let 
\[
V : \Hil_0 \rightarrow \Hil_1
\] 
be the isometry which embeds $\Hil_0$ into $\Hil_1$. This means that the operator $P_0 := V V^*$, defined on $\Hil_1$, is the projector on the subspace $\Hil_0$, whereas $V^* V = \one_0$ is the identity inside $\Hil_0$.  
We can define a new channel $\chan_0$ by restricting the channel $\chan$ to the subspace $\Hil_0$, which, physically, amounts to making sure that the initial state is prepared inside $\Hil_0$. Hence we define
\[
\chan_0(\rho) := \chan( V\rho V^*)
\]
whose dual is
\[
\chanh_0(A) =  V^* \chanh( A ) V
\]
This channel has its own correctable algebra:
\[
\mathcal A_0 = \{ A \in \mathcal B(\Hil_0) \;|\; [A, V^* E_k^* E_l V] = 0 \}.
\]
This algebra can be naturally embedded in $\mathcal B(\Hil_1)$ via the isometry $V$. Indeed, the map $A \mapsto V A V^*$ from $\mathcal B(\Hil_0)$ to $\mathcal B(\Hil_1)$ is a normal $*$-homomorphism. 
Note however that the identity on $\mathcal B(\Hil_0)$ is sent to the projector $VV^* = P$.

The algebra $V \mathcal A_0 V^*$ may or may not be a subalgebra of the correctable algebra $\mathcal A$. If it is not, then its effects are not correctable for the channel $\chan$ itself. However, we will see that it is one-to-one with a family of simultaneously correctable effects which do not form an algebra. Indeed, let $\mathcal R_0$ be the correction channel for $\chan_0$, which is a channel from $\mathcal B_t(\Hil_2)$ to $\mathcal B_t(\Hil_0)$, and define the set
\begin{equation}
\label{equ:simcorrsys}
\mathcal S_0 := \chanh(\mathcal R_0^*(\mathcal A_0)).
\end{equation}
Note that $\mathcal S_0$ is not equal to $\mathcal A_0$ because we have used $\chanh$ instead of $\chanh_0$ in Equation \ref{equ:simcorrsys}. In fact, we have
\begin{equation}
\mathcal A_0 = V^* \mathcal S_0 V.
\end{equation}
This subset $\mathcal S_0 \subset \mathcal B(\Hil_1)$ is in general not an algebra, but it is always an operator system. We claim that all the effects which are in $\mathcal S_0$ are simultaneously correctable. Indeed, for any effect $A = \chanh(\mathcal R_0^*(B)) \in \mathcal S_0$, where $B \in \mathcal A_0$, we have
\[
\begin{split}
\chanh(\mathcal R_0^*(V^* A V)) &= \chanh(\mathcal R_0^*(V^* \chanh(\mathcal R_0^*(B)) V))\\
 &= \chanh(\mathcal R_0^*(\chanh_0(\mathcal R_0^*(B))))\\
 &= \chanh(\mathcal R_0^*(B))\\
 &= A.\\  
\end{split}
\]
Hence, the correction map is
\[
\rho \mapsto V \mathcal R_0 (\rho) V^*
\]
which is a valid channel from $\mathcal B_t(\Hil_2)$ to $\mathcal B_t(\Hil_1)$.

Hence, for any subspace $\Hil_0 \subseteq \Hil_1$ we can construct an operator system $\mathcal S_0$ whose effects are all simultaneously correctable. In addition, it is clear that all the observables which are formed of effects in $\mathcal S_0$ are all simultaneously correctable. 

This proves the following:
\begin{thm}
\label{thm:corrsys}
For every subspace $\Hil_0 \subseteq \Hil_1$, the operator system $\mathcal S_0$ defined in Equation \ref{equ:simcorrsys} is such that all the observables $X$ with $X^*(\alpha) \in \mathcal S_0$ for all $\alpha$ are simultaneously correctable.
\end{thm}
Let us summarize how to obtain $\mathcal S_0$. We restricted the channel to the subspace $\Hil_0$, computed the correctable algebra $\mathcal A_0$ and the correction channel $\mathcal R_0$ for the restricted channel, and finally set $\mathcal S_0 = \chanh(\mathcal R_0^*(\mathcal A_0))$.
In fact we can be entirely explicit. Letting $V$ be the isomorphism embedding $\Hil_0$ into $\Hil_1$, $P := VV^*$ the projector on $\Hil_0$, $\chan_\lambda$ the regularized channel defined on the whole of $\mathcal B(\Hil_1)$, and
\[
K := (\chan_\lambda(P))^{-\frac{1}{2}}
\]
we have that the operator system
\[
\mathcal S_0 = \{ \chanh(K \chan_\lambda(VAV^*) K)\;|\; A \in \mathcal B(\Hil_0), [A, V^* E_i^* E_j V] = 0 \text{ for all $i$, $j$}\}
\]
is corrected on $\chan$ by the channel
\[
\mathcal R(\rho) = P \chan_\lambda^*(K \rho K) P.
\]
We will give an explicit example in Section \ref{section:exopsys}. 


Do these structures exhaust all the correctable observables for $\chan$? Some light could be cast on this question thank to theorem 3 of \cite{blume-kohout08}, the proof of which has unfortunately not been published at the time of writing. The authors state that, when $\Hil_1$ is finite-dimensional, the fixed point set of the dual of any channel must be made of elements of the form $A + \mathcal F^*(A)$ where $A$ belongs to a $*$-algebra $\mathcal A_0$ inside $\finops(\Hil_0)$, $\Hil_0$ a subspace of $\Hil_1$, and $\mathcal F^*$ is a fixed channel which is such that $P \mathcal F^*(A) P = 0$, where $P$ projects onto $\Hil_0$. 

If an operator system $\mathcal S$ is correctable for $\chan$, then it is fixed by $\chanh \circ \mathcal R^*$ for some channel $\mathcal R$, and, according to \cite{blume-kohout08}, made of elements of the form mentioned above: $A + \mathcal F^*(A)$. This means that if $V$ is the isometry embedding $\Hil_0$ into $\Hil_1$, we have $V^* \chanh(\mathcal R^*(A+\mathcal F^*(A))) V = V^* (A+\mathcal F^*(A))V = A$, which implies that the algebra $\mathcal A_0$ is correctable for $\chan$ restricted to $\Hil_0$, and therefore belongs to the correctable algebra for this restricted channel. This shows that $\mathcal S$ is of the form covered by Theorem \ref{thm:corrsys}, which therefore
exhausts all correctable observables, at least in finite dimension.

In the next section we will see a reason why, for most applications, it is more useful to make sure that the initial state is in the subspace $\Hil_0$ and work only with the correctable algebra $\mathcal A_0$, rather than working with the operator system $\mathcal S_0$ and no restriction on states.


\subsection{Simultaneously correctable channels}
\label{section:varchan}

In section \ref{section:qec} we motivated a situation where, of the channel elements $E_i$, only their span is known. It is already clear that the correctable algebra 
\[
\mathcal A = \{ A \;|\; [A,E_i^* E_j] = 0 \text{ for all $i$, $j$} \}
\]
only depends on the span of the elements $E_i$. Indeed, consider a channel $\chan$ with elements $E_i$ and a channel $\mathcal E'$ with elements $F_i = \sum_j \gamma_{ij} E_j$ where $\gamma_{ij}$ are arbitrary, provided that $\sum_i F_i^* F_i = \one$. Then it is clear that an operator which commutes with the products $E_i^* E_j$ for all $i$ and $j$ will also commute with operators $F_i^* F_j$ since they are just linear combinations of the former. This fact tells us that $\mathcal A$ is the correctable algebra for all channels whose elements are chosen in the span of the operators $E_i$. However this fact alone would not be very helpful if the correction channel itself depended on the particular choice of channel elements. Fortunately, it does not.

In fact, we have already exploited part of this freedom in defining $\mathcal R$. Indeed, remember that we defined it in terms of the map 
\[
\chan_\lambda(A) = \sum_i \lambda_i E_i A E_i^*
\]
defined on $\mathcal B(\Hil_1)$ (see Equation \ref{equ:chanlambda}). 
The sequence $\lambda$ was chosen so that the infinite sum in the expression for $\chan_\lambda$ is well defined for any operator. However the exact value of the components $\lambda_i$ did not matter in the proof that $\mathcal R$ corrects the algebra $\mathcal A$ for the channel $\chan$. In fact, the only important aspect of this channel was that its elements are linear combinations of the adjoints of the elements of $\chan$. 

The channel $\mathcal R'$ correcting $\mathcal E'$ would be defined in the same way in terms of the channel
\[
\begin{split}
\chan_\lambda'(A) &= \sum_i \lambda_i F_i A F_i^* \\
&= \sum_{ijk} \lambda_i \gamma_{ij} \overline \gamma_{ik} E_j A E_k^*\\
\end{split}
\]
which has also the right form for the corresponding correction channel 
\[
\mathcal R'(\rho) = (\chan_\lambda')^*((\chan_\lambda'(\one))^{-\frac{1}{2}}\,  \rho\,(\chan_\lambda'(\one))^{-\frac{1}{2}})
\]
to correct the channel $\chan$. 
We will not go through the proof that $\mathcal R'$ corrects $\chan$ on $\mathcal A$, since precisely the same steps can be followed as for $\mathcal R$ itself. 

Hence we have seen that  all the channels whose elements span the same space of operators will have the same correctable algebra, and be correctable through the same correction channel. This means that this theory can be applied to the case described in Section \ref{section:qec}, where the span of the elements is all that we know about the channel. 

There is a sense in which it is this fact which allows for the quantum errors to be understood as being {\em discrete} \cite{nielsen00}. Indeed, a standard error model for quantum computing is that where the system considered is a tensor product of {\em qubits}, namely two-dimensional quantum systems. The possible ``errors'' (i.e. possible channel elements of the noise) are supposed to be any operator acting on no more than $n$ subsystems, where $n$ is fixed. It is clear that this set of errors is continuous. However, for a finite number of qubits their span is separable (in fact finite-dimensional), which means that it suffices to choose a discrete set which spans the space and try to correct these only. 

This discussion applies to simultaneously correctable sets of observables characterized by an algebra, which are the correctable sharp observables. However, it does not apply to the classes of simultaneously correctable unsharp observables identified in the previous section. Indeed, in those cases the correctable operator systems $\mathcal S_0$ may be different for two channels whose elements span the same operator space. Indeed, remember that
\[
\mathcal S_0 = \chanh(\mathcal R_0^*(\mathcal A_0)).
\]
where $\mathcal A_0$ is the correctable algebra for the channel restricted to a subspace $\Hil_0$, and $\mathcal R_0$ the corresponding correction channel. Therefore, although both $\mathcal R_0$ and $\mathcal A_0$ would be the same for both channels, the set $\mathcal S_0$ in this expression depends explicitly of the action of the channel itself, and may be different in both cases.


\subsection{Nature of correctable channels}


The fact that $\mathcal A$ is the correctable algebra for the channel $\chan$, with correction channel $\mathcal R$, implies that the map $\chanh \circ \mathcal R^*$ acts simply as the identity on $\mathcal A$:
\[
(\chanh \circ \mathcal R^*) |_{\mathcal A} = \id_{\mathcal A}
\]
The following theorem elucidates what happens to the algebra $\mathcal A$ prior to its correction:
\begin{thm}
\label{thm:homo}
Let $\mathcal A$ be the correctable algebra for a channel $\chan$. Then $\chanh$ is a normal $*$-homomorphism of the algebra generated by the pre-image of $\mathcal A$. In particular, for any operators $B$, $B'$ such that $\chanh(B), \chanh(B') \in \mathcal A$, we have
\[
\chanh(BB') = \chanh(B) \chanh(B').
\]
\end{thm}
\proof Remember, that the projectors $P$ in the correctable algebra $\mathcal A$ satisfy $B E_i = E_i P$ for some operator $B$, and for all $i$ (see Equation \ref{equ:preseff3}). It can be directly checked that this is also true of the span of the projectors, which is almost the whole of the algebra $\mathcal A$, up to closure. Now consider two operators $B$ and $B'$ such that $A := \chanh(B)$ and $A' := \chanh(B')$ belong to the span of the projectors of $\mathcal A$. 
We know that $B E_i = E_i A$ and $B' E_i = E_i A'$. This implies that $B B' E_i = B E_i A' = E_i A A'$, from which it follows that
\[
\begin{split}
\chanh(B B') &= \sum_i E_i^* B B' E_i = \sum_i E_i^* E_i A A' = A A'\\
&= \chanh(B)\chanh(B').
\end{split}
\]
Since $\chanh$ is weak-$*$ continuous, this condition also applies to the weak-$*$ closure of the set of operators whose images are in the span of the projectors of $\mathcal A$. 
This shows that the above condition holds for every operators in the pre-image of $\mathcal A$. 
\qed

Note that the pre-image of $\mathcal A$ under $\chanh$ includes in particular the image of the dual of any correction channel $\mathcal R$. 


In fact, the correction channel defined in Equation \ref{equ:R} is itself a homomorphism. We saw in the above proof that for all operators $A$ in the span of the projectors of $\mathcal A$, we have $\mathcal R^*(A) E_i = E_i A$. This implies that
\[
\mathcal R^*(A) \chan_\lambda(\one) = \chan_\lambda(A).
\]
which means explicitly
\[
(\chan_\lambda(\one))^{-\frac{1}{2}} \chan_\lambda(A)(\chan_\lambda(\one))^{\frac{1}{2}} =\chan_\lambda(A). 
\]
or, simply,
\[
[(\chan_\lambda(\one))^{-\frac{1}{2}}, \chan_\lambda(A)] = 0.
\]
From the weak-$*$ continuity of $\chan_\lambda$, we have that this is true for all $A \in \mathcal A$. 
Using this fact, and also recalling that $[A, E_i^* E_j] = 0$ for all $A \in \mathcal A$, we have
\[
\begin{split}
\mathcal R^*(A) \mathcal R^*(A') &= (\chan_\lambda(\one))^{-\frac{1}{2}} \chan_\lambda(A) (\chan_\lambda(\one))^{-1} \chan_\lambda(A') (\chan_\lambda(\one))^{-\frac{1}{2}} \\
&= (\chan_\lambda(\one))^{-\frac{3}{2}} \chan_\lambda(A) \chan_\lambda(A') (\chan_\lambda(\one))^{-\frac{1}{2}} \\
&= (\chan_\lambda(\one))^{-\frac{3}{2}} \chan_\lambda(\one) \chan_\lambda(AA') (\chan_\lambda(\one))^{-\frac{1}{2}} \\
&= (\chan_\lambda(\one))^{-\frac{1}{2}} \chan_\lambda(AA') (\chan_\lambda(\one))^{-\frac{1}{2}} \\
&= \mathcal R^*(A A').
\end{split}
\]
Hence, we proved the following proposition:
\begin{prop}
The correction channel given by Equation \ref{equ:R} is a faithful representation of the correctable von Neumann algebra $\mathcal A$.
\end{prop}

This shows that the effect of a channel on its correctable algebra simply amounts to representing it in a different way on the Hilbert space. These results clarify certain aspects of \cite{kribs06x1}. We refer to \cite{beny08x4} for the precise relationship between our results and those of \cite{kribs06x1}.

\msection{Error correcting codes}

In Section \ref{section:qec} we mentioned that the purpose of quantum error correction was to find a ``code'' on which the channel can be inverted, without defining what we meant by a code. In the previous section, we started from the general assumption that a code should be a set of simultaneously correctable observables. We have then found that for any subsystem $\Hil_0$ of the source Hilbert space $\Hil_1$, we have a set of simultaneously correctable observables characterized by an operator system $\mathcal S_0$, or equivalently by the von Neumann algebra $\mathcal A_0$ which is such that $\mathcal A_0 = V^* \mathcal S_0 V$ where $V$ is the isometry embedding $\Hil_0$ into $\Hil_1$. The algebra $\mathcal A_0$ characterizes the sharp observables correctable for the channel restricted to the subspace $\Hil_0$. 

Hence, all the codes that we identified, on which the channel can be inverted, are, or correspond to, von Neumann algebras. In fact it easy to build abstract examples which yield any possible von Neumann algebra in this way, given that all von Neumann algebras can be defined as the commutant of an arbitrary set of operators. 


In order to understand the type of information represented by von Neumann algebras, we need to know how they look like. 


\subsection{Structure of von Neumann algebras}
\label{section:vnstruct}

Let us first summarize the representation theory of finite-dimensional von Neumann algebras, which are just $*$-algebras.


A concrete finite-dimensional $*$-algebra $\mathcal A$, represented by matrices, i.e. operators on a finite-dimensional Hilbert space, always has the form
\begin{equation}
\label{equ:finalgrep}
\mathcal A = \bigoplus_{k=1}^N \; \mathcal M_{n_k} \otimes \one_{m_k}
\end{equation}
where $\mathcal M_{n_k}$ denotes the full set of matrices on an $n_k$-dimensional Hilbert space, and $\one_{m_k}$ the identity on an $m_k$-dimensional Hilbert space. If the dimension of the algebra $\mathcal A$ is $D$ then we have 
\(
D = \sum_k n_k^2
\).
The direct sum of two matrix algebras must be understood as the algebra of block-diagonal matrices, with one block encoding the first algebra, and the other block the second algebra. Therefore the above means that, written as a matrix of blocks,
\[
 \mathcal A =
 \begin{pmatrix}
    \mathcal M_{n_1}  \otimes \one_{m_1} & 0 & \cdots & 0\\
    0 & \mathcal M_{n_2}  \otimes \one_{m_2} & \cdots & 0\\
    \vdots & \vdots & \ddots & \vdots\\
    0 & 0 & \cdots & \mathcal M_{n_N} \otimes \one_{m_N}  \\
  \end{pmatrix}. 
\]
In addition, tensoring a matrix algebra with the identity on another algebra means that we are considering matrices which are also block-diagonal, with as many blocks as there are elements on the diagonal of the identity matrix, but such that each blocks are all identical, not only in their size, but also in their content. 

For instance, any operator $A$ in the algebra $\mathcal A = (\mathcal M_{2} \otimes \one_2) \oplus (\mathcal M_{3})$ has the form
\[
A = \begin{pmatrix}
B & 0 & 0 \\
0 & B & 0 \\
0 & 0 & C \\
\end{pmatrix}
\]
for a two-by-two matrix $B$ and a three-by-three matrix $C$.

The block-diagonal structure of $\mathcal A$ is determined by the form of its {\em \ind{center}} $\mathcal Z(\mathcal A)$. The center is the set of operators inside the algebra which commute with all other elements of the algebra:
\[
\mathcal Z(\mathcal A) = \{ A \in \mathcal A \;|\; [A,B] = 0 \text{ for all $B \in \mathcal A$ } \}.
\] 
It is a commutative algebra. The center can also be written as the intersection of the algebra with its {\em \ind{commutant}} $\mathcal A'$ which is the algebra composed of all operators commuting with all elements of $\mathcal A$:
\[
\mathcal Z(\mathcal A) = \mathcal A \cap \mathcal A'.
\]
For instance, for an algebra of the form $\mathcal M_n \otimes \one_m$, we have 
\[
\mathcal Z(\mathcal M_n \otimes \one_m) = (\mathcal M_n \otimes \one_m) \cap(\one_n \otimes \mathcal M_m) \approx \mathbb C.
\]
A von Neumann algebra is said to be a {\em \ind{factor}} if its center is isomorphic to $\mathbb C$. Hence matrix algebras of the form $\mathcal M_n \otimes \one_m$ are factors.

More generally, if the representation of $\mathcal A$ is expressed as in Equation \ref{equ:finalgrep}, then the commutant is
\[
\mathcal A' = \bigoplus_{k=1}^N \;\one_{n_k} \otimes \mathcal M_{m_k} 
\]
and the center of $\mathcal A$ is
\begin{equation}
\label{equ:fincenter}
\mathcal Z(\mathcal A) = \bigoplus_{k=1}^N \;\mathbb C (\one_{n_k} \otimes \one_{m_k})
\end{equation}
which means that it is composed of diagonal matrices with only $N$ different eigenvalues. If $P_k$ is the projector on the $k$th block, then this means that a generic element $C \in \mathcal Z(\mathcal A)$ of the center is of the form
\[
C = \sum_k c_k P_k
\]
for arbitrary complex numbers $c_k$. 
The algebra $\mathcal A$ itself is block-diagonal in terms of the subspaces defines by the projectors $P_k$, in the sense that for all $A \in \mathcal A$,
\[
A = \sum_k P_k A P_k.
\]
Hence the center of the algebra essentially tells us what the blocks are in its representation.

For instance, consider again the algebra $\mathcal A = (\mathcal M_{2} \otimes \one_2) \oplus (\mathcal M_{3})$. Typical operators $A \in \mathcal A$, $A' \in \mathcal A'$ and $C \in \mathcal A$ have the form
\[
A = \begin{pmatrix}
B & 0 & 0 \\
0 & B & 0 \\
0 & 0 & C \\
\end{pmatrix}
\quad 
A' = \begin{pmatrix}
a \one_2 & b \one_2 & 0 \\
c \one_2 & d \one_2 & 0 \\
0 & 0 & x \one_3 \\
\end{pmatrix}
\quad
C = \begin{pmatrix}
a \one_2 & 0 & 0 \\
0 & a \one_2 & 0 \\
0 & 0 & x \one_3 \\
\end{pmatrix}
\]
where $a,b,c,d,x \in \mathbb C$, $B$ is any 2-by-2 matrix, and $C$ any 3-by-3 matrix. 



When $\mathcal A$ is infinite-dimensional, the direct sum must be replaced by a {\em \ind{direct integral}}.
This follows from the fact that the center can be any commutative algebra, which has the form
\[
\mathcal Z(\mathcal A) \approx L^\infty(\Omega)
\]
for some set $\Omega$ equipped with a measure. It is with respect to this measure that we can write 
\[
\mathcal A \approx \int^{\oplus}_\Omega \mathcal A(x) \, dx
\]
Where the generalized ``blocks'' $\mathcal A(x)$ are factors, i.e. have a trivial center. 
If $\Omega$ is finite then we must use a discrete measure, which gives us the direct sum in Equation \ref{equ:fincenter}.
Hence this integral can be intuitively understood as a continuous limit of the direct sum. 

Factors come in three main types. Up to now we have been using {\em \ind{type I factor}s}, which are always of the form $\mathcal B(\Hil)$ for some Hilbert space $\Hil$. Factors of type II or III are more exotic. An example of a factor of type II will be studied in Section \ref{section:extypetwo}.

\subsection{Standard codes and subsystem codes}
\label{section:oqec}


Traditionally, a {\em quantum error correcting code} is just a subspaces $\Hil_0$ of the initial finite-dimensional Hilbert space $\Hil_1$, which is assumed to be finite-dimensional \cite{nielsen00}.
The idea is that the channel $\chan$ is correctable for states in $\Hil_0$ if there is a channel $\mathcal R$ such that
\[
\mathcal R(\chan(\rho)) = \rho
\]
for all states $\rho$ which are mixtures of pure states in the subspace $\Hil_0$. 
If we introduce the isometry $V: \Hil_0 \rightarrow \Hil_1$ which embeds $\Hil_0$ into $\Hil_1$, this means that
\begin{equation*}
\mathcal R(\chan(V \rho V^*)) = V \rho V^*
\end{equation*}
for all $\rho \in \mathcal B_t(\Hil_0)$, which is equivalent to requiring the existence of a channel $\mathcal R'$ such that 
\begin{equation*}
\mathcal R'(\chan(V \rho V^*))  = \rho 
\end{equation*}
for all $\rho \in \mathcal B_t(\Hil_0)$. Indeed, it suffices to pick $\mathcal R'(\rho) = V^* \mathcal R(\rho) V$. 

 If we define $\chan_0(\rho) := \chan(V \rho V^*)$, this means that $\mathcal R' \circ \chan_0$ is the identity on $\mathcal B_t(\Hil_0)$, or equivalently that $\chanh \circ (\mathcal R')^*$ is the identity on $\mathcal B(\Hil_0)$, which, as we have shown matches our conception of correctability for the algebra $\mathcal B(\Hil_0)$.  Therefore we recover the framework of standard quantum error correction, for a code $\Hil_0$, when the correctable algebra is $\mathcal B(\Hil_0)$, and the channel is restricted to the subspace $\Hil_0$.   

In order to complete the comparison, let us check that our correctability condition reduces to the one introduced for standard codes~\cite{knill97}. The Knill-Laflamme condition states that a standard code represented by the subspace $\Hil_0$ is correctable for the channel $\chan$ with elements $E_i$ if there exists $\lambda_{ij} \in \mathbb C$ such that
\begin{equation}
\label{equ:qeccond}
V^* E_i^* E_j V = \lambda_{ij} \one \quad \text{ for all $i$, $j$}
\end{equation}
where $V$ embeds $\Hil_0$ into the source Hilbert space $\Hil_1$. 
In our framework, the correctable algebra must be precisely the commutant of the operators $V^* E_i^* E_j V$ for all $i$ and $j$. Since here they are all proportional to the identity on $\Hil_0$, the correctable algebra is indeed the whole algebra of operators on $\Hil_0$. 

A more general framework was also introduced which generalized the notion of a code to that of a {\em subsystem code} \cite{kribs05, kribs06}. In this approach one defines a code through a subspace $\Hil_0 \subseteq \Hil_1$ and a particular subsystem decomposition $\Hil_A \otimes \Hil_B = \Hil_0$ of this subspace. Again, let $V$ be the isometry embedding $\Hil_0$ into $\Hil_1$. 
We then say that the subsystem $\Hil_A$ is a correctable code if there is a channel $\mathcal R$ such that
\[
\mathcal R(\chan(V(\rho \otimes \tau) V^*)) = \rho \otimes \tau'
\]
for any states $\rho \in \mathcal B_t(\Hil_A)$, $\tau, \tau' \in \mathcal B_t(\Hil_B)$.  We want to show that this is equivalent to the case where the correctable algebra $\mathcal A$, in our framework, is any factor of type I, which in this case is
\[
\mathcal A = \mathcal B(\Hil_A) \otimes \one_A.
\]
That is, assuming that we are restricting the initial state to the subspace $\Hil_0$. In our language, this would mean that
\begin{equation}
\label{equ:vexev}
V^* \chanh(\mathcal R^*( X\otimes \one )) V = X \otimes \one
\end{equation}
for all $X \in \mathcal B(\Hil_A)$.
Indeed, suppose first that $\Hil_A$ is a subsystem code corrected by $\mathcal R$ , then we have that for all $X \in \mathcal B(\Hil_A)$, 
\[
\begin{split}
\tr(V^* \chanh(\mathcal R^*(X\otimes \one)) V (\rho\otimes\sigma)) 
&= \tr((X\otimes \one) \mathcal R (\chan(V \rho \otimes \sigma V^*)))\\
& = \tr(X\rho \otimes \tau) = \tr(X \rho) \tr(\tau) \\ 
&= \tr(X \rho) = \tr((X \otimes \one) (\rho \otimes \sigma)).
\end{split}
\]
This is true for all states $\rho \in \mathcal \mathcal B_t(\Hil_A)$ and all states $\sigma \in \mathcal B(\Hil_t(\Hil_B))$. By linearity it follows that $V^* \chanh(X\otimes \one)V = X \otimes \one$ for all $X \in \mathcal \mathcal B(\Hil_A)$.
Conversely, if Equation \ref{equ:vexev} is true for all $X$, then for all $\rho \in \mathcal \mathcal B(\Hil_0)$ we have
\[ 
\begin{split}
\tr(X\, \tr_B(\mathcal R(\chan( V \rho V^*))))
&= \tr( (X\otimes \one) \mathcal R(\chan(V \rho V^*))) \\
&= \tr(V^* \chanh(\mathcal R^*(X\otimes \one)) V \rho ) \\
&= \tr((X\otimes \one) \rho ) \\
&= \tr(X \,\tr_B(\rho)).
\end{split}
\]
Since the above equation is true for all $X$, we have $\tr_B(\mathcal R(\chan( V \rho V^*))) = \tr_B(\rho)$ for all $\rho \in \mathcal B_t(\Hil_0)$, which was shown in \cite{kribs06} to be equivalent to the definition of $\Hil_A$ being a noiseless subsystem for $\chan$. 

In this framework, the correctability condition reads \cite{kribs05}
\begin{equation}
\label{equ:oqeccond}
V^* E_i^* E_j V = \one \otimes \Lambda_{ij}
\end{equation}
for an arbitrary set of operators $\Lambda_{ij} \in \mathcal B(\Hil_B)$.  This means that the operators $V^* E_i^* E_j V$ for all $i$ and $j$ generate the sub-algebra $\one \otimes \mathcal B(\Hil_B)$ of $\mathcal B(\Hil_0)$, whose commutant is indeed $\mathcal B(\Hil_A) \otimes \one$: the correctable algebra defining the subsystem code. 

Note that a subsystem code can always also be identified as a standard code, provided that we can afford to put stronger constraints on the initial state. Indeed, consider the smaller subspace $\Hil_0'$ formed by the states inside $\Hil_0$ which are of the form $\ket{\psi} \otimes \ket{\phi_0}$, where $\ket{\phi_0}$ is fixed. This subspace is associated with the isometry $W = V \otimes \ket{\phi_0}$, for which we have
\[
W^* E_i^* E_j W = \bra{\phi_0} \Lambda_{ij} \ket{\phi_0} \one
\]
which is just the Knill-Laflamme condition for $\Hil_0'$. 







These results show that the standard codes, as well as the subsystems codes, correspond to the case where our correctable algebra is a finite-dimensional factor, which is always of type I.  
Our results yield two types of generalization over these codes. Firstly, we obtain a characterization of infinite-dimensional quantum codes and continuous classical codes, which will be discussed briefly in Section \ref{section:infdimcodes}. In addition, we can correct information which is not quantum nor classical, i.e. which is represented by an algebra which is not commutative nor a factor.

\subsection{Hybrid codes}
\index{hybrid codes}


We have seen that in our framework, the structure to be corrected, i.e. representing the code, can be any von Neumann algebra. 
A general von Neumann algebra with center $\mathcal Z(\mathcal A) = L^\infty(\Omega)$ is of the form
\begin{equation}
\label{equ:hybrid}
\mathcal A = \int_\Omega^\oplus \mathcal A(x) dx 
\end{equation}
where each $\mathcal A(x)$ is a factor. If the center is maximal, i.e. $\mathcal Z(\mathcal A) = \mathcal A$, then the algebra $\mathcal A$ is commutative and each factor $\mathcal A(x)$ is of dimension one, i.e. isomorphic to the complex numbers $\mathbb C$. If, on the other hand, the center is minimal, i.e. $\mathcal Z(\mathcal A) \approx \mathbb C$, then the set $\Omega$ contains only a single element $x_0$ and $\mathcal A = \mathcal A(x_0)$ is a factor.

If $\mathcal A$ is commutative, then it represents a classical system, which is clear from the fact that it has the form $L^\infty(\Omega)$. It is then natural to say that if it is a factor, it represents a ``pure'' quantum system. 

A physical system represented by an algebra, whose structure is given by the general form \ref{equ:hybrid}, can be understood as being partly quantum and partly classical. Indeed, we can consider the center $\mathcal Z(\mathcal A) = L^\infty(\Omega)$ as representing a classical system. For each possible ``state'' $x \in \Omega$ of this classical system, we have a pure quantum system represented by the factor $\mathcal A(x)$. 

For instance, a classical system, represented by $L^\infty(\Omega)$, next to a type I quantum system, with algebra $\mathcal B(\Hil)$, is represented by 
\[
\mathcal A = L^\infty(\Omega) \otimes \mathcal B(\Hil) \simeq \int^\oplus_\Omega \mathcal A(x) dx
\]
where each factor $\mathcal A(x)$ is a copy of $\mathcal B(\Hil)$. In the more general case however, the size and type of the algebra $\mathcal A(x)$ may depend upon $x$.

Generic operators $A,B \in \int_\Omega^\oplus \mathcal A(x) \, dx $ are of the form
\begin{equation}
\label{equ:hybrideffect}
A = \int_\Omega A_x \, dx \quad \text{and}\quad B = \int_\Omega B_x \, dx
\end{equation}
Where $\mathcal A_x, B_x \in \mathcal A(x)$ for all $x \in \Omega$. Their product is simply
\[
A B = \int_\Omega A_x B_x \, dx.
\]
An element of the center is of the form
\[
C = \int_\Omega \alpha(x) \one_x\, dx
\]
where $\alpha \in L^\infty(\Omega)$ and $\one_x$ is the identity on $\mathcal A_x$.

If $A$ in Equation \ref{equ:hybrideffect} is an effect, then each operator $A_x$ is also an effect which can be interpreted as a quantum proposition which is true conditionally on the classical system being in state $x$. 

If the finite-dimensional case, if a hybrid algebra
\[
\mathcal A = \bigoplus_k \mathcal M_{n_k} \otimes \one_{m_k}
\]
is correctable, then each factor $\mathcal A_k = \mathcal M_{n_k} \otimes \one_{m_k}$ represents a correctable subsystem code for states restricted to the subspace $\Hil_k$ projected onto by $P_k = \one_{n_k} \otimes \one_{m_k}$. Indeed, if $V$ is the isometry corresponding to $P_k$, then $V A V^*$ is the representation inside $\mathcal A$ of an operator $A \in \mathcal A_k$. Therefore
\[
V^* \chanh(\mathcal R^*(V A V^*)) V = V^* V A V^* V = A.
\]
Therefore, a finite-dimensional hybrid algebra can be understood as representing a family of orthogonal subsystem codes correctable simultaneously.





\subsection{Infinite-dimensional subsystem codes}
\label{section:infdimcodes}

The derivation of necessary and sufficient conditions for error correction of infinite-dimensional algebras is an interesting new result, given that all physical systems are naturally modeled by infinite-dimensional systems. In particular, it yields a formulation of quantum error correction for systems characterized by continuous variables \cite{braunstein03}. 

A code can be said to be purely quantum if it is represented by an algebra which is a factor. In the finite-dimensional case, we have seen in Section \ref{section:oqec} that factors represent subsystem codes characterized by Equation \ref{equ:oqeccond}, or Equation \ref{equ:qeccond}.  
Some authors assumed that this condition would hold unchanged in the infinite-dimensional case.
For instance, in \cite{braunstein98x1} the Knill-Laflamme condition was expressed for a channel with continuous elements $E_x$ as
\[
V^* E_x^* E_y V = \lambda(x,y) \one
\]
where $\lambda(x,y) \in \mathbb C$. 
Our results show that this condition is sufficient. Indeed, it implies that the commutant of the operators $V^* E_x^* E_y V$ for all $x$ and $y$ is the whole algebra $\finops(\Hil_0)$ on the subspace $\Hil_0 \subseteq \Hil_1$. However, in infinite dimension this condition is no longer necessary since it expresses that the code must be isomorphic to $\finops(\Hil_0)$, which is a factor of type I. Hence it misses the possibility of correcting more general factors. 
This means that our generalization of quantum error correction to infinite-dimensional systems introduces new types of quantum codes not previously considered, namely factors of type II and III. An example of correctable type II factor is given in Section \ref{section:extypetwo} below.

\msection{Examples}

Let us conclude this chapter by giving a few examples demonstrating features of the new type of codes that we obtained.

\subsection{Operator system}
\label{section:exopsys}

We can easily construct examples of correctable operator systems which are not algebras by taking any algebraic code correctable on a subspace $\Hil_0 \subseteq \Hil_1$, and using the correction channel to ``lift'' it to an operator system correctable without restriction on states, as explained in Section \ref{section:simultcorr}.
As an example, we will take the simplest standard code which corrects any random bit flip. This code is defined by a two-dimensional subspace $\Hil_0 \sim \mathbb C^2$ of a three-qubit Hilbert system $\Hil_1$. It is chosen such that the whole algebra of operators on the subspace is correctable for any channel with elements in the span of the ``error operators'' $\{\one, X_1, X_2, X_3\}$ where $X_i$ is a Pauli matrix acting on the $i$th qubit. The subspace $\Hil_0$ is defined by the isometry
\[
V\ket{i} = \ket{iii}
\]
where $i \in \{0,1\}$ and we wrote $\ket{iii} = \ket{i} \otimes \ket{i} \otimes \ket{i}$. We also define
\[
P := VV^*  = \proj{000} + \proj{111}.
\]
It is easy to check that a correction channel for this code is 
\[
\mathcal R^\dagger(A) = V A V^* + \sum_i X_i V A V^* X_i
\]
i.e., $V^*\chanh(\mathcal R^*(A))V = A$ for all $A \in \finops(\Hil_0)$. 

The algebra correctable on the code $\Hil_0$, as seen embedded in $\mathcal B(\Hil_1)$ is
\[
V \mathcal A_0 V^* = V \finops(\Hil_0) V^* = \Bigl\{ \sum_{ij} \alpha_{ij} \ket{iii}\bra{jjj} : \alpha_{ij} \in \mathbb C \Bigr\}
\]

In order to proceed, we need a specific error channel. We will pick one which corresponds to assigning a probability for the occurrence of each error in the set $\{\one, X_1, X_2, X_3\}$:
\[
\chan(\rho) = p_0 \rho + \sum_i p_i X_i \rho X_i.
\]
Writing $X_0 := \one$ for convenience, the operator system
\[
\mathcal S_0 = \chanh(\mathcal R^*(\mathcal A_0)) =
 \Bigl\{ \sum_{i,j=0}^1 \alpha_{ij} \sum_{k,l = 0}^3 p_{k} X_k X_l \ket{iii}\bra{jjj} X_l X_k : \alpha_{ij} \in \mathbb C \Bigr\}
\]
is correctable by $\mathcal R$ on all states. We should have $P \mathcal S_0 P = V \mathcal A_0 V^*$. This can be seen from the fact that $\ket{iii} = P \ket{iii}$ and $P X_k X_l P = \delta_{kl} P$. Explicitly separating the components respectively inside $V \mathcal A_0 V^*$ and orthogonal to $V \mathcal A_0 V^*$ we have 
\[
\mathcal S_0 = \Bigl\{ \sum_{ij} \alpha_{ij} \Bigl({ \ket{iii}\bra{jjj} + \sum_{k \neq l} p_{k} X_k X_l \ket{iii}\bra{jjj} X_l X_k }\Bigr) : \alpha_{ij} \in \mathbb C \Bigr\}.
\]

\subsection{Classical channels}
\label{section:excec}

Formally, we only derived our results for quantum channels. However we can deduce what happens in the case of a channel between two classical systems. For simplicity, let us consider the finite-dimensional case. 
A classical channel
\[
\pi: \finfun(\Omega_1) \rightarrow \finfun(\Omega_2).
\]
can be represented by a quantum channel 
\[
\chan: \mathcal \finops(L^2(\Omega_1)) \rightarrow \finops(L^2(\Omega_2))
\]
defined by
\[
\chan(\rho) = \sum_{ij} \pi_{ij} \bra{j} \rho \ket{j} \proj{i}.
\]
where
\[
\sum_{i} \pi_{ij} = 1.
\]
This means that the channel elements are
\[
E_{ij} = \sqrt{\pi_{ij}} \ket{i}\bra{j}.
\]
In order to make sure that our scheme is entirely classical, we want to check that the correction channel $\mathcal R$ is classical. In finite dimension, it can be constructed as
\[
\mathcal R(\rho) = \chanh(K\rho K)
\]
where $K := (\chan(\one))^{-\frac{1}{2}}$. Clearly $K$ is diagonal in our basis $\ket{i}$, and $\chanh$ maps diagonal operators to diagonal operators. Therefore $\mathcal R$ represents a classical channel. 

In order to find the correctable algebra, note that
\[
E_{ij}^\dagger E_{kl} = \delta_{ik} \sqrt{\pi_{ij} \pi_{il}}  \ket{j}\bra{l}.
\]
A classical effect $\alpha = \sum_k \alpha_k \proj{k}$ is correctable if and only if it commutes with all these operators, i.e. 
\begin{align*}
&& \sum_k \sqrt{\pi_{ij} \pi_{il}} \alpha_k \ket{k}\braket{k}{j}\bra{l} &= \sum_k \sqrt{\pi_{ij} \pi_{il}} \alpha_k \ket{j}\braket{l}{k}\bra{k} & & \forall i,j,l\\
\Longleftrightarrow& & \alpha_j \pi_{ij} \pi_{il} \ket{j}\bra{l}  &= \alpha_l \pi_{ij} \pi_{il}  \ket{j}\bra{l} & & \forall i,j,l\\
\Longleftrightarrow& & (\alpha_j - \alpha_l) \pi_{ij} \pi_{il}  &= 0   & & \forall i,j,l\\
\end{align*}
Hence, the effect $\alpha$ can be corrected if and only if $\alpha_j = \alpha_k$ for all the states $k$, $j$ which are such that there exists $i$ with $\pi_{ij} \neq 0$ and $\pi_{ik} \neq 0$. This simply means that two states cannot be distinguished from each other when there is a nonzero probability for a transition from a common state to both of them.

Let us see what the correction channel does explicitly. First, note that 
\[
K = (\chan(\one))^{-\frac{1}{2}} = (\sum_{ij} \pi_{ij} \proj{i})^{-\frac{1}{2}} = \sum_i (\sum_j \pi_{ij})^{-\frac{1}{2}} \proj{i}
\]
where the sum over $i$ is restricted to the the terms which are such that $\sum_j \pi_{ij} \neq 0$. 
Let us define $K_i := \sum_j \pi_{ij}$. 
The correction channel on a classical state $\rho = \sum_i \mu_i \proj{i}$ is
\[
\begin{split}
\mathcal R(\rho) &= \sum_i \frac{\mu_i}{\sum_j \pi_{ij}}  \chanh(\proj{i}) = \sum_{ik} \frac{\mu_i \pi_{ik} }{\sum_j \pi_{ij}}  \proj{k}.\\
\end{split}
\]
Expressed as a stochastic matrix, it has components
\[
\pi_{ij}^R = \frac{\pi_{ji} }{\sum_k \pi_{jk}}.
\]
Note that the stochastic matrix element $\pi_{ji}$ is the probability that the channel $\pi$ would output the state $j$ if the input was $i$. We see that this correction channel does the following: if it is fed with the state $j$, it randomly outputs any of the states $i$ for which $\pi_{ji} \neq 0$. Indeed, it has no way of knowing which one of these was the initial state. However it knows that it could not have been any other state. Clearly the correction works exactly only if there is no ambiguity, i.e. if $j$ could have come only from a single state $i$. This is why the correctable observables are those which do not distinguish between two states which have a non-zero probability of transitioning to the same output state $j$. 


\subsection{Failed teleportation}




We can view the standard quantum teleportation protocol \cite{bennett93} as an example of quantum error correction. Indeed, Bob must find a way to reconstruct Alice's quantum information from the classical bits she provides, and his half of the entangled pair.

Here we will use our framework to show that if some of the classical information gets lost in the way, Bob will only be able to reconstruct a {\em hybrid}. 


\begin{figure}
\begin{center}
\begin{pgfpicture}{0}{0}{5in}{3in}
 \pgfputat{\pgfxy(0,-3)}{
   \pgfputat{\pgfxy(0,0)}{
     \pgfbox[left,bottom]{\includegraphics[width=8cm]{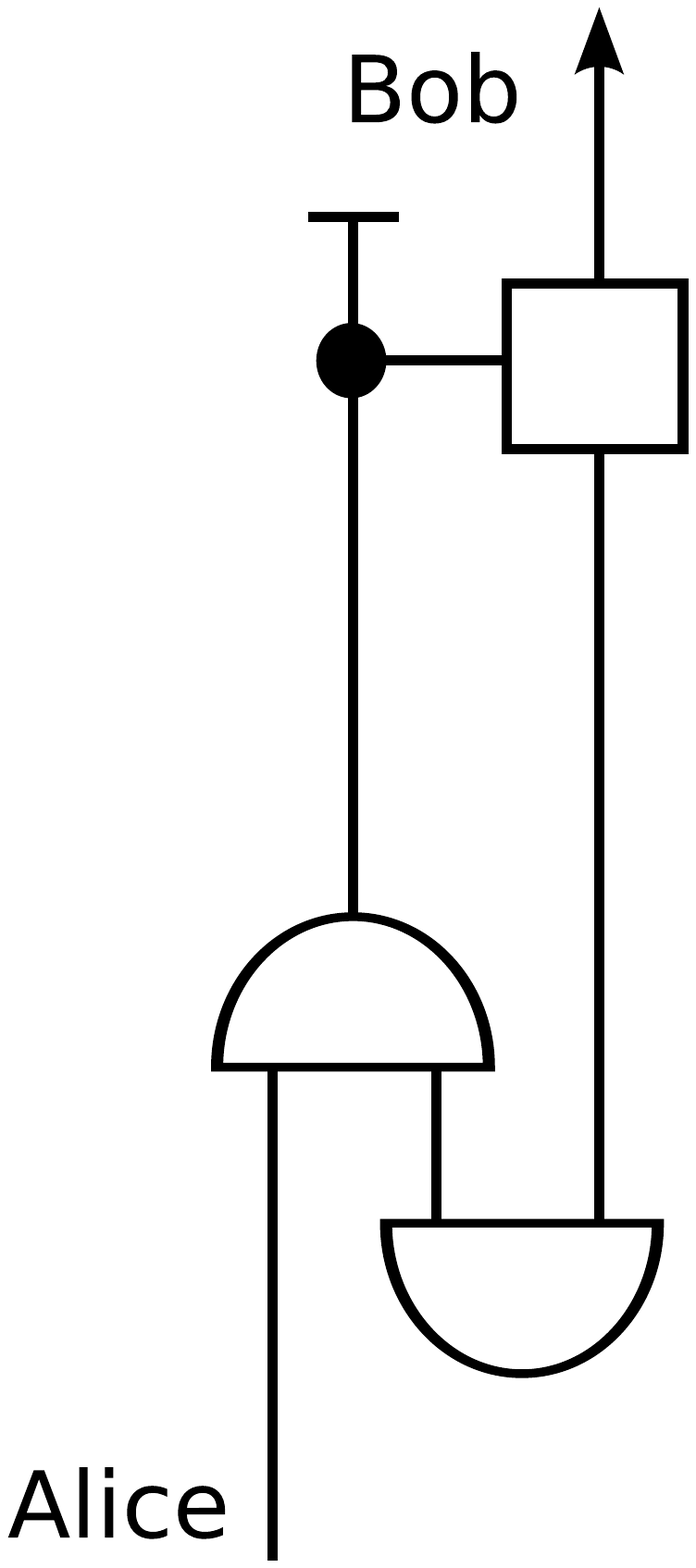}}
   }
   \pgfputat{\pgfxy(6,0)}{
     \pgfbox[left,bottom]{\includegraphics[width=8cm]{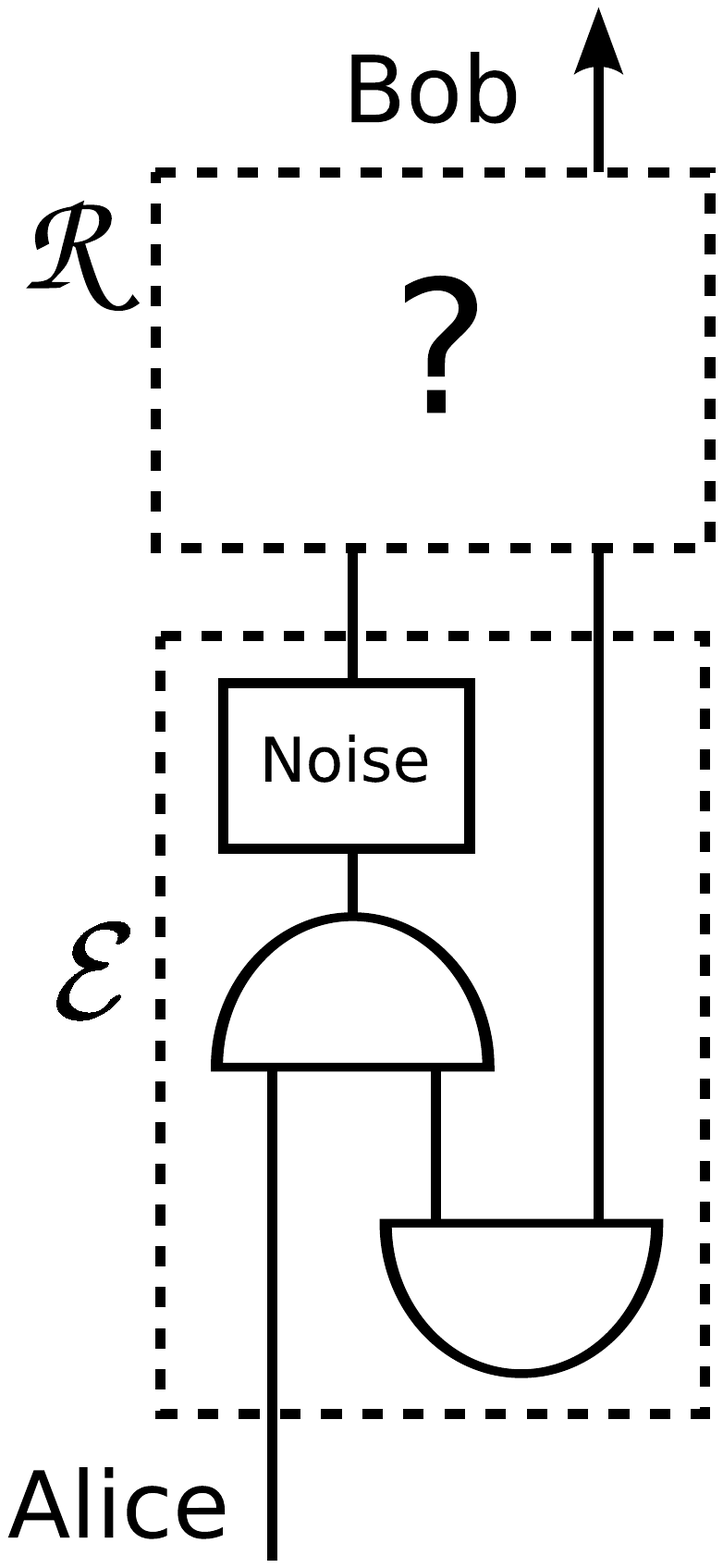}}
   }
}
\end{pgfpicture}
\end{center}
\caption[Noisy teleportation]{Left: standard teleportation circuit. Time runs from bottom to top. The input is Alice's quantum state to be teleported. The first half disk is a maximally entangled Bell state of two qubits. The second half disk represents a measurement in the Bell basis, the output of which is classical information (the outcome of the measurement). This classical information together with the second entangled qubit are given to Bob who decodes the information by applying a unitary transform on the qubits, conditioned on the classical information. Right: what should Bob do, and how much of the quantum information can he recover, if some of the classical information is lost in the way? 
}
\label{fig:teleportation}
\end{figure}

For teleportation to be possible, Bob and Alice must initially share a pair of entangled qubits. This pair is described by the state
\[
\ket{\psi} = \frac{1}{\sqrt 2}( \ket{0} \otimes \ket{0} + \ket{1} \otimes \ket{1}). 
\]
Beside her entangled qubit, Alice possesses an other qubit in an arbitrary state $\rho$ that she wants to transmit to Bob. The point of the protocol is that she is not allowed to send Bob any quantum information. 
This means that all she can do is to perform some measurement on her pair of qubits, and send the classical result to Bob (see Figure \ref{fig:teleportation}).

The measurement that she must perform is the one defined by the basis of four orthogonal maximally entangled states, which can be obtain by applying the Pauli matrices to one half of $\ket{\psi}$. Hence the observable's elements are
\[
X_i = (\one \otimes U_i) \proj{\psi} (\one \otimes U_i^*) = \proj{\psi_i}
\]
where $U_0 = \one$ and $U_i = \sigma_i$ for $i \in \{1,2,3\}$. Note that $U_i^* = U_i$. 
One can check by direct calculation that $\sum_i X_i = \one$.
These states are all maximally entangled because they are obtained by local transformations of a maximally entangled state. 
The key property of these states is the following:
\[
\ket{\psi_i} = (\one \otimes U_i) \ket{\psi} =  (U_i \otimes \one) \ket{\psi}.
\]


The channel from Alice to Bob is:
\[
\chan(\rho) =  (X \otimes {\rm id})(\rho \otimes \proj{\psi})
\]
where ${\rm id}$ is the identity channel on Bob's half of $\ket{\psi}$. To expand this, note that if we represent the classical target system of $X$ in terms of diagonal quantum states, we have, for an arbitrary state $\rho'$ of Alice's couple of qubits,
\[
X(\rho') = \sum_i \tr(X_i \rho') \proj{i} = \sum_i \ket{i}\bra{\psi_i} \rho' \ket{\psi_i}\bra{i}
\] 
Therefore,
\[
\begin{split}
\chan(\rho) &= \sum_i (\ket{i}\bra{\psi_i} \otimes \one)(\rho \otimes \proj{\psi} )(\ket{\psi_i}\bra{i} \otimes \one)\\
&= \frac{1}{2}\sum_{ijk} (\ket{i}\bra{\psi_i} \otimes \one)(\rho \otimes \ketbra{j}{k} \otimes \ketbra{j}{k}) (\ket{\psi_i}\bra{i} \otimes \one)\\
&= \frac{1}{2}\sum_{ijk} \ket{i}\bra{\psi_i} (\rho \otimes \ketbra{j}{k}) \ket{\psi_i}\bra{i}  \otimes \ketbra{j}{k}\\
&= \frac{1}{2}\sum_{ijk} \bra{\psi} ( U_i \rho U_i \otimes \ketbra{j}{k}) \ket{\psi}\; \proj{i}  \otimes \ketbra{j}{k}\\
&= \frac{1}{4}\sum_{ijknm} \bra{n} U_i \rho U_i \ket{m} \braket{n}{j} \braket{k}{m} \; \proj{i}  \otimes \ketbra{j}{k}\\
&= \frac{1}{4}\sum_{ijk} \bra{j} U_i \rho U_i \ket{k} \; \proj{i}  \otimes \ketbra{j}{k}\\
&= \frac{1}{4}\sum_{ijk} \proj{i}  \otimes \ket{j} \bra{j} U_i \rho U_i \ket{k} \bra{k}
= \frac{1}{4}\sum_{i}  \proj{i}  \otimes U_i \rho U_i.\\
\end{split}
\]
Once Bob receives this state, it is clear that he just has to measure the classical system, and apply the inverse of the corresponding unitary transformation in order to recover the state $\rho$.

From the point of view of quantum error correction, we can see that the qubit can indeed be recovered by computing the operators $E_i^*E_j$ where
\[
E_i = \frac{1}{2} \ket{i} \otimes U_i.
\]
Indeed, we have
\[
E_i^*E_j \propto \braket{i}{j} \otimes U_i^\dagger U_j = \delta_{ij} U_i^\dagger U_i = \delta_{ij} \one.
\]
This shows that the whole algebra of operators on the qubit is correctable. 

Note that we can teleport $n$ qubits simply by applying this protocol to each qubits in parallel, which requires also $n$ entangled pairs shared between Alice and Bob. In this case the channel from Alice to Bob is
\[
\chan_n(\rho) = \frac{1}{4^n} \sum_i \proj{i}  \otimes U_i^{(n)} \rho U_i^{(n)}
\]
where $\rho$ is now an $n$ qubits state, and $U_i$, $i=0,\dots,4^n-1$ are all possible tensor products of Pauli operators and identity operators. 

An interesting aspect of this teleportation protocol is that, in a way, the quantum state to be transmitted is entirely encoded in the classical information being transmitted. More precisely, $n$ qubits seem to be encoded in $2n$ classical bits. But how is this encoding done? We will try to clarify this question by answering a more precise one: what aspects of the quantum information do we loose if we destroy some of the classical information?

The most general way of destroying some of the classical information transmitted from Alice to Bob consists in applying a classical channel $\pi$ to it (see Figure \ref{fig:teleportation}). The channel elements of the classical channel are (see Section \ref{section:excec})
\[
F_{ij} = \sqrt{\pi_{ij}} \ketbra{i}{j}
\]
so that the overall channel from Alice to Bob has elements
\[
F_{ij} E_k = \frac{1}{2}\sqrt{\pi_{ij}} \ket{i} \braket{j}{k} \otimes U_k^{(n)} = \delta_{jk} \frac{1}{2}\sqrt{\pi_{ij}} \ket{i} \otimes U_j^{(n)}.
\]
The relevant operators for determining the correctable sharp observables are, dropping the unimportant factors $\frac{1}{2}$,
\[
\begin{split}
E_k^* F_{ij}^*  F_{i'j'} E_{k'} &\propto \delta_{jk}\delta_{j'k'} \sqrt{\pi_{ij} \pi_{i'j'}} \braket{i}{i'} \, U_j^{(n)} U_{j'}^{(n)} \\
&= \delta_{jk} \delta_{j'k'} \delta_{ii'} \sqrt{\pi_{ij} \pi_{ij'}} \, U_j^{(n)} U_{j'}^{(n)}.
\end{split}
\]
Therefore, the correctable sharp effects are those commuting with the operators
\[
\sqrt{\pi_{ij} \pi_{ik}} \, U_j^{(n)} U_{k}^{(n)}
\]
for all $i$, $j$, $k$. 

An effect may commute with one of these operators for one of two reasons. Either it is invariant under the corresponding transformation, or
\[
\pi_{ij} \pi_{ik} = 0.
\]
Remember from the classical error correction example (Section \ref{section:excec}), that this condition on $\pi$ means that the classical state $j$ cannot be distinguished from the state $k$ because of the noise. Hence, whenever the noise erases the distinction between two classical states $j$ and $k$, the observables that Bob can recover must be invariant under the unitary transformation $U_j^{(n)} U_{k}^{(n)}$. 
This result can be understood as implying that the classical information encodes how Bob's qubit must be transformed in order to recover Alice's state, which is in fact precisely how Bob effectively interprets it. If the information about which transformation to apply is lost, then the observables which can be recovered are those which do not distinguish between the outcomes of the two transformations. 

For instance, in the case $n=1$, suppose that we loose the distinction between $j=0$ and $k=3$, which corresponds to the unitary operators $\one$ and $\sigma_z$. The correctable observables must then commute with $\sigma_z$, which implies that they must be diagonal. Therefore they only represent one single bit of information from Alice's qubit.  
However, if we also loose the distinction between the states $1$ and $2$, the correctable information is exactly the same. Indeed, the new symmetry that must be imposed in this case is $\sigma_x\sigma_y \propto \sigma_z$. This shows that our error correction technique misses some information in the former case. 

Recall that the correctable information is characterized by the sharp preserved observables. 
In order to see what happens in more detail, we would need to look at the preserved unsharp observables. For simplicity, let us simply consider the preserved unsharp effects. Note that an effect of the hybrid quantum-classical system of Bob is of the form
\[
B = \sum_i \beta_i \proj{i} \otimes B_i.
\]
Its image in the Heisenberg picture is
\[
A = \frac{1}{4} \sum_{ij} \pi_{ij} \beta_i U_j B_i U_j = \frac{1}{4} [\beta_0 (B_0 + \sigma_z B_0 \sigma_z) + \beta_1 \sigma_x B_1 \sigma_x + \beta_2 \sigma_y B_2 \sigma_y].
\]
We can always choose $\beta_1 = \beta_2 = 1$, and $B_1$ and $B_2$ in order to obtain any effect for the second and third term. However, the term 
\[
B_0 + \sigma_z B_0 \sigma_z = 2 \sum_i \proj{i} B_0 \proj{i}
\]
is always diagonal. Therefore we can write all the preserved effects as
\[
A = \frac{1}{2} D + \frac{1}{2}B 
\]
where $B$ is an arbitrary effect, and $D$ is any diagonal effect. The further loss of the distinction between the states $1$ and $2$ would also force the second terms to be diagonal, so that only diagonal effects are then preserved. 

\subsection{Irrational rotation algebra}
\label{section:extypetwo}

Consider the algebra generated by two elements $\hat x$ and $\hat p$ satisfying the canonical commutation relations
\[
[\hat x, \hat p] = i \one.
\]
This algebra can be represented on $\Hil = L^2(\mathbb R)$, where the position operator $\hat x$ acts on a function $\psi \in L^2(\mathbb R)$ as $(\hat x \psi)(x) = x \psi(x)$ and the momentum $\hat p$ as $(\hat p \psi)(x) = i \frac{d}{dx} \psi(x)$. 

Suppose that this system interacts with an environment through a Hamiltonian of the form $H = \sum_i J_i \otimes K_i$, where the operators $J_i$ act on the system, and the operators $K_i$ on the environment. We have seen in Section \ref{section:qec} that this implies that the channel elements of the resulting channel on the system belong to the algebra generated by the operators $J_i$. 

We will now suppose that the interaction operators $J_i$ are of two forms. Some are periodic functions of $\hat x$, with period $L_x$, and others are periodic functions of $\hat p$, of period $L_p$. This implies that these functions are linear combinations of powers of the functions $x \mapsto e^{i \frac{2 \pi}{L_x} \,x}$ or $p \mapsto e^{i \frac{2 \pi}{L_p} \,p}$ respectively (their discrete Fourier components). For convenience, let us define
\[
\omega_x := \frac{2 \pi}{L_x} \quad \text{ and } \quad \omega_p := \frac{2 \pi}{L_p}
\]
The von Neumann algebra generated by the interaction operators is also generated by the two unitary operators
\[
U = e^{i \omega_x \,\hat x} \quad \text{and} \quad V = e^{i \omega_p \, \hat p}
\]
In addition, we assume that the real number
\[
\theta := \frac{\omega_x \omega_p}{2\pi} 
\]
is irrational. This number is important because it enters into the commutator of $U$ and $V$:
\[
UV = e^{2 \pi i \theta} V U.
\]

In principle, in order to find the correctable algebra, we need to find the operators commuting with the products $E_i^* E_j$ of the channel elements $E_i$. However, since we only know the span of these operators, we cannot exclude that $E_i = \one$ for some $i$. If this is the case, then these products include $E_i$ and $E_i^*$ for all $i$. Therefore, for an effects to be correctable, it needs to be in the commutant of the von-Neumann algebra generated by the operators $E_i$ for all $i$, which is the same as the von Neumann algebra generated by the interaction operators $J_i$. 


The operators
\[
U' =  e^{i \frac{\omega_x}{\theta} \,\hat x} \quad \text{and} \quad V' =  e^{i \frac{\omega_p}{\theta} \,\hat p}
\]
commute with both $U$ and $V$. 
To see that $U'$ commutes with $V$, simply note that:
\[
U' V = e^{i \frac{\omega_x \omega_p}{\theta} } V U' =  e^{i\frac{2\pi \theta}{\theta}} V U' = V U'.
\]
Similarly, $V'$ also commutes with both $U$ and $V$. 
In fact, the von Neumann algebra generated by $U'$ and $V'$ is the whole commutant of the algebra generated by the interaction operators $J_i$ \cite{faddeev95}. In addition, it happens to be a factor of type II, and, together, with its commutant they generate the whole of $\mathcal B(\Hil)$.

Therefore, this is an example of a correctable factor of type II. 
In fact, it is also {\em noiseless} \cite{knill00}, in the sense that the correction channel can be taken to be the identity channel, i.e. no active correction is needed. This happens simply because $\one$ was assumed to be among the channel elements. 
Indeed, we saw that it implied that the correction operators had to commute with the channel elements themselves. Hence, if $\chan$ is the channel, we have $\chanh(A) = \chanh(\one) A = A$ for all elements $A$ of the correctable algebra.

Let us see how we can understand this ``\ind{type II subsystem}'', and how it resembles, and differs from, the factors of type I with which we are familiar. 
If we were dealing with a factor of type I containing the identity, then the Hilbert space would take the form $\Hil = \Hil_1 \otimes \Hil_2$, so that our algebra would be simply $\mathcal B(\Hil_1) \otimes \one$. In this sense, a factor of type I defines a {\em subsystem} in the traditional sense. For instance, consider $\Hil = L^2(\mathbb R^2) = L^2(\mathbb R) \otimes L^2(\mathbb R)$. 
If $\psi \in L^2(\mathbb R^2)$, the operators in the first factor $\mathcal A = \mathcal B(L^2(\mathbb R)) \otimes \one$ are those which act only on the first component of $\psi$. For instance, the first factor is generated by the operators $(\hat x \psi)(x,y) = x\psi(x,y)$ and $(\hat p \psi)(x,y) = \frac{d}{dx}\psi(x,y)$. Note that here the set $\mathbb R^2$ on which the states are defined can be understood to be the set of joint eigenvalues to the position operators in $\mathcal A$ and $\mathcal A'$. 

Something similar happens for our factor of type II. Let $\mathcal A$ be the factor generated by $U$ and $V$, and $\mathcal A'$ its commutant, which is generated by $U'$ and $V'$. 
Let us see if we could see the elements of $\Hil$ as wavefunctions over the eigenvalues of $U$ and $U'$. First, note that the spectrum of both these operators is a circle (represented in the complex plane). These two operators being functions of the position operator $\hat x$, we may want to use the fact that the states of $\Hil$ can be represented as wavefunction over the spectrum of $\hat x$, i.e. elements of $L^2(\mathbb R)$. Indeed, we can naturally convert an eigenvalue $x$ of $\hat x$ into the eigenvalues 
\begin{equation}
\label{eq:extworep}
a = e^{i\omega_x x} \quad \text{ and } \quad b = e^{i \frac{\omega_x}{\theta} x }
\end{equation}
respectively of $U$ and $U'$. In fact, this relationship is invertible. Indeed, if we are given $a$ and $b$, then only a single real number $x$ will satisfy both these relations. Indeed, suppose that we had two different real numbers $x$ and $x'$ yielding the same values of $a$ and $b$. This would imply that they are related by $x-x' = 2 \pi n /\omega = 2 \pi m \theta /\omega$ for two integers $n$ and $m$. But this would imply $\theta = n/m$, which is not possible since we assumed $\theta$ to be irrational.  
This implies that for each state $\psi \in L^2(\mathbb R)$, we can define the function
\[
\widetilde \psi(a,b) := \psi(x)
\]
where $x$ is the unique real number related to $a$ and $b$ via the Equations \ref{eq:extworep}.
Note that this function $\widetilde \psi$ is defined only on the valid couples $(a,b)$ related to some $x \in \mathbb R$ via Equation \ref{eq:extworep}. However, due to the irrationality of $\theta$, these couples are dense in the unit torus. We can therefore think of $\widetilde \psi$ as being defined almost everywhere on the torus. 
We will indeed see that it can be interpreted, in a suitable sense, as the wavefunction of a particle on a two-dimensional \ind{torus}. 

The relation between the wavefunctions $\widetilde \psi(a,b)$ and the factors $\mathcal A$ and $\mathcal A'$ is given by the fact that they ``act'' respectively on the first and second arguments of $\psi$ respectively. Indeed, we have
\[
(U \widetilde \psi)(a,b) = (U \psi)(x) = e^{i\omega_x x} \psi(x) = a \, \widetilde \psi(a,b)
\]
which means that it acts just like the first component of the position of the particle. Similarly,
\[
(U' \widetilde \psi)(a,b) = (U' \psi)(x) = e^{i\frac{\omega_x}{\theta} x} \psi(x) = b \, \widetilde \psi(a,b)
\]
The action of $V$ is also easy to compute:
\[
(V \widetilde \psi)(a,b) = (V \psi)(x) = \psi(x + \omega_p) = \psi(a \,e^{2 \pi i \theta}, b).
\]
Hence, its effect is to rotate the first argument by the irrational angle $\theta$.
Similarly,
\[
(V' \widetilde \psi)(a,b) = (V' \psi)(x) = \psi(x + \omega_p/\theta) = \psi(a , b\,e^{2 \pi i \frac{1}{\theta}}).
\]

Although it looks like a particle on the torus, this system differs from it by the nature of the normalized states. Indeed, the norm is
\[
\|\widetilde \psi\|^2 = \int |\psi(x)|^2 \, dx = \int | \widetilde \psi( e^{i\omega_x x}, e^{i \frac{\omega_x}{\theta} x }) |^2 \,dx.
\]
What we have done is to take a standard particle in a one-dimensional space, and wrap its space around a torus in a dense trajectory. If we view the particle as a wavefunction $\widetilde \psi$ on the torus, its norm is an integral over this dense path.
This picture illustrates what the noise does. It disturbs only the first component of the position of this particle, but not the second.












\mchapter{Decoherence}%
\label{chapter:decoherence}

In this chapter we present three different models of \ind{decoherence}. The third one being, in a way, a combination of the first two. 
The first model (Section \ref{section:sharpdecoh}) focuses on sharp observables, and is essentially based on our results related to quantum error correction. It will serve as a guide for our intuition as to what happens in the more general case involving unsharp observables. The second model (Section \ref{section:sysdecoh}) focuses on the system itself, ignoring the environment, and defines the notion of an unsharp pointer observable. The third, more complete picture (Section \ref{section:fulldecoh}) considers the exchange of unsharp information between the system and the environment. These results were presented in \cite{beny08x1}.

Before we describe these models, we will come back on a point made in the introduction, and define what we mean by a classical limit.

\msection{Nature of a classical limit}%
\label{section:limit}





In nature, a classical system is a quantum system which happens to be in a context where its behaviour can be successfully modeled by a classical theory. In principle, neglecting computational difficulties, we ought to be able to model this system's behaviour using its full quantum description as well. This implies that we should have in principle two different models which both describe the same physical system: a quantum one, specified by an algebra of the form $\mathcal B(\Hil)$, and a classical one, defined by a commutative algebra $L^\infty(\Omega)$. Since the quantum theory is more fundamental, we should be able to translate every propositions of the classical model (i.e. every effects) into a proposition in the underlying quantum language. This means that we should have a map of the form
\[
\Gamma^*: L^\infty(\Omega) \rightarrow \mathcal B(\Hil)
\]
\index{$\Gamma$}
which maps effects to effects. In addition, we may expect that a certain knowledge about the quantum system should induce back a knowledge about the effective classical propositions. We followed the same arguments in Section \ref{section:observables} when we introduced observables, which yielded the requirement that the map must be linear, positive, unital and normal. The map $\Gamma^*$ is the \adjoint of a channel $\Gamma$ which maps quantum states into classical states, just like an observable:
\[
\Gamma: \mathcal B_t(\Hil) \rightarrow  L^1(\Omega).
\]

Hence, we expect that a classical limit, at the kinematical level, is given by an observable. 
Our model of decoherence will take quite literally the idea that a classical limit emerges from a ``measurement'' of the system by the environment. In those terms, $\Gamma$ is simply the observable being measured. We will refer to $\Gamma$ as the {\em \ind{pointer observable}}. 

\subsection{Quantization}
\label{section:quantization}

We know that $\Gamma$ does not only define a map from classical effects to quantum effects, but more generally a map from classical observables to quantum observables. Hence it can be understood as defining, in a somewhat restricted sense, a {\em \ind{quantization}} procedure.

In its most general form, a classical observable on a classical system defined by the algebra $L^\infty(\Omega)$ is a channel 
\[
\pi: L^1(\Omega) \rightarrow L^1(\Omega').
\]
For instance, suppose that $\Omega = \mathbb R^2$ is the phase-space for a single continuous variable system (say a single particle in a one-dimensional space). Then the position observable is the map which sends a probability distribution $\mu(q,p)$ over the phase-space $\mathbb R^2$ to its marginal $\mu' = \int \mu(q,p) \,dp$ (see also Figure \ref{figure:position} for the action of the dual map). But observables need not have only scalar values. For instance, there is a phase-space observable, which is simply the identity map on $L^1(\Omega)$. Clearly this ``phase-space'' observable always exists and has a special status: it gives full information about the classical system. 

Remember (from Section \ref{section:heisenberg}) that a quantum channel $\chan$ can be used indirectly to map an observable $X$ of the destination to the observable $Y = X \circ \chan$ of the source. The use of the dual map $\chanh$ is implicit in this expression. Dualizing the equation we obtain $Y^* = \chanh \circ X^*$, which means $Y^*(\alpha) = \chanh(X^*(\alpha))$ for each proposition $\alpha$. The expression $Y = X \circ \chan$ is convenient because it expresses in a straightforward way that observing $Y$ on the source is done by first applying the channel $\chan$ and then observing $X$. 

The pointer observable $\Gamma$ maps quantum states into classical states, and can therefore be used to map a classical observable $\pi$ into a quantum observable $X$ via the equation $X = \pi \circ \Gamma$. This means that measuring $X$ amounts to first measuring the observable represented by $\Gamma$, and then measuring the classical observable $\pi$ on the classical outcome of $\Gamma$. Equivalently we have $X^*(\alpha) = \Gamma^*(\pi^*(\alpha))$. 
Of course, $\Gamma$ itself is trivially the image of the special classical observable represented by the identity on $\Omega$: $\Gamma = \id \circ \Gamma$. 


If two classical observables are mapped to two quantum observables which do not commute, then the classical observable representing their joint measurement is mapped to an unsharp quantum observable. For instance, the classical observable ``$\id$'' mentioned above, which represents the joint measurement of all classical phase-space variables, is mapped to $\Gamma$.

Note that although $\Gamma$ maps any classical observable to a quantum one, and every quantum state to a classical one, it does not in general attain all classical states. Indeed, the classical states that it can represent by quantum states will always satisfy all the uncertainty relations. This is where the ``limit $\hbar \rightarrow 0$'' is involved. It comes as a second step which consists in artificially removing these constraints on states, hence assuming that the classical observations are too inaccurate to notice them.


Let us study this through a standard example.
Consider the case of a single free electromagnetic mode. This quantum system is characterized by the Hilbert space $\Hil = L^2(\mathbb R)$, and its classical limit by a phase-space $\Omega = \mathbb R^2$. 
We will denote the points of phase space by pairs $z = (p,q)$ where $q$ will be called the ``position'' and $p$ the ``momentum''.


In many physically important situations, we know that the classical pure states characterized by canonically conjugate pairs $z = (q,p) \in \Omega$ correspond to the quantum coherent states $\ket{z}$. 

Coherent states are pure states in the Hilbert space $\Hil = L^2(\mathbb R)$ which form a complete set, i.e.
\[
\int \proj{z} \, dz = \one
\]
where
\[
dz := \frac{dp \,dq}{2 \pi \hbar}.
\]
These states however are not orthogonal and do not form a basis. 
For our purpose, we need only say that they are generated by acting on a fiducial vector $\ket{\psi_0}$ with elements of the group of ``translations'' in phase-space:
\[
\proj{z} = U_z \proj{0} U_z^*
\]
where, using the quantum operators $\widehat q$ and $\widehat p$ satisfying $[\widehat q, \widehat p] = i\hbar$, we defined
\[
U_z := e^{- \frac{i}{\hbar} (q \widehat p + p \widehat q)}.
\]
This group is a representation of the group of translations and ``boosts'' (i.e. translations in momentum). Note that this set of unitary operators does not close into a group if we consider their action on vectors. Indeed, a non-trivial phase may pop up due to the non-commutativity of $\widehat p$ and $\widehat q$. However this phase is eliminated when the action on mixed states is considered, which is all that matters here.

In our language, this classical limit is defined by the observable $\Gamma$, which maps a quantum state $\rho$ to the classical state  $\mu := \Gamma(\rho)$ defined by
\[
\mu(z) = \bra{z} \rho \ket{z}.
\]
The effect of the \adjoint map $\Gamma^*$ is to send the classical effect $\alpha \in L^\infty(\Omega)$ to the quantum effect 
\[
\Gamma^*(\alpha) = \int_\Omega\, \alpha(z) \proj{z} dz.
\]
We can now play with this map and see how it transforms observables or states. 

For instance consider the classical position observable $Q$, which maps a phase-space probability distribution $\mu(q,p)$ to its marginal $\mu(p) = \int \mu(q,p) dp$. Its quantum version, $Q \circ \Gamma$, maps a quantum state $\rho$ to the marginal of the distribution $\mu(q,p) = \bra{q,p} \rho \ket{q,p}$. Its \adjoint, which operates on effects $\alpha \in L^\infty(\mathbb R)$, is given by
\[
\begin{split}
(Q \circ \Gamma)^*(\alpha) &= \int_\Omega \alpha(q) \proj{p,q} dz \\
 &=  \int \alpha(q) \left[{ \int\proj{p,q} \frac{dp}{2 \pi \hbar}}\right]\,dq  \\
 &=  \int \alpha(q) Q_q \,dq  \\ 
\end{split}
\]
where we have defined the ``effect density'' 
\[
Q_q := \int\proj{p,q} \frac{dp}{2 \pi \hbar}.
\]
These operators are, in fact, simple functions of the sharp position operator $\widehat q$.
Indeed, using the position representation, we have
\[
\braket{x}{0,p} = \bra{x}e^{-\frac{i}{\hbar} p \widehat q}\ket{\psi_0} = e^{-\frac{i}{\hbar} p \widehat q} \psi_0(x) =  e^{-\frac{i}{\hbar} p x} \psi_0(x).
\]
which we can use to compute
\[
\begin{split}
Q_q &= \int\proj{p,q} \frac{dp}{2 \pi \hbar}\\
&= \int \ket{x}\braket{x}{p,q}\braket{p,q}{x'}\bra{x'} \frac{dp}{2 \pi \hbar}\,dx\,dx'\\
&= \int \left[{ \int \braket{x}{p,q} \braket{p,q}{x'}\frac{dp}{2 \pi \hbar}  }\right] \ket{x}\bra{x'} \,dx\,dx'\\
&= \int \left[{ \int \braket{x-q}{p,0} \braket{p,0}{x'-q} \frac{dp}{2 \pi \hbar}  }\right] \ket{x}\bra{x'} \,dx\,dx'\\
&= \int \left[{ \int e^{-\frac{i}{\hbar} p (x-x')} \frac{dp}{2 \pi \hbar}}\right] \braket{x-q}{\psi_0} \braket{\psi_0}{x'-q} \, \ket{x}\bra{x'} \,dx\,dx'\\
&=  \int \braket{x-q}{\psi_0} \braket{\psi_0}{x-q} \ket{x}\bra{x} \,dx = f_q(\hat x)\\
\end{split}
\]
where $f_q(x) := |\bra{x} e^{-\frac{i}{\hbar} q \widehat p} \ket{\psi_0}|^2$.

If the fiducial vector $\psi_0(x)$ is localized at the origin, then we see that the observable $Q \circ \Gamma$, characterized by the operators $Q_q$ which play the role of continuous POVM elements, is an approximate version of the sharp position observable represented by $\widehat q$. In fact, it is likely that the set of operators $Q_q$ for all $q$ generate the algebra generated by $\widehat q$, which essentially identifies $\widehat q$. Of course the same is true for the quantization $P \circ \Gamma$ of the classical momentum observable $P$. 

More generally, consider a sharp real classical observable $\pi_f$ represented by a real (possibly unbounded) function $f$ on the phase-space $\mathbb R^2$. The channel $\pi_f$ is defined by
\[
(\pi_f^*(\alpha))(z) = \alpha(f(z))
\]
where $\alpha$ is an effect in $L^\infty(\mathbb R)$. 
When composed with $\Gamma^*$ it becomes the unsharp quantum observable $X_f$ with \adjoint
\[
X_f^*(\alpha) =  \Gamma^*(\pi_f^*(\alpha)) = \int \alpha(f(z))\, \proj{z} \, dz.
\]
It is clear that any classical observable can be given a quantum version without any ``ordering ambiguity''. 


Of course, calling this ``quantization'' is far from being fair, given that we only reinterpreted a known quantum theory with a known classical limit. 
However, the fact that most classical observables, even though they have no sharp quantum representation, must have an unsharp one is an important point to make. 
Also, we point out that when only the classical description of a system is known, there is a priori no reason to think that certain specific observables will have a sharp representation in the quantum theory. That is, unless one is guided by some specific experimental results involving, for instance, the quantization of the value of certain observable, or some uncertainty relations. Such results are lacking in the case of general relativity.


In addition, we note that, beside the choice of the dimension of our Hilbert space, or the type of our von Neumann algebra, the map $\Gamma$ really defines the quantum theory, given that it is what makes the link between abstract quantum observables, which are nothing more than pure mathematical objects, and concrete classical observable that we know how to measure and interpret. As we showed, specifying $\Gamma$ amounts to giving the quantization of all observables. In fact $\Gamma$ itself is the quantization of the phase-space observable, which represents all that there is to know about the classical theory, apart from the dynamics.

In the rest of this chapter we will attempt to give a realistic picture of the emergence of a classical limit defined by an observable $\Gamma$. The hope being that this provides clues as to how, given the classical theory and its environment, one may attempt to guess the nature of the map $\Gamma$ which essentially defines the quantum theory. 

\msection{Sharp decoherence}
\label{section:sharpdecoh}

The results presented in Chapter \ref{chapter:preserved} give us a thorough understanding of the ``sharp information'' preserved by a quantum channel, i.e. the information represented by sharp observables. 
We will now make use of these results 
to generalize the simple example of decoherence discussed in the introduction, in Section \ref{section:intro:decoherence}. Our analysis will be based essentially on the results presented in Section \ref{section:outinfo}.

We have seen that, if we are given a quantum channel $\chan$ which describes the evolution of an open quantum system, there is a complementary channel $\chan_c$ which describes the flow of information to the environment. The complementary channel is unique up to a unitary transformation of the environment. 
We have then shown that the sharp observables preserved by $\chan$ are precisely those which commute with all the effects (sharp and unsharp) preserved by $\chan_c$. 
It follows that the sharp observables preserved by both channels must all commute with each other and therefore form a commutative algebra $\mathcal C$. Concretely, $\mathcal C$ is the intersection of the algebras $\mathcal A$ and $\mathcal A_c$ correctable respectively for $\chan$ and $\chan_c$: $\mathcal C := \mathcal A \cap \mathcal A_c$. Since it is commutative, it characterizes classical information. This shows that the only {\em sharp} information which has been duplicated is classical. This can be understood as a version of the no-cloning theorem, with the addition that it tells us precisely which observable of the system represents the classical information which has been duplicated, namely the observable $\Gamma$ which generates the commutative algebra $\mathcal C$. Note that $\Gamma$ is defined up to a measure-preserving bijective transformation of its spectrum, i.e. the phase-space.

One possible interpretation is the following. Suppose that $\chan$ represents the time evolution of the system of interest. The algebra $\mathcal A_c$ represents properties of the system prior to the interaction. However, the part of $\mathcal A_c$ which is also preserved in the system, namely $\mathcal C$, also represents properties that the system possesses after the interaction. Therefore $\mathcal C$ represents the information gathered by the environment and which has predictive power about the future state of the system. 
In the next section we will show that each observable in $\mathcal C$ in fact implies the existence of correlations between the system and the environment after the interaction. 






\subsection{Correlations}
\label{section:correlations}

Consider a ``duplicated'' sharp discrete observable $X$ with effects $X_i^2 = X_i \in \mathcal C$. Since this observable is preserved by both $\chan$ and $\chan_c$, there exists an observable $Y$ of the system and an observable $Z$ of the environment which are such that $X_i = \chanh(Y_i)$, $X_i = \chan_c^*(Z_i)$ for all $i$.  Remember that this means that the information about the observable $X$ of the initial state of the system is represented by the observable $Y$ of the system after the interaction, and also by the observable $Z$ of the environment after the interaction. Hence the piece of information that it represents became redundant. We will show that the observables $Y$ and $Z$ are indeed correlated, and therefore also contain information about each other.

We can ``purify'' the channel $\chan$ with the isometry $V$ such that
\begin{equation}
\label{equ:pff}
X_i = \chanh(Y_i) = V^* (Y_i \otimes \one) V
\end{equation}
and
\begin{equation}
\label{equ:pfff}
X_i = \chan_c^*(Z_i) = V^* (\one \otimes Z_i) V.
\end{equation}
where $V = U (\one \otimes \ket{\psi})$, with the unitary operator $U$ representing the joint evolution of the two interacting systems, and $\ket{\psi}$ the initial state of the environment (see Figure \ref{fig:dilation}). These expressions are just the Heisenberg picture version of the relations $\chan(\rho) = \tr_{\text env}(U (\rho \otimes \proj{\psi}) U^*)$ and $\chan_c(\rho) = \tr_{\text sys}(U (\rho \otimes \proj{\psi}) U^*)$, where $\tr_{\text env}$ is the partial trace over the environment, and $\tr_{\text sys}$ the partial trace over the system. 

Consider the first equation. If we multiply the left hand side by $X_i^\perp$ on both sides, we obtain
\[
0 = X_i^\perp V^* (Y_i \otimes \one) V X_i^\perp.
\]
Since the right hand side is a positive operator, its square root must also be zero, i.e. 
\(
0 = (\sqrt{Y_i} \otimes \one) V X_i^\perp 
\)
or, by multiplying by $(\sqrt{Y_i} \otimes \one)$ on the left, and recalling that $X_i^\perp = \one -X_i$, 
\begin{equation}
\label{equ:redund1}
(Y_i \otimes \one) V (\one - X_i) = 0.
\end{equation}
Similarly we have $X_i^\perp = (\one - X_i) = V^* ((\one - Y_i) \otimes \one) V$, which, by the same argument, implies 
\begin{equation}
\label{equ:redund2}
((\one - Y_i) \otimes \one) V X_i = 0. 
\end{equation}
Combining Equations \ref{equ:redund1} and \ref{equ:redund2}, we obtain 
\[
(Y_i \otimes \one) V = V X_i.
\]
Applying the same reasoning to Equation \ref{equ:pfff} yields
\[
(\one \otimes Z_i) V = V X_i.
\]
Combing the two yields
\[
(Y_i \otimes Z_j) V = V X_i X_j = \delta_{ij} V X_i,
\]
from which we have
\begin{equation}
V^* (Y_i \otimes Z_j) V = \delta_{ij} X_i.
\end{equation}
This last equation is the result that we were seeking: it means that $Y$ and $Z$ are entirely correlated. Indeed, for any state $\rho$ of the system, 
\[
\tr( V \rho V^* (Y_i \otimes Z_j) ) = \tr( \rho V^* (Y_i \otimes Z_j) V ) = \delta_{ij} \tr( \rho X_i)
\]
is the joint probability characterizing the joint measurement of $Y$ and $Z$ after the interaction, and, as we can see thank to the Kronecker delta, the probability that both measurements yield different outcomes is zero.  
It is straightforward to generalize this analysis to the case of continuous observables. We would then obtain that
\[
V^* (Y^*(\alpha) \otimes Z^*(\beta)) V = X^*(\alpha) X^*(\beta) = X^*(\alpha \beta).
\]
Hence, for all states $\rho$,
\[
\tr( V \rho V^* (Y^*(\alpha) \otimes Z^*(\beta)) ) = \tr( \rho X^*( \alpha \beta ) ).
\]


\subsection{Predictability and objectivity}
\label{section:predobj}

These results show that the only sharp information that the environment learns about system, and which has any predictive power with respect to the outcome of some sharp measurement on the system after the interaction is classical. This classical information can be characterized by a single observable $\Gamma$: the pointer observable (defined up to a measure-preserving bijective transformation of phase-space).  Hence, in any interaction, some classical degrees of freedom of a quantum system are uniquely selected by the requirement that their information be sent to the environment, and at the same time, be useful to predict properties of the system after the interaction. From the point of view of an experimentalist who observes the system indirectly by gathering information from the environment, the effective classical system represented by $\Gamma$ evolves deterministically. The other degrees of freedom of the system can be simply ignored in this picture. 

We note that our use of the notion of preserved, or correctable, information allows for the information represented by the pointer state to have in principle any non-trivial evolution during the process of decoherence. Indeed, the fact that the information is preserved does not mean that it does not evolve. 

This provides a possible model of the process of decoherence. The interesting aspect of this picture is that it does not require any assumption about the interaction. This makes it universal. Nevertheless this model is limited by the fact that $\Gamma$ is sharp, and therefore cannot account for most classical limits. The next sections will address this problem. 

There is another possible way of interpreting the same technical results. First, we note that the two channels that we considered need not be the exact complement of each other. Indeed, the results hold if the second channel considered only maps the system to a subsystem of the environment. Consider the channel $\mathcal F$ defined by
\[
\mathcal F^*(A) := V^* (\one \otimes A \otimes \one) V
\]
where the second and third tensor factors correspond to two subsystems of the environment. Clearly, by comparing with Equation \ref{equ:pfff}, we have 
\[
\mathcal F^*(A) = \chan_c^*(A\otimes \one)
\]
which implies that all the information preserved by $\mathcal F$ is also preserved by $\chan_c$. Hence it is still true that the sharp observables preserved by both $\mathcal F$ and $\chan$ form a commutative algebra. In fact this is true of any couple of channels from the system to two different subsystems, be it part of the future state of the system or of the environment. 

Given this fact, let us consider two channels $\mathcal F_1$ and $\mathcal F_2$ from the system to two different subsystems of the environment. The information about the system preserved by both observables is classical and characterized by a single observable $\Gamma$ of the system. Therefore, a classical limit is selected simply by the condition that it be represented {\em redundantly}\index{redundancy} in the environment. As stressed by Ollivier {\it et al.} \cite{ollivier04, ollivier05} (see also \cite{blume-kohout05}), redundancy of representation is an important aspect of classical information, because it guarantees the objectivity of this information. Indeed, it allows two different observers to indirectly obtain the same information about the system, without disturbing each other's observation in any way. This slightly different picture of the process of decoherence has the advantage of allowing for the emergence of a classical system subject to noise. Indeed, the requirement of redundancy uniquely selects a classical limit without requiring the evolution of the effective classical system to be deterministic. This will be analyzed in greater details for the more general model introduced in Section \ref{section:fulldecoh}. For now, let us just mention that this particular model reproduces the main result of \cite{ollivier04}, namely the selection of a sharp pointer observable through the requirement that it be represented redundantly in the environment. However our notion of redundancy is stronger because it does not only involves correlations at a given time, but it also requires that this information represents a property that the system possessed at an earlier time.




\subsection{Examples}

Let us show how this model reduces to the simple example mentioned in the introduction.  
We considered the channel $\chan$ defined by
\[
\chan(\rho) = \sum_i \proj{i} \rho \proj{i}
\]
where the vectors $\ket{i}$ form an orthonormal basis of a Hilbert space $\Hil$. 
The correctable algebra $\mathcal A$ for this channel is the commutant of the algebra generated by the operators $\ket{i}\braket{i}{j}\bra{j} = \delta_{ij} \proj{i}$. These operators simply generate the commutative algebras of operators diagonal in the basis $\ket{i}$. This algebra is its own commutant, hence we see that the correctable algebra is already commutative, and characterized by the discrete sharp observable $\Gamma$ with elements $\Gamma_i = \proj{i}$. 

In order to find the complementary channel, note that if $V$ is an isometry resulting from the dilation of $\chan$, then $\proj{i} = (\one \otimes \bra{\phi_i}) V$, for some basis $\ket{\phi_i}$ of the environment (Equation \ref{equ:stinekraus}). This implies that the elements of the complementary channel are 
\[
\begin{split}
F_i &= (\bra{i} \otimes \one) V = \sum_j (\bra{i} \otimes \proj{\phi_j}) V \\
& = \sum_j \ket{\phi_j} (\bra{i} \otimes \bra{\phi_j}) V =  \sum_j \ket{\phi_j}\bra{i} (\one \otimes \bra{\phi_j}) V\\
&= \sum_j \ket{\phi_j}\braket{i}{j}\bra{j} = \ket{\phi_i}\bra{i}.
\end{split}
\]
Therefore the sharp information preserved by the complementary channel is characterized by the commutant the operators $F_i^* F_j = \ket{i}\braket{\phi_i}{\phi_j}\bra{j} = \delta_{ij} \proj{i}$, which is identical to the correctable algebra for $\chan$. Hence in this case $\mathcal A = \mathcal A_c$, so that $\mathcal C = \mathcal A \cap \mathcal A_c = \mathcal A$.
This shows that the environment gathers information precisely about the observable $\Gamma$ which is preserved in the system. In this situation, the eigenstates $\ket{i}$ of $\Gamma$ are the {\em pointer states} \index{pointer states} usually referred to in the literature \cite{zurek81, zurek03}.

For a slightly more general example, consider the channel
\begin{equation}
\label{equ:ex2}
\chan(\rho) = \sum_i  U P_i \rho P_i U^* 
\end{equation}
for a unitary operator $U$ and an complete family of orthogonal projectors $P_i$. The correctable algebra is given be the commutant of the operators $P_i U^* U P_j = P_i P_j = \delta_{ij} P_i$. This algebra is composed of all matrices which are block-diagonal in terms of the subspaces defined by the projectors $P_i$. Hence
\begin{equation}
\label{equ:ex2alg}
\mathcal A \approx \bigoplus_i \mathcal M_{n_k}
\end{equation}
where $n_k = \tr(P_k)$. If any of the dimension $n_k$ is larger than $1$, this algebra contains quantum information. 

In order to compute the elements of the complementary channel, we will use a basis $\{\ket{\psi^i_j}\}_{ij}$ of the system which is an aggregate of basis of the complementary subspaces defined by the projectors $P_i$, i.e. $P_i = \sum_j \proj{\psi^i_j}$.
The elements of the complementary channel then are, for some basis $\ket{\phi_i}$ of the environment,
\begin{equation}
\label{equ:comchanT}
F_{ij} = \sum_k \ket{\phi_k}(\bra{\psi^i_j} U^*) U P_k = \ket{\phi_i}\bra{\psi^i_j}.
\end{equation}
Therefore, the algebra correctable for the complementary channel is the commutant of the operators
\[
F_{ij}^* F_{kl} = \ket{\psi^i_j}\braket{\phi_i}{\phi_k}\bra{\psi^k_l} = \delta_{ik} \ketbra{\psi^i_j}{\psi^i_l}.
\]
These operators precisely span $\mathcal A$. Therefore $\mathcal A_c = \mathcal A'$, which implies that the intersection of $\mathcal A$ and $\mathcal A_c$ is the whole center of $\mathcal A$:
\[
\mathcal C = \mathcal A \cap \mathcal A_c = \mathcal Z(\mathcal A)
\]
which is the commutative algebra of operators of the form
\begin{equation}
\label{equ:A}
A = \sum_n \lambda_i P_i
\end{equation}
for any $\lambda_i \in \mathbb C$. Hence the sharp information preserved in the system and transmitted to the environment is characterized by the discrete pointer observable $\Gamma$ with elements $\Gamma_i = P_i$. 

Note that in this case, the concept of pointer state is not adapted to a description of the situation. Instead, we have a phenomenon of \ind{environment-induced superselection} rules (\ind{einselection}) \cite{zurek82}. Here $\Gamma$ is the superselection {\it charge}\index{superselection charge}. 

We see that the unitary evolution characterized by the operator $U$ plays no role in this analysis. This is because it does not remove or add any information to the system. This demonstrates an important feature of our framework, which is that it automatically identifies the original nature of the preserved information while neglecting any reversible change in its representation.


We can also look at how the decoherence gradually sets in, when time is introduced in the previous example.
Consider the Hamiltonian
\[
H = A \otimes B
\]
where 
\(
A := \sum_{i=1}^N \lambda_i \, P_i
\)
with $\lambda_i = \frac{i}{T}$, and $B$ generates cyclic shifts in the basis $\ket{\phi_i}$, $i=0,\dots,N$ of the environment, i.e. 
\[
e^{-\imath B}\ket{\phi_i} = \ket{\phi_{(i+1 \;{\rm{mod}}\; N)}}.
\]
Note that we will be using the symbol $\imath = \sqrt{-1}$ for notational convenience. 
The eigenstates of $B$ are the discrete Fourier transforms of the basis, namely $\ket{j} = \frac{1}{\sqrt N} \sum_k e^{\imath \omega k j} \ket{\phi_k}$, where $\omega = \frac{2 \pi}{N}$. 
The unitary operator for an interval of time $t$ is
\[
\begin{split}
U_t &= e^{-\imath t H} = e^{-\imath t \sum_i \lambda_i P_m \otimes B} \\
&= \sum_{n>=0} \frac{(-\imath t)^n}{n!} \sum_m \lambda_i^n P_i \otimes B^n \\
&= \sum_i P_i \otimes \sum_{n>0} \frac{(-\imath t)^n}{n!} \lambda_i^n B^n \\
&= \sum_i P_i \otimes e^{-\imath t \lambda_i B}\\
\end{split}
\]
Note that 
\[
\begin{split}
e^{-\imath t \lambda_i B} \ket{\phi_0} &= \sum_{j} e^{-\imath t \lambda_i \omega j}\ket{j} \braket{j}{\phi_0} = \frac{1}{\sqrt N}\sum_{j} e^{-\imath t \lambda_i \omega j}\ket{j} \\
&= \frac{1}{N} \sum_{j k} e^{-\imath \omega j (t \lambda_i - k)}  \ket{\phi_k} \\
\end{split}
\]
Therefore, if the initial state of the environment is $\ket{0}$, the resulting isometry is
\[
\begin{split}
V_t &= U_t (\one \otimes \ket{\phi_0}) = \sum_i P_i \otimes e^{-\imath t \lambda_i B} (\one \otimes \ket{\phi_0})\\
&= \sum_i P_i \otimes (e^{-\imath t \lambda_i B} \ket{\phi_0}) = \frac{1}{N} \sum_{ijk} e^{-\imath \omega j (t \lambda_i - k)} P_i \otimes \ket{\phi_k}\\
\end{split}
\]
The elements of the channel $\chan_t(\rho) = \tr_E(V_t \rho V_t^*)$, describing the evolution of the system, can be chosen to be 
\[
E_k(t) = (\one \otimes \bra{k}) V_t = \frac{1}{\sqrt N} \sum_{i} e^{-\imath \omega k t \lambda_i} P_i.
\]
In particular, for $t = \frac{1}{\lambda_1} = T$, we have
\[
\begin{split}
\chan_t(\rho) &= \sum_k E_k(t) \rho E_k^*(t) = \sum_{ij} \Bigl[{\frac{1}{N}  \sum_k e^{-\imath \omega k (i-j)}}\Bigr] P_i \rho P_j\\
&= \sum_{ij} \delta_{ij} P_i \rho P_j = \sum_i P_i \rho P_i.
\end{split}
\]
Hence at time $T$ we recover the channel defined in Equation \ref{equ:ex2} with $U = \one$. 

Let us now study what happens before time $T$. At time $t < T$, the algebra of sharp observables preserved in the system is given by the commutant of the operators
\[
\begin{split}
E_k^*(t)E_{k'}(t) &= \frac{1}{N}\sum_{i} e^{-\imath \omega (k-k') t \lambda_i}  P_i\\
\end{split}
\]
which is the algebra $\mathcal A$ defined in Equation \ref{equ:ex2alg}. Indeed, every single one of these operators for $k \neq k'$ generates the algebra spanned by the projectors $P_i$, of which $\mathcal A$ is the commutant. To see this it suffices to note that for fixed $k \neq k'$, the coefficients $e^{-\imath \omega (k-k') t \lambda_i}$ are distinct for all $i$. This shows that the sharp information preserved in the system at all time is the same. However, as we will see, the unsharp preserved information gets degraded until only the sharp preserved information is left. 

In order to obtain the elements of the complementary channels, we choose as before a basis $\ket{\psi^i_j}$ of the system which is compatible with the projectors $P_i$, i.e. so that $P_i = \sum_j \proj{\psi^i_j}$. The elements of the complementary channels then are
\[
F_{ij}(t) =  (\bra{\psi^i_j} \otimes \one) V_t = \frac{1}{N} \sum_{nk} e^{-\imath \omega n (t \lambda_i - k)} \ket{\phi_k}\bra{\psi^i_j}
\]
from which we have
\[
\begin{split}
F_{ij}^*(t)F_{i'j'}(t) &= \frac{1}{N^2}\sum_{n n' k} e^{\imath \omega n (t \lambda_i - k)} e^{-\imath \omega {n'} (t \lambda_{i'} - k)} \ket{\psi^i_j} \bra{\psi^{i'}_{j'}}\\
 &=  \frac{1}{N}\sum_{n n'} e^{\imath \omega n t \lambda_i} e^{-\imath \omega {n'} t \lambda_{i'}} \frac{1}{N} \sum_k e^{- \imath \omega (n-n') k}  \ket{\psi^i_j} \bra{\psi^{i'}_{j'}}\\
 &=  \frac{1}{N}\sum_{n} e^{\imath \omega n t (\lambda_i - \lambda_{i'})}\ket{\psi^i_j} \bra{\psi^{i'}_{j'}}.\\
\end{split}
\]
We want to show that for any time $0 \le t < T$, these operators generate the whole matrix algebra.
One can check that the coefficient $\gamma_{ii'} = \sum_n e^{\imath \omega n t (\lambda_i - \lambda_{i'})}$ is zero if and only if $t(\lambda_i - \lambda_{i'})$ is an integer. Since $\lambda_i - \lambda_{i'} = \frac{t}{T} (i-i')$, this cannot happen if $t/T$ is irrational. Otherwise, if $t/T = p/q$ for some integers $p$ and $q < N$, then $\gamma_{ii'} = 0$ when $i - i' = q$. If this happens, we are missing the matrix elements of the form $\ket{\psi^i_j} \bra{\psi^{i-q}_{j'}}$ in the span of the operators $F_{ij}^*(t)F_{i'j'}(t)$ for all $i,i',j,j'$. However, this element can be recovered by the product $\ket{\psi^i_j} \bra{\psi^{i-1}_j} \cdot \ket{\psi^{i-1}_j} \bra{\psi^{i-2}_{j}} \cdot \dots \cdot \ket{\psi^{i-q+1}_j} \bra{\psi^{i-q}_{j'}}$. From the previous argument, each element in this sum is always present in our span, unless $t/T$ is an integer. Therefore, when $t < T$, there is no non-trivial sharp observable commuting with all the operators of the form $F_{ij}^*(t)F_{i'j'}(t)$, and hence no non-trivial sharp observable flowing to the environment. 

However there are non-trivial unsharp observables flowing to the environment. In fact, we can see how one of these preserved unsharp observable becomes sharper with time until it equals one of the projectors $P_i$. Indeed, let us find an observable of the environment that one should measure at time $T$ in order to reproduce the statistics of the observable with elements $P_i$. At time $T$ the elements of the complementary channel are given by Equation \ref{equ:comchanT}. Using these, one can check that the sharp observable with elements $Q_k := \proj{\phi_k}$ does the job. Indeed,
\[
\chan_c^*(Q_k) = \sum_{ij} \ket{\psi^i_j}\braket{\phi_i}{\phi_k} \braket{\phi_k}{\phi_i}\bra{\psi^i_j} = \sum_j \ket{\psi^k_j}\bra{\psi^k_j} = P_k.
\]
Measuring the same observable of the environment at an earlier time $t$ yields information about the observable of the system with elements
\[
\begin{split}
A_m &= (\chan_{t})_c^*(Q_m)= (\chan_{t})_c^*(\proj{\phi_m}) \\
&= \frac{1}{N^2} \sum_{ijnkn'k'}  e^{\imath \omega n' (t \lambda_i - k')} e^{-\imath \omega n (t \lambda_i - k)}  \ket{\psi^i_j}\braket{\phi_{k'}}{\phi_m} \braket{\phi_m}{\phi_k}\bra{\psi^i_j}\\
 &= \frac{1}{N^2} \sum_i | \sum_n e^{-\imath \omega n (t \lambda_i - m)}|^2  \sum_j\ket{\psi^i_j}\bra{\psi^i_j}\\
 &= \sum_i \gamma_{im} P_i\\
\end{split}
\]
where we have introduced
\[
\begin{split}
\gamma_{im} &:= \frac{1}{N^2} | \sum_n e^{-\imath \omega n (t \lambda_i - m)}|^2\\
\end{split}
\] 
As $t \rightarrow T$, $\gamma_{im}$ converges to $\delta_{im}$, which implies that $A_m$ converges to $P_m$. 

\msection{Classical set of observables}
\label{section:classicalset}

The picture presented in the previous section is fundamentally tied to sharp observables, and cannot describe the emergence of a classical limit characterized by an unsharp observable. In particular this means that it cannot describe the emergence of a non-trivial phase-space. In Section \ref{section:fulldecoh} we will see that this picture can in fact be generalized to unsharp pointer observables. However it will be instructive to first consider a simpler generalization. For the moment we will ignore the environment, and attempt to understand what it means for a channel to destroy any information but that represented by an unsharp observable. 

In the previous section, 
we used the fact that the set of preserved sharp observables form a sub-algebra, and that, when this algebra is commutative, it represents an effective classical system. This idea does not work with unsharp observables, which are not associated with any algebra. Therefore we need to generalize what we mean by a classical set of observables. 



We want to understand when the set of observables preserved by a channel, or even any set of observables, can be said to characterize classical information. We have already defined classical information as that represented by a physical system whose set of effects belong to a commutative algebra.
Hence, it is a priori natural to think of a classical set of quantum observables as that which is characterized by a commutative sub-algebra of the quantum algebra. However, we have seen in Section \ref{section:limit}, that there exists much more general ways for a classical system to be represented by a quantum system. In general, the representation can be given by a channel mapping quantum states into classical states, or equivalently by its dual which translates classical effects into quantum effects. This map, which is simply an observable, does not in general preserve the full structure of the commutative algebra. Note that when it does, i.e. when it represents a sharp observable, it singles out a commutative sub-algebra of the quantum algebra. If however it is not a homomorphism, then it may map sharp observables into unsharp ones, and commutative pairs of observables into non-commutative ones, which implies a certain loss of information. However this loss of information allows for fundamentally different quantum representations of a classical system, as seen in Section \ref{section:limit}.

Consider an observable 
\[
\Gamma: \mathcal B_t(\Hil) \rightarrow L^1(\Omega)
\]
which defines how our quantum system with Hilbert space $\Hil$ imperfectly represents the classical system with phase-space $\Omega$. This maps also represents every observables of the classical system by an observable of the quantum system. Indeed, consider a classical observable
\[
\pi:  L^1(\Omega) \rightarrow L^1(\Omega')
\]
which is also simply a stochastic map between the two classical systems $\Omega$ and $\Omega'$.
This observable is represented by the quantum observable
\[
X = \pi \circ \Gamma.
\]
Measuring $X$ amounts to first measuring $\Gamma$, and then measuring $\pi$ on the classical system $\Omega$ representing the outcome of the quantum measurement. 

Therefore, to the classical limit $\Gamma$ is associated the set of observable of the form $X = \pi \circ \Gamma$, for all classical observables $\pi$. Note that this is simply the set $\preserved \Gamma$ of observables {\it preserved} by $\Gamma$. 
\begin{diagram}
  {\mathcal B_t(\Hil)} &\rTo^{\quad \Gamma \quad}  & {L^1(\Omega)}\\
  & \rdTo_{X} & \dTo_{\pi} \\
  & & L^1(\Omega')
\end{diagram}

We will also say that these observables are {\it \ind{coarse-grainings}} of $\Gamma$ since they amount to measuring $\Gamma$ and then forgetting about some aspects of the classical result by applying the stochastic map $\pi$. In this sense, each of these observables represent less information than $\Gamma$.

We will take this to be the prototype of a classical set of observable. In fact we will say that a set of observable is {\it classical}\index{classical set of observable} if it belongs to such a set, i.e. if they are all coarse-grainings of a single observable $\Gamma$. 

\begin{dfn}
\label{def:classical}
We say of a set $\mathcal O$ of observables that it is {\em classical} if there is an observable $\Gamma$ and a family of stochastic maps $\pi_X$, $X \in \mathcal O$ such that
\[
X = \pi_X \circ \Gamma \quad \text{for all $X \in \mathcal O$}.
\]
This is equivalent to saying that the observables $X \in \mathcal O$ are all preserved by $\Gamma$, i.e. 
\[
\mathcal O \subseteq \preserved \Gamma.
\]
\end{dfn}
\index{$\preserved \Gamma$}

Note that it would be tempting to characterize this set of observables by a set of effects, i.e. the set of coarse-grainings of $\Gamma$ which take value in the set $\Omega' = \{0,1\}$. But the counter-example used in Section \ref{section:preservedeff} shows that this is not possible. 

As an example, consider the set of observables defined by a commutative sub-algebra $\mathcal A \subseteq B_t(\Hil)$. In this case $\Gamma$ can be chosen to be any sharp observable whose spectral projectors generate the commutative sub-algebra $\mathcal A$. More concisely, $\Gamma$ can be represented by a self-adjoint operator $\widehat \Gamma$, whose bounded functions generate $\mathcal A$. 
If $\Gamma$ takes value in $\Omega_\Gamma$, i.e. $\Gamma:  \mathcal A_* \rightarrow L^1(\Omega_\Gamma)$, then the \adjoint map $\Gamma^*$ is a homomorphism of $L^\infty(\Omega_\Gamma)$ onto $\mathcal A$. This means than a sharp observable $X$ represented by a self-adjoint operator $\widehat X \in \mathcal A$ is a coarse-graining of $\Gamma$ in the following way. We know that there is a real function $f$ such that $\widehat X = f(\widehat \Gamma)$. The function $f$ is a mapping from the spectrum $\Omega_\Gamma$ of $\Gamma$ to that of $X$. This means that the effects $X^*(\chi_\omega)$ of $X$ satisfy $X^*(\alpha) = \Gamma^*(\alpha \circ f)$. We then define the stochastic map $\pi$ by $\pi^*(\alpha)(x) := \alpha(f(x))$ for all $x \in \Omega_\Gamma$, 
so that we have $X^*(\alpha) = \Gamma^*(\alpha \circ f) = \Gamma^*(\pi^*(\alpha))$ for all $\alpha$, i.e. $X = \pi \circ \Gamma$.



For instance, if $\Gamma$ is sharp and {\em discrete} with elements $\Gamma_i^2 = \Gamma_i$, then all the operators in the algebra $\Gamma$ that it generates are represented by self-adjoint operators of the form $\widehat X = \sum_{i \in \Omega_\Gamma} \lambda_i \Gamma_i$. It is then clear that measuring $\Gamma$ is as good as measuring the observable represented by $\widehat X$. All that one has to do to simulate a measurement of $\widehat X$ is to forget the distinction between the outcomes $i$ and $j$ in the case that $\lambda_i = \lambda_j$, an operation which can be represented by the stochastic matrix $\pi$ defined as follow. Let $\Omega$ be the set of distinct values taken by the coefficients $\lambda_i$ for all $i$, which is the spectrum of $\widehat X$.
The stochastic matrix $\pi$ is defined by $\pi_{i\lambda_i} = 1$ for all $i \in \Omega_\Gamma$ and $0$ for all other components.




\subsection{Non-contextuality}

Our definition of a classical set of observable is further justified by that fact that they represent an effective theory which can be simulated by a \ind{non-contextual hidden variable} model, in the sense defined in \cite{spekkens05}. 

Suppose that we have a physical system defined by a set of observables $\mathcal O$ and a set of states $\astates$. A non-contextual hidden variable model associates to each observable $X \in \mathcal O$ a classical observable $\pi_X$, and, to each state $\rho \in \astates$, a classical states $\mu_\rho$, in such a way that, when combined, these classical states and observables yield the same probability distributions as the quantum states and observables they represent. 

The idea is that the model would be contextual\index{contextuality} with respect to preparations if the classical state was a function not only of $\rho$ but also of the observable $X$, which would mean that the particular state we must choose depends on the particular experiment that we are going to perform. Correspondingly, it would be contextual with respect to measurements if the classical observable to choose depended on the quantum state.

It is known that quantum theory cannot be given a non-contextual model.  However, a classical set of quantum observables can. Indeed, this is precisely what the observable $\Gamma$ which characterizes our classical set does. Indeed, it maps any quantum state $\rho$ to a probability distribution $\mu_\rho = \Gamma(\rho) \in L^1(\Omega)$. In addition, each observable $X$ of the classical set is associated with a classical observable $\pi_X$ which is such that $X = \pi_X \circ \Gamma$. This relation implies that the probability distribution $X(\rho)$ is equal to one given by the effective classical theory, i.e. $\pi_X(\mu_\rho)$. Indeed, 
\[
\pi_X(\mu_\rho) = \pi_X(\Gamma(\rho)) = X(\rho).
\] 


\subsection{Functional coexistence}


There exists a concept of {\it coexistence} \index{coexistent observables} between POVMs \cite{lahti01} which generalizes the notion of commutativity for PVMs. 

Remember that an observable $X$ defines a map $X^*$ from classical effects to quantum effects. When this map is restricted to sharp classical effects, which are all of the form $\chi_\omega$ for some subset $\omega \subseteq \Omega$ of the phase-space $\Omega$, then it defines a POVM, which is a map from subsets $\omega$ to quantum effects. The effects of the form $X^*(\chi_\omega)$ can therefore be said to be in the range of the POVM. A set of POVMs is called {\em coexistent} \index{coexistent observables} if the union of their range are all within the range of a single POVM. 
To show that the observables in our ``classical'' set are coexistent, we first need to prove a Lemma. 
We say that an observable $X$ is a {\em \ind{marginal}} of 
\[
Y: \mathcal B_t(\Hil) \rightarrow L^1(\Omega_1)\otimes L^1(\Omega_2) = L^1(\Omega_1 \times \Omega_2)
\]
if $X^*(\alpha) = Y^*(\alpha \otimes \lone)$, which means that measuring $X$ amounts to measuring $Y$ and then discarding the $\Omega_2$ component of the result.
The following then is true:
\begin{lem}
Observables in a classical set are all marginals of a single observable. 
\end{lem}
\proof Suppose that we have a set of observables $\mathcal O$, which are all coarse-grainings of $\Gamma: \mathcal B(\Hil) \rightarrow L^1(\Omega)$. For clarity, we will label the observables in $\mathcal O$ by a set $\Sigma$, i.e. $\mathcal O = \{X_\sigma \;|\; \sigma \in \Sigma\}$. For all $\sigma \in \Sigma$ there exists a channel $\pi_\sigma: L^1(\Omega) \rightarrow L^1(\Omega_\sigma)$, where $\Omega_\sigma$ is the set of values of $X_\sigma$, such that 
\(
X_\sigma = \pi_\sigma \circ \Gamma.
\)
Consider the channel
\[
{\rm copy}: L^1(\Omega) \rightarrow \bigotimes_{\Sigma} L^1(\Omega) = L^1(\prod_\Sigma \Omega) 
\]
defined by ${\rm copy}(\mu) = \bigotimes_{\Sigma} \mu$. This channel produces as many copies of the state of the classical system $\Omega$ as there are elements in the set $\Sigma$. 
We use this to define the new observable
\[
\widetilde \Gamma :=  \bigl({ \bigotimes_{\sigma \in \Sigma} \pi_\sigma }\bigr) \circ {\rm copy} \circ \Gamma
\]
which amounts to measuring $\Gamma$, then duplicating the output as many times as there are elements in $\Sigma$, and finally applying $\pi_\sigma$ to the $\sigma$'s copy.
It is then clear that each observable $X_\sigma$ is a marginal of $\widetilde \Gamma$, namely it amounts to measuring $\widetilde \Gamma$ and then discarding all but the $\sigma$'s component of the outcome. \qed

This almost directly shows that all coarse-grainings of a single observable $\Gamma$ are coexistent. Indeed, using the same objects as in the above proof, we have $X_\sigma^*(\chi_\omega) = \widetilde \Gamma^*(\chi_{\widetilde \omega})$, where $\widetilde \omega$ contains all the elements of $\prod_{\Sigma} \Omega$ which have their $\sigma$'s component inside $\omega$.  

In fact, this shows that the coarse-grainings of  $\Gamma$ are more than just coexistent. They are also {\em functionally coexistent}\index{functional coexistence}, which means that $\widetilde \omega$ is related to $\omega$ by a function $f: \prod_\Sigma \Omega \rightarrow \Omega$, i.e. $\widetilde \omega = f^{-1}(\omega)$. Here the function is $f(x) = x_\sigma$, and $f^{-1}(\omega)$ denotes the pre-image of $\omega$ under $f$.  In fact, it is also clearly true that functionally coexistent observables form a classical set. Therefore we have the following:
\begin{prop}
Observables form a classical set if and only if they are functionally coexistent.  
\end{prop}

\msection{Decoherent channels}
\label{section:sysdecoh}


Equipped with our definition of a classical set of observables, we can now define a {\it fully decoherent} channel $\chan$ as one whose preserved set of observables $\preserved \chan$ is classical, i.e. 
\[
\preserved \chan \subseteq \preserved \Gamma
\]
for some observable $\Gamma$. The observable $\Gamma$ then is the {\em pointer observable} characterizing the decoherence process. 
\begin{dfn}
A channel $\chan$ is said to be {\em fully decoherent} in terms of the pointer observable $\Gamma$ if all its preserved observables are coarse-grainings of $\Gamma$. 
\end{dfn}
It may be convenient to visualize this definition by combining the commutative diagrams which state that $X$ is preserved by both $\chan$ and $\Gamma$: 
\begin{diagram}
  {\mathcal B_t(\Hil_1)} & \rTo^{ \quad  \chan \quad}  & {\mathcal B_t(\Hil_2)}\\
  \dTo^{ \Gamma} & \rdTo_{X} & \dTo^{Y} \\
   {L^1(\Omega)} & \rTo_{\pi}  & L^1(\Omega')
\end{diagram}
This can be read as follows. Imagine that the channel $\chan$ and the classical limit $\Gamma$ are given, and disregard the diagonal $X$.  The condition for $\Gamma$ to be a pointer observable selected by $\chan$ is that for every observable $Y$, $\pi$ exists to complete the square.

In general, if $\preserved \chan \neq \preserved \Gamma$, the pointer observable $\Gamma$ is not unique. For instance, it could be that the channel preserves no information at all, which would mean that the preserved observables are only those whose effects are proportional to the identity. In this case, absolutely any observable could serve as the pointer observable, according to the above definition. The idea is that in this case any observable would yield a consistent classical interpretation of this information, albeit admitting only maximally mixed quantum states. 
In general, the description based on a pointer observable $\Gamma$ contains more information than what has actually been preserved by the channel. Therefore, the arbitrariness in the choice of the pointer observable corresponds to the arbitrariness of the added information in the classical description. In fact we may expect that, in realistic systems, the differences between various choices of the pointer observable vanish when the accuracy with which the observer can resolve the classical observables is small compared to $\hbar$.

\subsection{Example: entanglement-breaking channels}
\label{section:exebc2}

As an example, consider all channels with rank-one elements, i.e. of the form 
\[
\chan(\rho) = \sum_i \lambda_i \ketbra{\psi_i}{\phi_i} \rho \ketbra{\phi_i}{\psi_i}
\]
where the states $\ket{\psi_i}$ and $\ket{\phi_i}$ are arbitrary, apart from the condition that $\chanh(\one) = \one$, which implies that
\[
\sum_i  \lambda_i \proj{\phi_i} = \one.
\]
This is the general form of the \ind{entanglement-breaking channel}s in finite-dimensions \cite{horodecki03}.
Channels of this form are fully decoherent, with discrete pointer observable $\Gamma$ defined by its elements 
\[
\Gamma_i = \lambda_i \proj{\phi_i}.
\]
Indeed, an observable $X$ preserved by $\chan$ has the form
\[
X^*(\alpha) = \sum_i \lambda_i \ketbra{\phi_i}{\psi_i} Y^*(\alpha) \ketbra{\psi_i}{\phi_i}
\]
for some observable $Y$. Therefore,
\[
\begin{split}
X^*(\alpha) &= \sum_i \bra{\psi_i} Y^*(\alpha) \ket{\psi_i}\, \Gamma_i = \Gamma^*(\pi^*(\alpha))\\
\end{split}
\]
where the classical channel $\pi$ is defined by
\[
\pi^*(\alpha)_i = \bra{\psi_i} Y^*(\alpha) \ket{\psi_i}.
\]
To see that this map $\pi$ is a valid channel, note first that it is manifestly positive and unital. In addition, it is the \adjoint of the map
\[
\pi(\mu)_i = Y(\sum_i \mu_i \proj{\psi_i}) 
\]
on states, and is therefore normal. Note that such channels were already studied in Section \ref{section:exmeas}. In particular, the sets of effects represented in Figure \ref{fig:diamonds} are examples of classical sets.

If the sum is replaced by an integral over a continuous set of states, the resulting channel has the same properties. An example of such a channel is given in Section \ref{section:cohdecoh} below.


\subsection{Covariant decoherent channels}
\label{section:covariance}

The above considerations about the freedom in the choice of the pointer observable $\Gamma$ can be made more concrete in the case where the channel $\chan$ is {\em \ind{covariant}} with respect to a unitary representation $U: g \in G \rightarrow U_g$ of a compact \ind{Lie group} $G$.
The channel $\chan$ is said to be covariant \index{covariant channel} with respect to this group if there exists another unitary representation $V: g \rightarrow V_g$ which is such that, for all $g \in G$,
\[
\chan(U_g \rho U_g^*) = V_g \chan(\rho) V_g^*.
\]

Now suppose that this channel is fully decoherent, in the above sense, in terms of some pointer observable 
\[
\Gamma_0: \mathcal B_t(\Hil) \rightarrow L^1(\Omega)
\]
This means that for all observables $X$ there exists a classical observable $\pi_X$ such that for all effects $\alpha$,
\[
\chanh(X^*(\alpha)) = \Gamma_0^*(\pi_X^*(\alpha)) 
\]
but then we also have, for all $g \in G$,
\begin{equation}
\label{equ:covindobs}
\begin{split}
\chanh(X^*(\alpha)) &= \chanh(V^*_g V_g X^*(\alpha) V^*_g V_g) = U^*_g \chanh( V_g X^*(\alpha) V^*_g) U_g\\
& = U_g^* \Gamma^*(\pi_{gX}^*(\alpha)) U_g
\end{split}
\end{equation}
where, by $gX$ we mean the observable defined by
\[
(gX)^*(\alpha) := V_g X^*(\alpha) V_g^*
\]
Equation \ref{equ:covindobs} means that the preserved observables are not only coarse-grainings of $\Gamma$, but are also coarse-grainings of any observable of the form 
\[
\Gamma_g^*(\alpha) := U^*_g \Gamma^*_0(\alpha) U_g.
\] 
This shows an example of ambiguity in the choice of the pointer observable. However, it may be that the pointer observable $\Gamma$ is such that $\Gamma_g$ is equivalent to $\Gamma$ up to a relabelling of its values $\Omega$. This would mean that $\Gamma$ itself is covariant with respect to $G$ in the sense that there exists an action $\lambda$ of $G$ on $\Omega$ which is such that $\Gamma_g(\alpha) = \Gamma_0(\alpha \circ \lambda_g)$.
In fact we can always build a pointer observable $\overline \Gamma$ which is covariant in this way. It is, in a sense, and average of each $\Gamma_g$ over the Haar measure. More precisely, we define $\overline \Gamma$ by
\[
\overline \Gamma^*(\alpha) := \int \Gamma^*_g(\alpha_g) \, dg = \int U^*_g \Gamma^*_0(\alpha_g) U_g \, dg.
\]
where $\alpha \in L^\infty(G)\otimes L^\infty(\Omega)$, $\alpha_g(x) := \alpha(g,x)$ for all $x \in \Omega$. The invariant Haar measure $dg$ on $G$ is normalized such that $\int dg = 1$.
This observable is covariant for the group $G$ with respect to the action $\lambda$ on $G \times \Omega$ defined by the right-action
\[
\lambda_h(g,x) := (gh^{-1}, x) 
\]
for all $h \in G$ and $x \in \Omega$. Indeed, we have
\[
\begin{split}
U^*_h \overline \Gamma^*(\alpha) U_h &= \int U^*_{gh} \Gamma^*_0(\alpha_g) U_{gh} \, dg = \int U^*_{g} \Gamma^*_0(\alpha_{gh^{-1}}) U_{g} \, dg\\
 &= \int U^*_{g} \Gamma^*_0(\alpha_{g} \circ \lambda_h ) U_{g} \, dg = \overline \Gamma(\alpha \circ \lambda_h).\\
\end{split}
\]
We still have to show that, indeed, all observables preserved by $\chan$ are coarse-grainings of $\overline \Gamma$, so that it can serve as a pointer observable. Given an observable $X$ taking value in the set $\Omega_X$,
\[
\begin{split}
\chanh(X^*(\alpha)) &= \int \chanh(X^*(\alpha)) \, dg\\
 &= \int U^*_g \Gamma^*(\pi_{gX}^*(\alpha)) U_g\, dg\\
 &= \overline \Gamma^*(\Pi_X^*(\alpha)) \\
\end{split}
\]
where we have defined the classical observable 
\[
\Pi_X:  L^1(G)\otimes L^1(\Omega) \rightarrow  L^1(\Omega_X)
\]
by 
\[
\Pi_X^*(\alpha)(g,x):= \pi_{gX}^*(\alpha)(x). 
\]



\subsection{Example: coherent states}
\label{section:cohdecoh}

Let us give an example of a channel decoherent in terms of an unsharp pointer observable.  We will use the setting of a Hilbert space $\Hil$ which contains coherent states $\ket{z} =\ket{q,p}$ generated from a fiducial vector $\ket{\psi_0}$, as in Section \ref{section:quantization}. We consider the channel $\chan$ defined by
\begin{equation}
\label{equ:cohchannel}
\chan(\rho) = \int \proj{z} \rho \proj{z} dz
\end{equation}
or
\[
\chan^*(A) = \int \proj{z} A \proj{z} dz.
\]
This channel is of the form of the class of examples studied in Section \ref{section:exmeas}, apart from the fact that the sum has been replaced by an integral. An observable $X$ which is preserved by this channel is of the form
\[
X^*(\alpha) = \int \proj{z} Y^*(\alpha) \proj{z} dz.
\]
for some observable $Y$. It is immediately clear that this observable is a coarse-graining of the coherent-state observable $\Gamma$ studied in Section \ref{section:quantization}. Indeed, remember that
\[
\Gamma^*(\alpha) = \int \alpha(z) \proj{z} dz. 
\]
Hence, we have
\[
\begin{split}
X^*(\alpha) &= \int \bra{z} Y^*(\alpha) \ket{z} \proj{z} dz = \int (\pi^*(\alpha))(z)\, \proj{z}  dz\\
&= \Gamma^*(\pi^*(\alpha)).
\end{split}
\]
where we have defined the stochastic map $\pi$ by
\[
\pi^*(\alpha)(z) = \bra{z} Y^*(\alpha) \ket{z}.
\]
Note that $\pi^*$ is positive and unital. In addition it is normal, because it is the \adjoint of
\[
\pi(\mu) = Y(\int \mu(z) \proj{z} dz) 
\]
which is well-defined given that $\rho := \int \mu(z) \proj{z} dz$ is trace-class. Indeed, for a basis $\ket{i}$, we have $\sum_i \bra{i} \rho \ket{i} = \int \mu(z)\, dz = 1$. 

Therefore we have in this example that $\preserved \chan \subseteq \preserved \Gamma$. However, we do not have equality between these sets. Indeed, $\Gamma$ itself is not preserved by the channel. Instead, the closest we can get is an observable $\widetilde \Gamma$ of the form
\(
\widetilde \Gamma := \Gamma \circ \chan
\)
which, more explicitly, has the form
\[
\begin{split}
\widetilde \Gamma^*(\alpha) &= \chanh(\Gamma^*(\alpha)) =  \int \alpha(z)\,\ket{z} \braket{z}{z'}\braket{z'}{z} \bra{z} \,dz \,dz'\\
&= \int \alpha(z)\,|\braket{z}{z'}|^2 \proj{z} \,dz \,dz'.
\end{split}
\]
This observable can be understood as an approximate form of the observable $\Gamma$. For instance, if the fiducial state $\ket{\psi_0}$ that we used to define the coherent states is a Gaussian then so is $|\braket{z}{z'}|^2$, which shows that if $\alpha(z)$ is peaked around $z$, then $\widetilde \Gamma^*(\alpha)$ is given by a Gaussian smearing of coherent states around $z$, instead of the coherent state $\proj{z}$ itself. 
However, $\widetilde \Gamma$ itself is not sufficient to characterize $\preserved \chan$. Indeed, $\preserved {\widetilde \Gamma}$ does not contain the whole of $\preserved \chan$. 

Note that this channel $\chan$ is clearly covariant with respect to the group of translations and boosts in phase-space, which was used to generate the coherent states. However, we cannot use this fact to generate other pointer states from $\Gamma$, given that $\Gamma$ is already covariant with respect to this group.

In conclusion, the channel defined by Equation \ref{equ:cohchannel} yields an example of how a realistic classical limit can emerge dynamically from decoherence. We note that the emergence of phase-space through decoherence has been studied before \cite{zurek92, zurek03}, however this is the first model which allows in principle for the possibility of a derivation of the structure of the emergent phase-space. Indeed, all other approaches attempt to show that a given classical description is consistent, without excluding other possible classical descriptions. 
Although our description may also have a certain ambiguity in the choice of the pointer observable $\Gamma$, the main properties of the classical description should come from the features encoded in the well-defined set $\preserved \chan$. 

Another advantage of this picture is that it enables us to study the transfer of information to the environment, as we will show in the next section.  





\msection{Decoherence from broadcasting}
\label{section:fulldecoh}

In Section \ref{section:outinfo}, we showed that sharp preserved observables which are duplicated twice must form a classical set. We exploited this fact in Section \ref{section:sharpdecoh} in order to provide a model of decoherence in terms of a sharp pointer observable. In the previous section we familiarized ourselves with the idea of an unsharp pointer observable. We will now try to generalize the result of Section \ref{section:sharpdecoh} to unsharp observables in order to have a full-fledged picture of decoherence. However, we will see that in the case of unsharp observables, it is not sufficient to duplicate them twice in order to make sure that they form a classical set. A counter example will be provided in Section \ref{section:nonclasscloning}. Instead, we will show that the unsharp observables which are duplicated an arbitrary number of times must form a classical set. Our proof however will assume that we are dealing with a finite-dimensional quantum system.

We consider a channel $\chan$ from the system with Hilbert space $\Hil_A$ to an infinite number of subsystems $\Hil_{B_1},\Hil_{B_2},\Hil_{B_3},\dots$ of the environment. 
\begin{equation}
\label{equ:broadchan}
\chan: \mathcal B_t(\Hil_A) \rightarrow \mathcal B_t(\Hil_{B_1} \otimes \Hil_{B_2} \otimes \dots)
\end{equation}
If needed, the infinite tensor product Hilbert space can be made separable by keeping only basis elements with a finite number of factors different from a special state selected in each $\Hil_{B_i}$ \cite{neumann39}. 
The flow of information from $A$ to the $i$th system $B_i$ is described by the channel
\[
\chan_i:  \mathcal B_t(\Hil_A) \rightarrow \mathcal B_t(\Hil_{B_i})
\]
defined by
\begin{equation}
\label{equ:partialchans}
\chan_i(\rho) = \tr_{i}'\chan(\rho)
\end{equation}
where $\tr_i'$ is the partial trace over all subsystems but the $i$th. 
We want to characterize the information which is preserved by all these channels at the same time. 
This information is encoded in the set of observables which are preserved by all the channels: 
$$
\mathcal I = \bigcap_i \preserved {\chan_i}
$$ 
We will say that the observables in this set are {\em broadcast}\index{broadcast observables}.
We can show that any countable set $\{X_i\} \subset \mathcal I$ is classical, in the sense that all the observables in it are coarse-grainings of a single observable (see Section \ref{section:classicalset}). Indeed, let $Y_i$ be the observable such that $X_i = Y_i \circ \chan_i$, which exists thank to the assumption that $X_i$ is preserved by $\chan_i$. Then consider the observable
$$
\Gamma := (Y_1 \otimes Y_2 \otimes \dots) \circ \chan.
$$
It is clear that all the observables $X_i$ are marginals of $\Gamma$. Indeed, measuring $\Gamma$ and then discarding all classical output subsystems but the $i$th is equivalent to measuring  $Y_i \circ \chan_i = X_i$. This proves that $\{X_i\} \subset \preserved {\Gamma}$. Note that any coarse-graining or convex combination of these observables $X_i$ is also in $\preserved {\Gamma}$. 

If the set $\mathcal I$ was separable with respect to a certain topology $\mathbb T $, then we would have that $\{X_i\}$ can be chosen to be dense in $\mathcal I$, so that 
\[
\mathcal I \subseteq \overline {\preserved {\Gamma}} 
\]
where the horizontal bar denotes closure with respect to the topology $\mathbb T$. 
This would show that a single observable $\Gamma$ suffices to simulate the broadcast observables in $\mathcal I$ to arbitrary precision (with respect to $\mathbb T $), and therefore that the set $\mathcal I$ is classical, and characterized by the pointer observable $\Gamma$, in this somewhat more general sense.

The set $\mathcal I$ can be made separable under the following assumptions
\begin{enumerate}
\item It includes only observable taking value in a single set $\Omega$. For instance, one could pick the disjoint union
\(
\Omega = \mathbb R \cup \mathbb N
\)
in order to include both continuous and discrete observables.
\item The source Hilbert space $\Hil_A$ (on which the observables in $\mathcal I$ are defined), is finite-dimensional.
\end{enumerate}
The finite dimensionality of $\Hil_A$ in fact guarantees that the whole set of observables of the form $X: \mathcal B_t(\Hil_A) \rightarrow L^1(\Omega)$ is separable. Indeed, consider a basis $\sigma_i$ of $\mathcal B_t(\Hil)$. Each observable $X$ is entirely characterized by its components $X(\sigma_i) \in L^1(\Omega)$. Given that there is a finite number of them, and that $L^1(\Omega)$ itself is separable, the set of all these observables is also separable. The topology $\mathbb T$ which is involved here is defined by the metric
\begin{equation}
\label{equ:distance}
d(X,Y) = \max_i \|X(\sigma_i) - Y(\sigma_i)\|_1 
\end{equation}
where $\|\cdot\|_1$ is the norm in $L^1(\Omega)$.


This proves the following:
\begin{thm}
\label{thm:fundamental}
Let $\chan$ be a channel of the form specified by Equation \ref{equ:broadchan}, with the source Hilbert space $\Hil_A$ finite-dimensional. Then the set $\mathcal I_\Omega$ of observables with values in a given measure space $\Omega$, which are preserved by all the partial channels $\chan_i$, can be approximated arbitrarily well by coarse-grainings of a single observable $\Gamma$, in terms of the distance defined by Equation \ref{equ:distance}.
\end{thm}

We will look at the consequences of this result in Section \ref{section:discussion}. But first, let us examine the question of the correlations induced by unsharp preserved observables. 









\subsection{Correlations}


In Section \ref{section:correlations} we showed that a sharp observable which is duplicated twice implies the existence of exact correlations between the two destination systems. For unsharp observables, the correlation will in general not be exact. However, there are cases where the correlation is exact even for genuinely unsharp observables. To show this, let us give a slightly more general form of the result derived in Section \ref{section:correlations}.


Consider a discrete observable $X$ with effects $X_i$. Suppose that for each $i$, $P_i$ is a projector satisfying 
\[
P_i^\perp X_i P_i^\perp = 0.
\]
We will assume that this observable $X$ is preserved by two channels $\chan_1$ and $\chan_2$ which are given by two different partial traces of the same channel $\chan$. For instance they could be any pairs of channel $\chan_i$ defined in Equation \ref{equ:partialchans}. Hence, there exists observables $Y$ and $Z$ which are such that $X_i = \chanh_1(Y_i)$, $X_i = \chanh_2(Z_i)$ for all $i$.  

We have
\begin{equation}
\label{equ:pff2}
X_i = \chanh (Y_i \otimes \one)  = \chanh (\one \otimes Z_i)
\end{equation}
If we multiply by $P_i^\perp$ on both sides in each term, we obtain
\[
0 = P_i^\perp \chanh (Y_i \otimes \one)  P_i^\perp =  P_i^\perp \chanh (\one \otimes Z_i)  P_i^\perp.
\]
If $E_k$ are the elements of the channel $\chan$, then using Lemma \ref{lemma:trick} we get
\[
(Y_i \otimes \one) E_k = (\one \otimes Z_i) E_k = E_k P_i
\]
Together, these two equations yield
\[
(Y_i \otimes Z_j) E_k = (Y_i \otimes Z_j) E_k (P_i P_j)^n
\]
where $n$ is an arbitrary integer. In the limit for $n \rightarrow \infty$, we get a new projector $P_{ij}$ on the right-hand side which projects on the intersection of the subspaces corresponding to $P_i$ and $P_{j}$. Hence, if our observable $X$ is such that $P_i$ and $P_j$ project on subspaces with trivial intersections, then
\[
(Y_i \otimes Z_j) E_k = (Y_i \otimes Z_j) E_k \delta_{ij} P_i
\]
which implies that whenever $i \neq j$,
\[
\chanh (Y_i \otimes Z_j) = 0.
\]
Hence $Y$ and $Z$ are fully correlated.
Note that we had to assume that the supports of the effects $X_i$ all have trivial intersections. This happens for instance if $X_i \propto \proj{\psi_i}$ and the states $\ket{\psi_i}$ are all distinct (but not necessarily orthogonal).

More generally, we do not expect that a duplicated unsharp observable will always induce exact correlations between the target systems. However, we expect that the amount of correlations between $Y$ and $Z$ is related to the information content of the observable $X$, as defined in Section \ref{section:obsinfo}. 

\subsection{Non-classical two-fold broadcasting}
\label{section:nonclasscloning}

We showed that sharp observables must form a classical set when duplicated twice. For unsharp observables however, we showed that they form a classical set under the assumption of an infinite quantity of copies (Theorem \ref{thm:fundamental}). Can we arrive at the same conclusion under a weaker assumption, in the unsharp case? Here we give an example which shows that, unlike for the sharp case, two copies are not sufficient\footnote{This example was suggested to the author by A. Winter.}. 
 
Consider the isometry $V$ which embeds a three dimensional Hilbert space $\Hil$ into the antisymmetric subspace of the tensor product $\Hil \otimes \Hil$ as 
\[
V = \sum_{ijk} \frac{1}{\sqrt{2}}\,\epsilon_{ijk} \ket{ij} \bra{k}
\]
where $\epsilon_{ijk}$ is the totally antisymmetric tensor (in three dimensions), and $\ket{i}$, $i \in {0,1,2}$ are an orthonormal basis of $\Hil$, and $\ket{ij} := \ket{i} \otimes \ket{j}$. If we trace out one of the space $\Hil$ after the action of the isometry, we obtain a channel whose \adjoint is
\[
\begin{split}
\chanh(A) &= V^* (A \otimes \one) V\\
&= \frac{1}{2} \sum_{ijk i'j'k'} \epsilon_{ijk} \epsilon_{i'j'k'} \ket{k}\bra{ij}  (A \otimes \one) \ket{i'j'} \bra{k'}\\
&= \frac{1}{2} \sum_{ijk i' k'} \epsilon_{ijk} \epsilon_{i'jk'} \ket{k}\bra{i} A \ket{i'}\bra{k'} \\
&= \frac{1}{2} \sum_{ik i' k'} (\delta_{ii'} \delta_{kk'} - \delta_{ik'} \delta_{ki'}) \ket{k}\bra{i} A \ket{i'}\bra{k'} \\
&= \frac{1}{2} \sum_{ik} \ket{k}\bra{i} A \ket{i}\bra{k} - \ket{k}\bra{i} A \ket{k}\bra{i}  \\
&= \frac{1}{2} (\tr(A) \one - A^T)\\
\end{split}
\]
where $A^T$ is the transpose of $A$ seen as a matrix with respect to the basis $\ket{i}$. By construction, it is clear that this map is completely positive and unital, and that
\[
\chan = \chan_c
\]
since tracing-out the other destination subsystem yields exactly the same channel. 
Therefore, the whole set of observables preserved by $\chan$ is also preserved by $\chan_c$ and is therefore duplicated. 

Let us suppose that the set of preserved observable is classical, i.e. that there exists an observable $\Gamma$ such that all preserved observables are coarse-grainings of $\Gamma$. We want to reach a contradiction.

First, let us show that the channel $\chan$ is covariant for the whole unitary group $\mathcal U(\Hil)$. 
Indeed, for any unitary $U$ on $\Hil$ we have
\[
\begin{split}
\chanh(U^* A U) &= \frac{1}{2} (\tr(U^* A U) \one - (U^* A U)^T)\\
 &= \frac{1}{2} (\tr(A) \one - U^T A^T \overline U)\\
 &= \frac{1}{2} (\tr(A) \one - U^T A^T \overline U)\\
 &= \overline U^* \chanh(A) \overline U
\end{split}
\]
where $\overline U$ is the complex conjugate of $U$, which is also a unitary operator. In fact, $g \mapsto \overline U_g$ is also a representation of the group. 


As seen in Section \ref{section:covariance}, this implies that there is a covariant pointer observable $\overline \Gamma$, defined by
\[
\overline \Gamma^*(\alpha) = \int U^* \Gamma^*(\alpha_U) U \, dU.
\]
where $\alpha_U(x) := \alpha(U,x)$ for all $U \in \mathcal U(\Hil)$ and $x \in \Omega$, the set in which $\Gamma$ takes value. 
this observable $\overline \Gamma$ is covariant in terms of the representation $\lambda$ on $\mathcal U(\Hil) \times \Omega$ defined by
\[
\lambda_U(V,x) = (VU^*,x)
\]
for any unitary operators $U, V$ and any $x \in \Omega$.


In fact we can build a covariant pointer observable which is more generic that $\overline \Gamma$. First, note that each matrix element $\bra{i} \Gamma^*(\alpha) \ket{j}$ defines a normal linear map from $L^\infty(\Omega)$ to $\mathbb C$, i.e. it can be associated with an element $\mu_{ij}$ of the pre-dual $L^1(\Omega)$ such that
\[
\bra{i} \Gamma^*(\alpha) \ket{j} = \int_{\Omega} \mu_{ij}(U,x) \,\alpha(U,x)\, dx.
\]
For each $x$ we can build the matrix 
\[
\Gamma_x := \sum_{ij} \mu_{ij}(x) \ketbra{i}{j}.
\]
We then have
\[
\Gamma^*(\alpha) = \int_{\Omega} \alpha(x) \Gamma_x \, dx
\]
The operators $\Gamma_x$ are the continuous equivalent of the elements of a discrete POVM. 
If we diagonalize these operators we obtain $\Gamma_x = \sum_i \lambda_i^x \proj{i}_x$ where $\lambda_i^x$ is the eigenvalue of $\Gamma_x$ with eigenstate $\ket{i}_x$. We pick unitary operators $U_i^x$ which are such that $\ket{i}_x = U_i^x \ket{0}$, where $\ket{0}$ is some fixed arbitrary state. 

Recall that given our assumptions, for each observable $X$ there is a stochastic map $\pi$ such that $\chanh(X^*(\alpha)) = \overline \Gamma^*(\pi^*(\alpha))$. Expanding the right-hand side of this equation, we obtain
\[
\begin{split}
\chanh(X^*(\alpha)) &= \int U^* \Gamma^*((\pi^*(\alpha))_U) U \, dU.\\
&= \int (\pi^*(\alpha))(U,x)  U^* \Gamma_x  U \, dU dx.\\
&= \int (\pi^*(\alpha))(U,x)  \sum_i \lambda_i^x U^* U_i^x \proj{0} (U_i^x)^*  U\, dU dx\\
&= \int \sum_i (\pi^*(\alpha))(U_i^x V, x)  \lambda_i^x V^* \proj{0} V\, dV dx\\
&= \int \left[{  \sum_i \int (\pi^*(\alpha))(U_i^x V, x) \lambda_i^x\, dx }\right] V^* \proj{0} V\, dV \\
&= \Delta^*( \Pi^*(\alpha) )
\end{split}
\]
where we have defined the observable 
\[
\Delta : \mathcal B(\Hil) \rightarrow  L^1(\mathcal U(\Hil))
\]
by 
\[
\Delta^*(\alpha) := 3 \int \alpha(U) U^* \proj{0} U \, dU
\]
and the stochastic map (or classical observable) 
\[
\Pi: L^1(\mathcal U(\Hil))  \rightarrow  L^1(\Omega_X)
\]
by
\[
(\Pi^*(\alpha))(U) :=  \frac{1}{3} \sum_i \int (\pi^*(\alpha))(U_i^x U, x)  \lambda_i^x\, dx.
\]



Since all preserved observable must be coarse-grainings of $\Delta$, then in particular, we must have that
\begin{equation}
\label{equ:impossible}
\chanh(\proj{i}) = \frac{1}{2}(\one - \proj{i}) = 3 \int \alpha(U) \, U^* \proj{0} U \, dU.
\end{equation}
for some effect $\alpha$ in $L^\infty(\mathcal U(\Hil))$. But this is not possible. Indeed, we would have
\[
 \int \alpha(U) \, \bra{i} U^* \proj{0} U \ket{i} \, dU = 0
\]
which implies that for all $U$, either $\alpha(U) = 0$ or $\bra{0} U \ket{i} = 0$. 
Hence the integral in Equation \ref{equ:impossible} is supported on the set of unitary operators $U$ which are such that $\bra{0} U \ket{i} = 0$. But this set is of measure zero. This means that we must have $\chanh(\proj{i}) = 0$, which contradicts the hypothesis. Therefore the set of preserved observables in this example is not classical, even though it has been duplicated twice. 



\subsection{Example: symmetric broadcasting}
\label{section:exsymclone}

We consider the optimal fully symmetric quantum \ind{cloning machine} introduced in \cite{gisin97}. 
This is a channel which approximately copies $n$ to $m$ qubits without discrimination in favour of any state, and with optimal fidelity. In the case $n=1$, each individual channel $\chan_k^{(m)}$ resulting from tracing out all destinations qubits but one, is of the form
\[
\chan_k^{(m)}(\rho) = \alpha_m \rho + (1-\alpha_m) \, \tr(\rho)  \frac{1}{2}\one.
\]
In the limit of an infinite number of copies ($m \rightarrow \infty$), the parameter $\alpha$ tends to $\alpha_\infty = \frac{1}{3}$:
\[
\chan_k(\rho) := \chan_k^{(\infty)}(\rho) = \frac{1}{3} \rho + \frac{2}{3} \tr(\rho) \frac{1}{2}\one.
\]
Since these channels are identical for every $k$, the information preserved by one of them is also preserved by all the other channels. Therefore $\mathcal I = \preserved {\chan_k}$ for any $k$. 
Note that $\chan_k^*$ is invertible. This means that we can apply Proposition \ref{prop:invertible}, which states that the observables preserved by $\chan_k$ are those whose effects are all preserved. 


We will show that the preserved effects $\chanh_k(\qprops \Hil)$ are all coarse-grainings of any of the symmetric informationally complete (SIC) POVMs \cite{renes04}. A POVM is said to be informationally complete if its elements $X_i$ span the space of density operators, which implies that its statistics entirely determines the state. Indeed, the component of a state $\rho$ associated with the basis element $X_i$ is given by the scalar product $\tr(\rho X_i)$ which is also the probability for the $i$th outcome in a measurement of $X$. 

For a qubit, such a POVM must have at least four elements. Indeed, the Hermicity condition on states selects a four dimensional subspace of the space of matrices. Note that the condition that the trace be unity, although it does reduce the dimension of the manifold of states to three, does not do so in a linear fashion. An informationally complete POVM is further said to be symmetrical if the scalar product $\tr(X_i X_j)$ is the same for all pairs of distinct POVM elements. In addition, one usually assumes that the elements $X_i$ are of rank one. 

For a qubit, a SIC-POVM $\Gamma$ has elements $\{\Gamma_0, \Gamma_1, \Gamma_2, \Gamma_3\}$ which are proportional to projectors onto four pure states $\ket{\psi_i}$. If we represent these states in the Bloch sphere, i.e. in terms of their traceless components $\lambda^i_j = \tr((\proj{\psi_i} - \frac{1}{2}\one) \sigma_j)$ where $\sigma_j$, $j=1,\dots,3$ are the Pauli operators, these states correspond to the vertices of a regular tetrahedron inscribed in the Bloch sphere. 

The set of effects preserved by $\Gamma$ is 
\[
\Gamma^*(\cprops \Omega) = \{ \sum_{i \in \Omega} \alpha_i \Gamma_i \;|\; 0 \le \alpha_i \le 1\}.
\]
where $\Omega = \{0, 1, 2, 3\}$. 
Let us picture this convex set in the space of Hermitian matrices spanned by the Pauli basis $\{\one, \sigma_1, \sigma_2, \sigma_3\}$. 
Note first that the boundary of the set $\qprops \Hil$ of all the quantum effects is a double cone with tips at $0$ and $\one$. The intersection of the two cones is the Bloch sphere in the 3-sub-manifold of trace-one operators (i.e. with component $\frac{1}{2}$ in the direction specified by the identity $\one$). The subset $\Gamma^*(\cprops \Omega)$ is made of two pyramids whose tips are also $0$ and $\one$ and whose edges are on the surface of the cones and intersect the Bloch sphere where the pure states $\ket{\psi_i}$ are. (This is the four-dimensional version of the $n=3$ case represented in Figure \ref{fig:diamonds}.)

This set contains the set $\chanh_k(\qprops \Hil)$ of effects preserved by $\chanh_k$: 
\[
\chanh_k(\qprops \Hil) \subset \Gamma^*(\cprops \Omega).
\]
Indeed, the action of the channel $\chanh_k$ consists in reducing the radius of the two cones representing $\qprops \Hil$ by a factor one-third, which creates two smaller cones which are precisely inscribed into the two pyramids defining $\Gamma^*(\cprops \Omega)$. This fact follows from knowing that the largest sphere inscribed in a regular tetrahedron (which is the shape of the base of these four-dimensional pyramids) has radius equal to one-third the distance from the center to any of the vertices.

Hence all the effects preserved by $\chan$ are also coarse-grainings of $\Gamma$. 
Since the operators $\Gamma_i$ are linearly independent, $\Gamma^*$ is invertible, which implies that any observable made of effects preserved by $\Gamma$ is also preserved by $\Gamma$. Hence, any observable which is preserved by the channel $\chan_k$ has its effects in $\chanh_k(\qprops \Hil) \subset \Gamma^*(\cprops \Omega)$ and is therefore a coarse-graining of $\Gamma$. 

This shows that for this example, 
\[
\mathcal I \subset \preserved {\Gamma}
\]
for any SIC-POVM $\Gamma$. 
Note that all the SIC-POVMs are given by applying an arbitrary unitary transformation to any given one. 
In this example, the channel $\chan$ is covariant with respect to the whole unitary group. Indeed,
\[
\chan(U \rho U^*) = \frac{1}{3} U \rho U^* + \frac{2}{3} \tr(\rho) \frac{1}{2}U U^* = U \chan(\rho) U^*.
\]
This is the reason why any unitary transformation of our pointer-observable $\Gamma$ is also a pointer observable. 

\subsection{Example: iterated interactions}

Consider an interaction defined by the isometry $V: \Hil \rightarrow \Hil \otimes \Hil_{E}$, where $\Hil_{E}$ denotes a subsystem of the environment, and $\Hil$ is finite-dimensional. This interaction can be iterated in a way which adds a new subsystem of the environment at each iteration. For instance, the third iteration is given by the isometry 
\[
V^{(3)}: \Hil \rightarrow \Hil \otimes \Hil_{E_1} \otimes \Hil_{E_2}\otimes \Hil_{E_3}
\]
where $\Hil_{E_3} \simeq \Hil_{E_2} \simeq  \Hil_{E_1} \equiv \Hil_{E}$, which is defined by
\[
V^{(3)} \ket{\psi} = ((V \otimes \one_{E_2}) V\otimes \one_{E_1}) V \ket{\psi}.
\]
An example would be a system colliding with different particles, one after the other. The system $\Hil_{E_n}$ corresponds to the $n$th particle it has interacted with.

The channel $\chan$ describing the evolution of the system during each interaction is given by
\[
\chanh(A) := V^* (A\otimes \one_{E}) V
\]
and has for complement 
\[
\chanh_c(B) := V^* (\one \otimes B) V.
\]
These two channel can be used to describe the channel 
\[
\chan_k: \mathcal B_t(\Hil) \rightarrow \mathcal B_t(\Hil_{E_k})
\]
from the initial state of the system to the final state of the $k$th particle of the environment:
\[
\chan_k(\rho) = \chan_c(\chan^{k-1}(\rho))
\]
where, by $\chan^{k-1}$ we mean $k-1$ iterations of the channel $\chan$. 
It is clear from the definition that the observables preserved by $\chan^n$ are also preserved by $\chan^m$ whenever $m \le n$. Therefore, the observables preserved by $\chan_m$ are also preserved by $\chan_n$ whenever $m \le n$. This implies that the observables $\mathcal I$ preserved by all the channels $\chan_n$, $k \le n$ are also preserved by $\chan_k = \chan_c \circ \chan^{k-1}$. What is $\mathcal I$ in the limit $k \rightarrow \infty$? 


 Let us assume that $\Hil$ is finite-dimensional. Then we know\cite{kuperberg02} that there is an increasing sequence of integers $\{ k_i \}_{i=1}^\infty \subseteq \mathbb N$ which is such that in the limit $i \rightarrow \infty$, $(\chan^{k_i})^*$ tends to a projective unital completely positive map $\mathcal P$ which projects on the fixed point set ${\rm Fix} \, \chanh$. This shows that the observables preserved by the channels $\chan^k$ for all $k \in \mathbb N$ are those which are fixed by $\chan$, i.e. such that
\[
X = X \circ \chan.
\]
Note that this is a case where the preserved observables are entirely characterized by the preserved effects. 

An effect $A$ which is preserved by all the channels $\chan^k$, $k \in \mathbb N$ must be such that
\[
A = \mathcal P(\chan_c^*(B))
\]
for some effect $B$. This implies in particular that $A \in {\rm Fix} \, \chanh$. 
But what can we deduce about $\chan_c^*(B)$? 

For simplicity, let us suppose that $\chan$ is unital, i.e. $\chan(\one) = \one$. Then we know \cite{holbrook03}, that ${\rm Fix} \,\chanh$ is a $*$-algebra, which we will call $\mathcal A$. 
Clearly it is also trivially a correctable algebra for $\chan$. Therefore we know from Corollary \ref{cor:outgoing} that all the effects in $\chan_c^*( \qprops {\Hil_{E}} )$ must belong to the commutant $\mathcal A'$:
\[
\chan_c^*( \qprops {\Hil_{E}} ) \subset \mathcal A'.
\]


We have seen in Section \ref{section:vnstruct} that the center $\mathcal Z(\mathcal A)$ is characterized by a complete orthogonal family of subspace $\Hil_k$, which we will associate with isometries $V_k$ and projectors $P_k = V_k V_k^*$. A generic element $A \in \mathcal A$ is of the form
\[
A = \sum_k V_k (A_k \otimes \one_k) V_k^*
\]
where $A_k \otimes \one_k$ is an operator on $\Hil_k$. Similarly, an element $B \in \chan_c^*( \qprops {\Hil_{E}} ) \subseteq \mathcal A'$ is of the form
\[
B = \sum_k V_k (\one_k \otimes B_k) V_k^*.
\]

In addition, since $\mathcal P$ fixes $\mathcal A$, its channel elements $E_i$ must all belong to the commutant $\mathcal A'$ \cite{lindblad99}. This can be seen by studying its effects on projectors inside $\mathcal A$, and using Lemma \ref{lemma:trick}. 
This implies that it is of the form
\[
\mathcal P(A) = \sum_{ijk} V_k (\one_k \otimes (E_k^i)^*) V_k^* A V_j (\one_j \otimes E_j^i) V_j^*
\]
Hence we can deduce that an effect $A$ preserved by all channels $\chan_k$ is of the form
\[
\begin{split}
A &= \mathcal P(B) = \sum_{ik} V_k (\one_k \otimes (E_k^i)^*) (\one_k \otimes B_k) (\one_k \otimes E_k^i) V_k^*\\
&= \sum_{ik} V_k (\one_k \otimes (E_k^i)^* B_k E_k^i ) V_k^*\\
\end{split}
\]
where we assumed that $B \in \chan_c^*( \qprops {\Hil_{E}} ) \subset \mathcal A'$. 
This shows that $A \in \mathcal A'$. But we also know that $A \in \mathcal A$ since it is in the image of $\mathcal P$. 
Hence $A \in \mathcal A \cap \mathcal A' = \mathcal Z(\mathcal A)$. 

Therefore, in the case where $\chan$ is unital, the observables in $\mathcal I = \bigcap_k \preserved {\chan_k}$ all belong to a commutative algebra: the center of the algebra $\mathcal A = {\rm Fix} \, \chanh$. 




\msection{Discussion}
\label{section:discussion}

Let us summarize the picture of decoherence suggested by Theorem \ref{thm:fundamental}. We considered a quantum system undergoing an open evolution, i.e. interacting with its environment. Nothing forces the information preserved in the system to be classical, unless the evolution is a ``fully decoherent'' channel of the form studied in Section \ref{section:sysdecoh}. However, we have seen that the information which is broadcast redundantly to the environment is characterized by a classical set of observables $\mathcal I \subseteq \preserved \Gamma$.  
This process is the physical realization of a measurement of the pointer observable $\Gamma$ on the system by the environment. 
The redundancy guarantees the objectivity of the information stored in the environment, as argued in \cite{ollivier04}. Furthermore, if this information is also preserved in the system itself during the interaction, then the information contained in the environment is correlated with that contained in the final state of the system (even if it evolved). Therefore this information has predictive power and can characterize the state of a deterministic effective classical model of the system. 

In this picture, the pointer observable can be defined even if it does not represent information which is preserved in the system. In this case, the emergent classical degrees of freedom represent a classical system subject to noise. Alternatively, the process of decoherence in this situation can be seen as a destructive or partially destructive measurement. 

Note that if $\mathcal I = \preserved {\Gamma}$ this process represents an exact measurement of $\Gamma$ by the environment, whereas $\mathcal I \subsetneq \preserved {\Gamma}$ implies that the environment does not gain full information about $\Gamma$. In this case, $\Gamma$ represents only an approximate description of the information present in the environment. This also means that there exists an ambiguity in the pointer observable due to the arbitrariness of the extra information contained in $\preserved \Gamma$.

Finally, we observed that even if the measurement of $\Gamma$ by the environment is complete, the corresponding classical limit may have some forbidden states. 
Indeed, if $\Gamma$ represents an approximate simultaneous measurement of non-commuting observables, the classical states in its image are constrained to respect the corresponding uncertainty relations. However we expect these quantum constraints on the effective classical states to be irrelevant for phase-space observations coarser than $\hbar$. 







\subsection{Universality of classicality}



The main advantage of this picture is that it relies on very little physical assumptions. For instance, no particular model of interaction has been assumed. 
We did however, make one rather unusual assumption which requires some comments, namely 
the idea that only redundant information matters at macroscopic scales.
This idea seems intuitively reasonable as redundancy of information evokes the amplification of a signal. In addition, we cannot think of any classical degree of freedom which is not represented redundantly in a way or another, given that large systems are always made of many particles which share the essential properties of the object. As an example, we can think of the position of a rigid body. Certainly, knowing the position of any of its molecule would suffice to localize it. 

In fact, since what it means to be macroscopic has never been defined unambiguously, it is tempting to think of this assumption as a mere definition. Indeed, we could say that \ind{macroscopic physics} is that which deals with highly redundant information. In this view, that which is ``macro-'' in ``macroscopic'', i.e. large, is the number of equivalent representations of the same information. If we accept this postulate, then Theorem \ref{thm:fundamental} shows that macroscopic information is always classical, hence making the universality of classicality at macroscopic scales a tautology. A problem with this view is that the our notion of redundancy relies entirely on a preferred decomposition of the environment into subsystems. For our interpretation to make sense, the decomposition should be such that a classical observer is only able to measure observables which are {\em local} with respect to this decomposition. 


This view stands in contrast to the standard approach on this question, which consists in the study of realistic models under a class of assumptions as broad as possible, with the aim of showing that they are indeed fully decoherent, in the sense that their evolution destroys all but classical information \cite{braun01}.  However such studies have a point: decoherence is omnipresent even in microscopic experiments, which is why building a quantum computer is so difficult.

These two views may be reconciled if we consider that quantum experiments are about the control of a microscopic system by a macroscopic one, which is by definition subject to decoherence. Macroscopic systems are those which are extremely efficient at spreading information. It is therefore no surprise that when put in contact with quantum systems, classical instruments cause a decoherence which may otherwise be absent.

\subsection{Outlook: dynamics and constraints}



To conclude, we will say a few words about the possibility of integrating dynamics into our model of decoherence, which is an open question. 


Let us consider the ``broadcasting'' channel $\chan$ defined by Equation \ref{equ:broadchan}. As already explained in the beginning of Section \ref{section:discussion}, in order to obtain a dynamical effective classical theory, we need to assume that the information represented by the set $\mathcal I$ is also preserved in the system. We will suppose that the evolution of the system itself is given by the channel $\chan_A$. 
This channel represents only the relation between two specific moments in time. However, if $\chan$ and $\chan_A$ are obtained from a unitary interaction generated by a Hamiltonian, they naturally depend on the time at which we decide to trace out the system or the environment respectively. In fact, we need not trace-out both at the same time. For instance, we could use the channel $\chan$ as it is at asymptotic times, assuming that the environment permanently keeps a record of the relevant information about the system.

Recall that the channel $\chan$ to the environment selects a classical model characterized by the set of observables $\mathcal I$, or its approximate representation given by the pointer observable $\Gamma$ which is such that $\preserved \Gamma \supseteq \mathcal I$.
We suppose that, in addition, $\chan_A(t)$ preserves $\mathcal I$ for all time $t$.
This means that for any observable $X \in \mathcal I$, there exists a family of observables $Y(t)$ which are such that $X = Y(t) \circ \chan_A(t)$. The time-dependant observable $Y(t)$ represents how the information about the initial property $X$ evolves in time. 
A problem is that in general, $Y(t)$ has no reason to belong to the set $\mathcal I$ or to its idealization $\preserved \Gamma$. Therefore, the map $X \mapsto Y(t)$ cannot be understood as a ``classical'' transformation within $\mathcal I$ in any straitforward manner. This indicates that, in order to be able to compare observables at different times and identify an effective classical dynamics, some extra assumptions or inputs are needed.

In principle, what we expect is that the environment is continuously monitoring the same observables of the system. However it is not easy to see how to model this phenomenon in our framework. Indeed, the time-dependant channel $\chan(t)$ from the system to the environment only describes the information that the environment at time $t$ contains about the initial state of the system. Of course we could instead consider discrete successive measurement of the system by different parts of the environment, which should in fact be close the what really happens. However, this would seem to remove one of the advantages of our picture, namely its ability to describe a continuous decoherence process.

We will leave these questions open for further work. For now, let us introduce another possible approach, which consists in considering time as a variable of the system with no special status.

There are two ways of interpreting the phase-space $\Omega$ of a classical system. It can be seen as representing the state of the system at a given time, but it can also be understood as representing all physically allowed histories of that classical system. Indeed, for simple unconstrained systems, the elements of $\Omega$ are one-to-one with possible time histories. If, in addition, the system described by the phase-space $\Omega$ contains a physical {\em clock}, then the external time becomes superfluous and can be considered as gauge.



The same should be true in quantum mechanics. A quantum state $\ket{\psi} \in \Hil$ can be understood as describing our general knowledge of the system, without any assumption about time. This is particularly apparent in the Heisenberg picture, in which time simply labels the possible observations. 
This point of view yields a different light on a classical limit of the form
\(
\Gamma: \mathcal B_t(\Hil) \rightarrow L^1(\Omega),
\) 
which now maps quantum histories to classical histories. 

Let us therefore suppose that we have such a ``timeless'' quantum system described by the  Hilbert space $\Hil$. Since our decoherence framework does not make any explicit mention of time, we can still consider an environment $B$ with subsystems $B_1, B_2, \dots$  and a channel
\[
\chan: \mathcal B_t(\Hil) \rightarrow \mathcal B_t(\Hil_{B_1} \otimes \Hil_{B_2} \otimes \dots)
\]
which represents how information about $\Hil$ happens to be contained in the environment, and expect $\Gamma$ to emerge as described in Section \ref{section:fulldecoh}.

In a sense, the map $\Gamma$ directly takes into account the dynamics. Note however that the classical set of histories $\Omega$ may need to be larger than what we really expect. Remember that in our picture the limit $\hbar \rightarrow 0$ is taken only at a later stage, when the constraints on the classical states $L^1(\Omega)$ imposed by the quantum uncertainty relations are ignored. In the case where the limit of whole quantum histories is taken into account, it is likely that the classical limit that we get for finite $\hbar$, which is characterized by $\Gamma$, yields mixtures of trajectories which differ by powers of $\hbar$, and hence explore paths which are not characterized by the ideal classical phase-space. 

There is another problem with this picture, which is that, for realistic interacting theories, the structure of the set of physical solutions is never fully understood. Therefore we do not know the space $\Hil$ to start with. In fact, this is what equations of motions, or more generally field equations, are about: they allow one to express local properties of the physical system without having to understand its global behaviour. Equations of motion are local in time, while full field equations are also local in space. 

Note also that field equations do not only describe the evolution of the field in time, but may also describe spacial constraints that the field must satisfy. In fact we can view the equations of motions themselves as constraint on possible histories, which allows to unify the picture. 


This fact is especially vivid in the case of general relativity, where no physical notion of time exists prior to solving all of the equations which govern the theory. In this sense, if by ``dynamics'' we refer to the process of solving the equations of motion, then the dynamics of general relativity has nothing to do with time. Instead, one start from the set of all ``kinematical'' histories, defined by all possible metrics (and matter fields) on a fixed spacetime manifold, and then solves Einstein's equations in order to obtain the ``physical'' histories. Einstein's equations can be interpreted as specifying a constraint on the kinematical histories. In fact it is not difficult to formulate any classical theory in a similar way, given that the equations of motions can always be interpreted as specifying some constraint on general histories of the usual phase-space variables. 

Instead of giving a classical example, let us directly show that the same can be done in quantum mechanics. We will follow Rovelli \cite{rovelli90, reisenberger02} and show how the dynamics of a non-relativistic quantum particle can be formulated as a constraint, implemented as a projector on a ``kinematical'' extended Hilbert space. 
Consider a single, one-dimensional, non-relativistic particle with standard Hamiltonian 
\[
H = \frac{\widehat p^2}{2m} + \phi(\widehat x).
\]
In the position representation, the time-evolution of a particle with initial state $\psi_0 \in \Hil = L^2(\mathbb R)$ is given by Schr\"odinger's equation
\[
i\hbar \frac{\partial \psi_t(x)}{\partial t} = - \frac{\hbar^2}{2m} \frac{\partial^2 \psi_t(x)}{\partial x^2} + \phi(x) \psi_t(x).
\]
This equation can be reinterpreted as an equation on $\psi \in \Hil_{K} := L^2(\mathbb R^2)$:
\[
i\hbar \frac{\partial \psi(x,t)}{\partial t} + \frac{\hbar^2}{2m} \frac{\partial^2 \psi(x,t)}{\partial x^2} + \phi(x) \psi(x,t) = 0.
\]
What this is saying is that $\psi$ must belong to the kernel of an operator $\widehat C$ on $L^2(\mathbb R^2)$ which is defined by
\[
\widehat C := \widehat E - \frac{\widehat p^2}{2m} - \phi(\widehat x).
\]
where $\widehat E := i\hbar \frac{\partial}{\partial t}$ and $\hat p := i\hbar \frac{\partial}{\partial x}$. 
Technically, although $\widehat C$ has $0$ in its spectrum, its kernel is empty because the solutions to Schr\"odinger's equation cannot be normalized in $L^2(\mathbb R^2)$. Indeed, we know that it preserves probabilities in time, which means that the square of $|\psi(x,t)|$ integrated over $x$ is the same at any fixed time $t$, which implies that the integral over $t$ will diverge. There are various ways to go around this problem. Here, let us simply suppose that we are working within a finite interval of time. Therefore, we redefine $\Hil_{K}$ as 
\[
\Hil_{K} = L^2(\mathbb R \times [0,T]).
\]
This does not change anything about the properties of Schr\"odinger's equation, but it has the advantage of putting all the (time-limited) solutions inside $\Hil_{K}$. The drawback is that $\widehat C$ here is not self-adjoint, however it does have a kernel which is all we need from it. Let $P$ be the projector on the kernel of $\widehat C$ inside $\Hil_{K}$. There is an isometry $V: \Hil \rightarrow \Hil_{K}$ such that $V^* V = \one_\Hil$, and $V V^* = P$. Indeed, we know from the properties of Schr\"odinger's equation that the space of solutions is isomorphic to $\Hil = L^2(\mathbb R)$. The map $V^*$ represents the ``effect'' of solving Schr\"odinger's equation: it projects on the space of solutions. 

Let us give an example of a possible classical limit for this dynamical system. To this construction, we add $\Gamma_0$, the coherent-state observable on $\Hil = L^2(\mathbb R)$ introduced in Section \ref{section:limit}. 
If $\Gamma_0$ really defines the classical limit of the theory, we may hope that it indirectly maps the quantum evolution defined by the isometry $V$ to an effective classical evolution. 
Let us write $\mathcal V$ for the channel defined by $V$, i.e. $\mathcal V(\rho) := V \rho V^*$ for a state $\rho$ of $\Hil$. This channel $\mathcal V$ can be understood as mapping our knowledge of the initial state into a knowledge of the whole history of the particle. 

In order to obtain a proposition for the classical limit of the dynamical system, we first need to make some assumption about the classical limit of time itself. 
Note that we introduced time as a quantum observable which commutes with both the position and momentum operators. Indeed, time is a subsystem of its own since $\Hil_{K} = L^2(\mathbb R) \otimes L^2([0,T])$. 
Let $\Gamma_T$ be the map which sends a state $\rho$ of the time subsystem $L^2([0,T])$ into the classical time state $\mu(t) = \bra{t} \rho \ket{t}$, where $\ket{t}$ represents the formal eigenstates of the time operator $\widehat T$. Overall we obtain an extended classical limit $\Gamma_{K} := \Gamma_0 \otimes \Gamma_T$, which operates between quantum and classical {\it histories}:
\[
\Gamma_{K} : \mathcal B_t(\Hil_{K}) \rightarrow L^1(\mathbb R^2 \times [0,T])
\]
It simply sends a state $\rho$ of $\Hil_{K}$ to the probability distribution
\begin{equation}
\label{equ:history}
\mu(q,p,t) = (\bra{q,p} \otimes \bra{t})\, \rho \,(\ket{q,p} \otimes \ket{t}) 
\end{equation}
where $\ket{q,p}$ are coherent states. 
This map alone 
has nothing to do with the dynamics contained in the isometry $\mathcal V$. However suppose that the quantum states that we use actually satisfy Schr\"odinger's equation, i.e. $\rho = V \rho_0 V^*$, where $\rho_0$ is a state of $\Hil$. The resulting map 
\[
\Gamma := \Gamma_{K} \circ \mathcal V
\]
 does not directly give us a distribution over the classical phase-space, but instead a probability distribution over classical histories:
\[
\Gamma: \mathcal B_t(\Hil) \rightarrow L^1(\mathbb R^2 \times [0,T]).
\]

Note that, although this map $\Gamma$ is technically a valid abstract classical limit, the resulting classical system is not deterministic in terms of the variable $t \in [0,T]$. 
This is due to the fact that this picture assumes a process of decoherence which does not interfere with the exact quantum evolution of the system. It amounts to applying the map $\Gamma_0$ on the state of an evolving quantum system independently at each time, therefore erasing the information which would be required to predict the evolution of an initial state. This shows that this cannot be a realistic model of decoherence for this particular system. 

\newpage

\pagestyle{plain}

\renewcommand{\bibname}{References \phantomsection\addcontentsline{toc}{chapter}{\textbf{References}}}


\bibliographystyle{utphys} 


\bibliography{pig} 



\newpage
\phantomsection
\newpage
\markboth{Index}{}
\addcontentsline{toc}{chapter}{\textbf{Index}}
\printindex  

\end{document}